%% file: ap1006.tex
\documentclass{llncs}

\newcommand{\suc}{suc}

\newcommand{\bi}{\begin{itemize}}
\newcommand{\ei}{\end{itemize}}
\newcommand{\be}{\begin{enumerate}}
\newcommand{\ee}{\end{enumerate}}
\newcommand{\bd}{\begin{description}}
\newcommand{\ed}{\end{description}}
\newcommand{\tab}{\hspace*{1cm}}
\newcommand{\mc}[2]{\multicolumn{#1}{l}{#2}}
\newcommand{\mca}[2]{\multicolumn{#1}{@{}l@{}}{#2}}

\newcommand{\1}[1]{\!#1\!}

\newcommand{\sortdef}{\doteq}

\newcommand{\tpl}[1]{\langle #1 \rangle}
\newcommand{\COMMENT}[1]{}

\newcommand{\ra}{\rightarrow}
\newcommand{\la}{\leftarrow}
\newcommand{\lra}{\leftrightarrow}
\newcommand{\raa}{\longrightarrow}

\newcommand{\Ra}{\Rightarrow}
\newcommand{\La}{\Leftarrow}
\newcommand{\Lra}{\Leftrightarrow}

\newcommand{\deref}{\uparrow}

\newcommand{\app}{+\!\!\!\!+}

\newcommand{\N}{I\!\!N}
\renewcommand{\phi}{\varphi}
\newcommand{\bigmid}{\rule[-0.07cm]{0.04cm}{0.4cm}\hspace*{0.1cm}}

\renewcommand{\leq}[0]{\leqslant}
\renewcommand{\geq}[0]{\geqslant}
\newcommand{\abs}[1]{\mathop{\#} #1}

\newlength{\outerparindent}
\newcommand{\T}{{\cal T}}
\newcommand{\F}{{\cal F}}
\newcommand{\V}{{\cal V}}

\renewcommand{\S}{{\cal S}}
\renewcommand{\L}{{\cal L}}
\renewcommand{\O}{{\cal O}}
\newcommand{\NF}{{\cal N}}
\newcommand{\jTH}{{\cal TH}}
\newcommand{\TY}{{\cal TY}}
\newcommand{\A}{{\cal A}}

\newcommand{\eqc}[1]{{\underline{#1}}}

\spnewtheorem{algorithm}{Algorithm}[theorem]{\bfseries}{\itshape}
\newcommand{\rulename}[1]{({\em #1\/})}

\usepackage{amssymb}

\newcommand{\eqr}[1]{\ref{#1}}
\newcommand{\eqR}[1]{#1}
\newcommand{\eqd}[1]{\label{#1}}

\renewcommand{\arraystretch}{1.1}
\renewcommand{\baselinestretch}{1.1}

\newlength{\thmspace}
\newcounter{thm}

\newcommand{\THM}[2]{%
	\par%
	\vspace{\thmspace}%
	\refstepcounter{thm}%
	{#1~\thethm. }%
	#2%
	\par%
	\vspace{\thmspace}%
	}

\newcommand{\PRF}[2]{%
	\par%
	{#1}%
	#2%
	\par%
	\vspace{\thmspace}%
	}

\newcommand{\EXAMPLE}[1]{\THM{\bf Example}{#1}}
\newcommand{\ALGORITHM}[1]{\THM{\bf Algorithm}{#1}}
\newcommand{\COROLLARY}[1]{\THM{\bf Corollary}{#1}}
\newcommand{\LEMMA}[1]{\THM{\bf Lemma}{#1}}
\newcommand{\THEOREM}[1]{\THM{\bf Theorem}{#1}}
\newcommand{\DEFINITION}[1]{\THM{\bf Definition}{#1}}
\newcommand{\PROOF}[1]{\PRF{\it Proof. }{#1}}

\begin{document}

\titlepage

\title{Implementing Anti--Unification Modulo Equational Theory
}

\author{Jochen Burghardt\inst{1} and Birgit Heinz\inst{2}}

\institute{GMD Berlin, jochen@first.gmd.de,
http://www.first.gmd.de/persons/Burghardt.Jochen.html
\and
TU Berlin, heinz@cs.tu-berlin.de,
http://www.cs.tu-berlin.de/\~heinz
}

\maketitle

\vfill

{\sf

\begin{tabular}[t]{@{}ll@{\hspace*{0.5cm}}l@{}}
Technical Report	\\
Arbeitspapiere der GMD 1006	\\
June 1996	\\
ISSN 0723--0508	\\
\end{tabular}
\hfill
\begin{tabular}[t]{@{}ll@{\hspace*{0.5cm}}l@{}}
\mca{2}{GMD -- Forschungszentrum}	\\
\mca{2}{Informationstechnik GmbH}	\\
\mca{2}{D--53754 Sankt Augustin}	\\
Tel. & *49--2241--14--0	\\
Fax & *49--2241--14--2618	\\
Telex & 889469 gmd d	\\
\mca{2}{http://www.gmd.de}	\\
\end{tabular}

}

\newpage

\setcounter{page}{3}

\begin{abstract}
We present an implementation of E--anti--unification as defined
in \cite{Heinz.1995}, where tree--grammar descriptions of
equivalence classes of terms are used to compute generalizations
modulo equational theories. 
We discuss several improvements, including an efficient
implementation of variable--restricted E--anti--unification from
\cite{Heinz.1995}, and give some runtime figures about them. 

We present applications in various areas, including lemma
generation in equational inductive proofs, intelligence tests,
diverging Knuth--Bendix completion,
strengthening of induction hypotheses,
and theory formation about finite algebras.
\end{abstract}

%{
%
%%\renewcommand{\baselinestretch}{0.9}
%\small
%
%\clearpage
%
%\tableofcontents
%
%}
%
%\clearpage
%
%\listoffigures

\clearpage
\section{Introduction}
\label{Introduction}

An important task in the field of artificial
intelligence is generalization.
To the extent that
a generalization approach allows us to incorporate certain
background knowledge, it opens up applications in various fields of
computer science.
Inductive logic programming, for example, is concerned with
generalization wrt.\ Horn--logic theories, with potential
applications in automated scientific discovery, knowledge
discovery in databases, automatic programming, and other areas. 

\cite{Heinz.1995} gives several algorithms for generalizing terms wrt.\
equational background theories; they are applied to lemma
generation in equational induction proofs.
In this paper, we describe the implementation of these algorithms and
give some further applications.

As a motivating example,
consider the sequence $0$, $1$, $4$, $9$, represented by terms
$0$, $\suc(0)$, $\suc^4(0)$, $\suc^9(0)$, respectively. 
Trained people with a knowledge of addition and multiplication
of natural numbers will easily recognize that square numbers
computable by the scheme $x*x$ are the desired solution for
finding the continuation of the numbers. However, syntactical
anti--unification merely leads to the term $y$, which is too
general to be used for computing a continuation of the above
sequence. 

A useful mechanism for computing schemes is anti--unification
extended to take account of equational theories. Addition and
multiplication of natural numbers can be specified by the
equations of Theory (1) in Fig.~\ref{Background Equational
Theories Used}. 
In Fig.~\ref{Anti--Unification Modulo Equational Theory (1) from
Fig.}, we demonstrate that $x*x$ is one of the generalizations
resulting from anti--unification of the above terms $0$, 
$\suc(0)$, $\suc^4(0)$, $\suc^9(0)$ modulo equational Theory (1),
since each of them equals an instance of $x*x$.

\begin{figure}
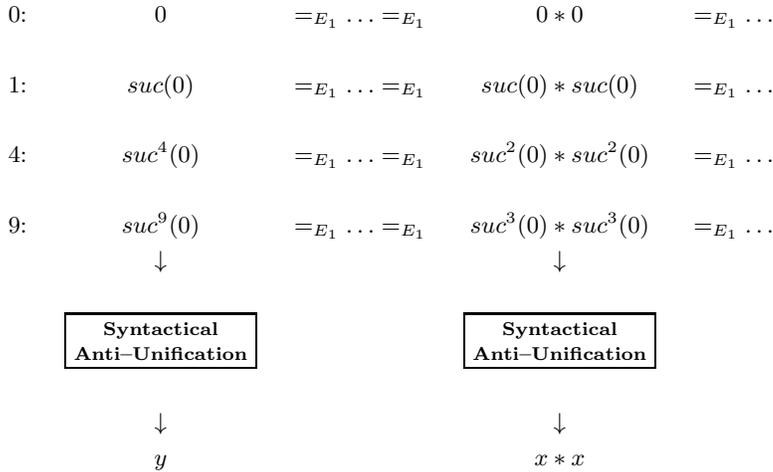

\begin{center}
\begin{tabular}{@{}r@{\hspace*{0.5cm}}c@{\hspace*{0.5cm}}
	c@{\hspace*{0.5cm}}c@{\hspace*{0.5cm}}l@{}}
$0$: & $0$ & $=_{E_1} \ldots =_{E_1}$ & $0*0$ & $=_{E_1} \ldots$ \\
\\
$1$: & $\suc(0)$ & $=_{E_1} \ldots =_{E_1}$
	& $\suc(0)*\suc(0)$ & $=_{E_1} \ldots$ \\
\\
$4$: & $\suc^4(0)$ & $=_{E_1} \ldots =_{E_1}$
	& $\suc^2(0)*\suc^2(0)$ & $=_{E_1} \ldots$	\\
\\
$9$: & $\suc^9(0)$ & $=_{E_1} \ldots =_{E_1}$
	& $\suc^3(0)*\suc^3(0)$ & $=_{E_1} \ldots$	\\
& $\downarrow$ && $\downarrow$	\\
\\
& \fbox{\scriptsize\bf \shortstack{Syntactical\\Anti--Unification}}
	&& \fbox{\scriptsize\bf
	\shortstack{Syntactical\\Anti--Unification}} \\
\\
& $\downarrow$ && $\downarrow$ \\
& $y$ && $x*x$	\\
\end{tabular}
\end{center}

\caption{Anti--Unification Modulo Equational Theory (1) from Fig.\
	\protect\ref{Background Equational Theories Used}}
\label{Anti--Unification Modulo Equational Theory (1) from Fig.}
\end{figure}

Some problems arise if generalizations of terms wrt.\  equational
theories have to be computed:

\begin{itemize}
\item There may exist many generalizations allowing the elements of a
	sequence of terms to be computed.
\item The set of generalizations wrt.\ equational theory is usually
	infinite (see, however, Fig.~\ref{Some Enumerated Series Laws}
	in Sect.~\ref{Series Guessing -- Results and Runtimes}).
\item Hence, only approaches enumerating its elements can be provided.
\item Depending on the application area, the set of possible
	generalizations may contain useless or undesired computation
	schemes.
\end{itemize}

In \cite{Heinz.1995}, an approach to generalization modulo canonical
equational theories, called anti--narrowing, has been developed. This
approach simply allows all generalizations of terms modulo canonical
theories to be enumerated; useless generalizations can only be
eliminated {\em after} their computation using corresponding criteria.

Instead of enumerating all generalizations of terms modulo equational
theory, a compact, finite representation of the set of all
generalizations is desirable for several reasons, e.g.\ to enable
all useless generalizations to be eliminated at one go.

\cite{Heinz.1995} therefore investigates
a second approach to E--anti--unification based on regular tree
grammars\footnote{Called ``sorts'' in this paper, following
	\cite{Comon.1990}}.
At the heart of the approach is an algorithm that takes two sorts
or regular tree languages $s,s'$,
and computes a sort or regular tree language
$\theta$ of all syntactical generalizations of
terms from $s$ and $s'$:

$$\L(\theta) = \{ t \sqcap t' \mid t \in \L(s), t' \in \L(s') \} .$$

In order to E--anti--unify two terms modulo a given background
equational theory $E$, their equivalence classes modulo $E$ are
represented by sorts $s,s'$; the corresponding $\theta$ then
contains all E--generalizations of both terms. 

In this paper, we describe the implementation of the sort approach to
E--anti--unification including technical optimizations and run-time
measurements.
We present several applications:
\bi
\item generation of lemma candidates in blocked situations of an
	inductive proof (Sect.~\ref{Lemma Generation})
\item construction of series--formation laws
	(Sect.~\ref{Series Guessing})
\item some other potential applications, which are only sketched
	(Sect.~\ref{Other Potential Applications}).
\ei

Section~\ref{Definitions and Notations} introduces some
necessary formal definitions and notations.
Section~\ref{E--Anti--Unification} presents the implementation
of E--Anti--Unification. Section~\ref{Sort Enumeration}
discusses how to enumerate the terms of a computed sort.
Section~\ref{An E--Anti--Unification Example} gives an elaborate
example of E--anti--unification. Sections~\ref{Lemma Generation}
to~\ref{Other Potential Applications} present the
above--mentioned applications. 
In Sect.~\ref{Equational Theories
of Finite Algebras},
we show that for each finite algebra we can always generate a closed
representation of all its quantifier--free and variable--bounded
theorems.
The PROLOG source code of our implementation is listed in
Appendix \ref{PROLOG Source Code}.

\clearpage
\section{Definitions and Notations}
\label{Definitions and Notations}

\DEFINITION{
Let $\V$ be an enumerable set of variables,
$\F$ a finite set of function symbols, each with fixed arity;
$ar(f)$ denotes the arity of $f \in \F$.
Terms are built from $\V \cup \F$;
$\T$ denotes the set of terms.

We assume familiarity with the classic definitions and notations of
terms.
We say that a term $t$ starts with a function symbol $f$
if $t = f(t_1,\ldots,t_n)$ for some terms $t_1,\ldots,t_n$.
$vars(t)$ describes the set of variables occurring in a term
$t \in \T$.
A term is called linear if no variable occurs in it twice.
A term $t$ is called a ground term if $t$ does not contain any
variables.
}

\DEFINITION{
Lists are built from $[\;]$ and $cons$, written as $(.)$ as in PROLOG;
$length(L)$ denotes the number of elements of the list $L$;
$head(L)$ denotes its first element.
The list--comprehension notation
$[t \mid p(t)]$
denotes a list of all $t$, such that $p(t)$ holds,
in some arbitrary order.

$\tpl{t_1,\ldots,t_n}$ denotes an $n$-tuple;
$\pi_i^n(\tpl{t_1,\ldots,t_n}) := t_i$
denotes its $i$-th projection.

$:=$ means equal by definition.
$A \subset B$ means that $A$ is a subset of $B$, or $A$ is equal to
$B$.

$\abs{A}$ denotes the number of elements of a finite set $A$.
$\phi[A] := \{ \phi(a) \mid a \in A\}$ denotes the image of the set $A$
under the mapping $\phi$;
$\phi^{-1}[A] := \{ a \mid \phi(a) \in A \}$
denotes the inverse image of $A$ under $\phi$.
For a set $A$, we denote its $n$-fold cartesian product by $A^n$.
}

\DEFINITION{
Substitutions are defined as usual;
$\{x_1 \la t_1,\ldots,x_n \la t_n\}$ denotes a substitution that
maps each variable $x_i$ to the term $t_i$.
Sometimes we also use a set--comprehension--like notation:
$\{ x \la t \mid p(x,t)\}$ denotes a substitution that maps each $x$ to
$t$ such that $p(x,t)$ holds.
$dom(\sigma) := \{ x \in \V \mid \sigma x \neq x \}$
denotes the domain of $\sigma$.

$\sigma$ is called a linear substitution
if $\sigma \tpl{x_1,\ldots,x_n}$ is a linear term,
where $dom(\sigma) = \{x_1,\ldots,x_n\}$.
$\sigma$ is called a flat substitution
if $\sigma x \in \V$ for all $x \in \V$.
$\sigma$ is called a renaming substitution
if it is a bijective mapping from variables to variables,
and hence is both linear and flat.

$t$ is called an instance of $t'$,
and $t'$ an anti--instance of $t$,
if a substitution $\sigma$ exists such that $t = \sigma t'$.
If $\sigma$ is a renaming, $t$ is also called a variant of $t'$.
}

\DEFINITION{
An equational theory $E$ is a finite set of equations $t_l = t_r$.
The relation $(=_E)$ is defined as usual as the smallest
reflexive, symmetric, and transitive rewrite relation
that contains $E$.

The equivalence class of a term $t$ mod.\ $(=_E)$ is denoted
by $\eqc{t} := \{ t' \in \T \mid t' =_E t\}$.
$\T /_E$ denotes the algebra of equivalence classes.
We assume that in each equivalence class $\eqc{t}$,
some term $nf(t)$ is distinguished, which we call the normal form of
$t$.
Let $\NF := nf[\T]$ be the set of all normal forms.
$\eqc{t}$, $nf(t)$, and $\NF$ depend on $E$.

Note that we do not require $(=_E)$ to be computable by a confluent
and noetherian term--rewriting system.
However, we require that each equivalence class $\eqc{t}$ be a
regular tree language.
}

\DEFINITION{
Let $\S_{NAME}$ be an enumerable set of sort names, and
let the sets $\V$, $\F$, and $\S_{NAME}$ be pairwise disjoint.
Let $\S$ denote the set of all sort expressions;
a sort expression is a sort name or has one of the forms
\bi
\item[] $s_1 \mid s_2$, where $s_1, s_2 \in \S$;
\item[] $f(s_1,\ldots,s_n)$, where $f \in \F$, $s_1,\ldots,s_n \in \S$;
	or
\item[] $x$ for $x \in \V$.
\ei
A sort definition is of the form $s_n \sortdef s_e$,
where $s_n$ is a sort name and $s_e$ is a sort expression.
We say that $s_n$ is defined by $s_e$.

Given a system of sort definitions where each occurring sort name
has exactly one definition, we define their semantics as their least
fixed point.
We denote the semantics of a sort expression $s$ by $\L(s) \subset \T$,
which has the following properties:

$$
\begin{tabular}[t]{@{}rl@{\hspace*{0.5cm}}l@{}}
$\L(s_1 \mid s_2)$ & $= \L(s_1) \cup \L(s_2)$	\\
$\L(x)$ & $= \{x\}$ & for $x \in \V$	\\
$\L(f(s_1,\ldots,s_n))$ & $= \{f(t_1,\ldots,t_n) \mid t_i \in \L(s_i)\}$
	& for $f \in \F$	\\
$\L(s_n)$ & $= \L(s_e)$ & for $s_n \sortdef s_e$	\\
\end{tabular}
$$

Observe that $\T \subset \S$, and $\L(t) = \{t\}$
for each $t \in \T$.
The empty sort is denoted by $\bot$.
A sort definition of the form
$s \sortdef
f_1(s_{11},\ldots,s_{1n_1}) \mid \ldots \mid
f_m(s_{m1},\ldots,s_{mn_m})$
is said to be in head normal form,
where $s,s_{ij} \in \S_{NAME}$,
$f_i \in \F \cup \V$, and $n_i = ar(f_i)$.
For proof--technical reasons, we define a system of sort definitions
as being in normal form if each sort definition has either the form
\bi
\item $s \sortdef s_1 \mid \ldots \mid s_n$ with
	$s, s_1,\ldots,s_n \in \S_{NAME}$, or
\item $s \sortdef f(s_1,\ldots,s_n)$ with
	$s, s_1,\ldots,s_n \in \S_{NAME}$,
\ei
and if no cycles

$$
\begin{tabular}[t]{@{}crcl@{}}
$s_1$ & $\sortdef \ldots \mid$ & $s_2$ & $\mid \ldots$	\\
$s_2$ & $\sortdef \ldots \mid$ & $s_3$ & $\mid \ldots$	\\
$\ldots$	\\
$s_{n-1}$ & $\sortdef \ldots \mid$ & $s_n$ & $\mid \ldots$	\\
$s_n$ & $\sortdef \ldots \mid$ & $s_1$ & $\mid \ldots$	\\
\end{tabular}
$$

occur.
Each system of sort definitions can be transformed into head normal
form, and into normal form, maintaining its semantics.
Any system of sort definitions
in head normal form, or in normal form, has
exactly one fixed point.
}

\THEOREM{
\eqd{ind}
Assume a system of sort definitions in normal form.
Let $p$ be a family of unary predicates, indexed over the set of all
defined sort names.
Show for each defined sort name $s$:

	$$
	\begin{tabular}[t]{@{}ll@{\hspace*{0.5cm}}l@{}}
	$\forall t \in \T \;\; p_s(t) \lra$
		& $p_{s_1}(t) \vee \ldots \vee p_{s_n}(t)$
		& if $s \sortdef s_1 \mid \ldots \mid s_n$ \\
	$\forall t \in \T \;\; p_s(t) \lra$
		& $\exists t_1,\ldots,t_n \in \T \;\;
		t = f(t_1,\ldots,t_n)
		\wedge p_{s_1}(t_1) \wedge \ldots \wedge
		p_{s_n}(t_n)$
		& if $s \sortdef f(s_1,\ldots,s_n)$	\\
	$\forall t \in \T \;\; p_s(t) \lra$
		& $t = x$
		& if $s \sortdef x$	\\
	\end{tabular}
	$$

Then, $\forall t \in \T \;\; t \in \L(s) \lra p_s(t)$
holds for each defined sort name $s$.
}

\DEFINITION{
A term $t$ is called a generalization of $t_1$ and $t_2$
iff there exist substitutions $\sigma_1$ and $\sigma_2$
such that $\sigma_1 t = t_1$ and $\sigma_2 t = t_2$.
$t$ is called the most specific generalization
iff each generalization $t'$ of $t_1$ and $t_2$
is an anti--instance of $t$.
The most specific generalization $t$ of two terms $t_1,t_2$
always exists and is unique up to renaming.
We sometimes use the notation $t_1 \sqcap t_2 := t$.

The above definitions can be extended to generalization of $n$ terms;
we write $t_1 \sqcap \ldots \sqcap t_n$ to denote the most specific
generalization of $t_1,\ldots,t_n$.
}

% dual to unification,
% upper semi--lattice

\DEFINITION{ \label{E--au}
A term $t$ is called an $E$-generalization of terms $t_1,t_2$
iff there exist substitutions $\sigma_1 $ and $\sigma_2$
such that $\sigma_1 t =_E t_1$
and $\sigma_2 t =_E t_2$.

As in to unification, a most specific
$E$-generalization of arbitrary terms does not usually exist.
A set $G \subset \T$ is called a correct set of $E$-generalizations
of $t_1,t_2$ iff each member is an $E$-generalization of $t_1,t_2$.
$G$ is called complete if, for each $E$-generalization $t$ of $t_1,t_2$,
$G$ contains an instance of $t$.
$G$ is called complete wrt.\ linear generalizations
if for each linear term $t$,
which is an $E$-generalization of $t_1,t_2$,
$G$ contains an instance of $t$.
}

The following algorithm
can be traced back to the early seventies
\cite{Plotkin.1970,Plotkin.1971,Reynolds.1970}.
It takes two terms $t,t'$ and computes the
syntactical generalization $t \sqcap t'$.

\ALGORITHM{ \eqd{sg}
Let $V$ be an infinite set of new variables
and $\phi: \T \times \T \ra V$ an injective mapping.
\begin{enumerate}
\item Define $sg(f(t_1,\ldots,t_n),f(t'_1,\ldots,t'_n)) :=
	f(sg(t_1,t'_1),\ldots,sg(t_n,t'_n))$.
\item Define $sg(f(t_1,\ldots,t_n),f'(t'_1,\ldots,t'_m)) :=
	\phi(f(t_1,\ldots,t_n),f'(t'_1,\ldots,t'_m))$,
	if $f \neq f'$.
\end{enumerate}
}

Since syntactical anti--unification is unique only up to renaming,
the mapping $\phi$ is used to fix one concrete variable naming that
is the same in all subterms.
In the example in Fig.~\ref{Example: Computation of Syntactical
Anti--Unification},
the use of $\phi$ ensures that both occurrences of $sg(3,4)$ yield the
same result, viz.\ the variable $v_{34}$.

\begin{figure}
\begin{center}
\begin{tabular}[t]{@{}ll@{}}
& $sg(3+y*3,4+x*4)$	\\
$=$ & $sg(3,4) + sg(y*3,x*4)$	\\
$=$ & $sg(3,4) + sg(y,x)*sg(3,4)$	\\
$=$ & $v_{34} + v_{yx}*v_{34}$	\\
\end{tabular}
\end{center}
\caption{Example: Computation of Syntactical Anti--Unification}
\label{Example: Computation of Syntactical Anti--Unification}
\end{figure}

\LEMMA{ \eqd{sg1}
Algorithm \eqr{sg} can be extended to anti--unify $N$ terms
simultaneously, requiring $\phi: \T^N \ra \V$.
For any such $\phi$
and any finite $V \subset \V$,
we may define the substitutions
$\sigma_i$ for $i=1,\ldots,N$
by $\sigma_i \phi(t_1,\ldots,t_N) := t_i$.
We have $dom(\sigma_i) = V$ for all $i$.

The result of Alg.~\eqr{sg} then satisfies the following correctness
property:
\\
$\sigma_i sg([t_1,\ldots,t_N]) = t_i$
for $i=1,\ldots,N$
provided we choose $V \supset vars(sg([t_1,\ldots,t_N]))$.
}

\clearpage
\section{E--Anti--Unification}
\label{E--Anti--Unification}

In this section, we describe the implementation of
E--anti--unification based on sorts. 

Section~\ref{Linear Generalizations} restates the algorithm
$rsg$ from \cite{Heinz.1995} for computing a sort containing all
linear generalizations; Sect.~\ref{Nonlinear Generalizations}
discusses how to subsequently modify this sort in order to get
all nonlinear generalizations as well. To obtain the sort of all
generalizations, Alg.~\ref{hsg} from Sect.~\ref{Linear
Generalizations} is run as a first phase, then the algorithm
from Sect.~\ref{Nonlinear Generalizations} is run as the
second phase; cf.\ also Sect.~\ref{An E--Anti--Unification
Example}. 

In Sect.~\ref{Variable--Restricted E--Anti--Unification}, the
algorithm $rsg_V$ for computing the sort of all generalization
terms that contain only variables from a given set $V$ is
restated from \cite{Heinz.1995}. Note that all linear {\em and}
nonlinear generalizations are computed in one phase only, thus
saving a large amount of computation time; cf.\
Sect.~\ref{E--Anti--Unification -- Runtimes},
Fig.~\ref{Technical Optimizations}. 

Section~\ref{Optimized Argument Selection} discusses a technical
optimization that applies to both $rsg$ and $rsg_V$ and helps to
avoid many useless recursive calls.
Sect.~\ref{E--Anti--Unification -- Runtimes} presents some figures
for runtime measurements and the improvement factors of technical
optimizations.

\subsection{Modeling Equivalence Classes as Sorts}
\label{Modeling Equivalence Classes as Sorts}

The algorithms from Sect.~\ref{Linear
Generalizations}/\ref{Nonlinear Generalizations} and from
Sect.~\ref{Variable--Restricted E--Anti--Unification} both
require the representation of the equivalence classes of the
input terms as sorts. In this paper, we do not treat this issue
in detail; rather, we drew up the respective sort definitions
manually.

\begin{figure}
\begin{center}
\begin{tabular}[t]{@{}|*{10}{@{}r}|l@{$\;$}l|l|@{}}
\hline
0&1&2& & &5&6&7& &9& $x+0$ & $= x$ &	\\
0&1&2& & &5&6&7& &9& $x+\suc(y)$ & $= \suc(x+y)$ &	\\
\hline
 & & & & & &6&7& & & $x-0$ & $= x$ &	\\
 & & & & & &6&7& & & $\suc(x)-\suc(y)$ & $= x-y$ &	\\
\hline
 &1&2& & &5& &7& &9& $x*0$ & $= 0$ &	\\
 &1&2& & &5& &7& &9& $x*\suc(y)$ & $= x*y+x$ &	\\
\hline
 & & & & & & & & &9& $0/y$ & $= 0$ &	\\
 & & & & & & & & &9& $\suc(x)/\suc(0)$ & $= \suc(x)$ &	\\
 & & & & & & & & &9& $\suc(x)/\suc(\suc(y))$
	& $= \suc( \; (x-\suc(\suc(y))) \; / \; \suc(\suc(y)) \;)$ & \\
\hline
 & &2& & & & & & & & $dup(0)$ & $= 0$ & $duplicate$	\\
 & &2& & & & & & & & $dup(\suc(x))$ & $= \suc(\suc(dup(x)))$ &	\\
\hline
 & & & & &5& & & & & $if(0,x,y)$ & $= y$ & $if\_then\_else$ \\
 & & & & &5& & & & & $if(\suc(z),x,y)$ & $= x$ &	\\
\hline
 & & & & &5& & & & & $ev(0)$ & $= \suc(0)$ & $is\_even$	\\
 & & & & &5& & & & & $ev(\suc(0))$ & $= 0$ &	\\
 & & & & &5& & & & & $ev(\suc(\suc(x)))$ & $= ev(x)$ &	\\
\hline
 & & & &4& & & & & & $len([\;])$ & $= 0$ & $length$	\\
 & & & &4& & & & & & $len(x.y)$ & $= \suc(len(y))$ &	\\
\hline
 & & &3&4& & & &8& & $app([\;],x)$ & $= x$ & $append$	\\
 & & &3&4& & & &8& & $app(x.y,z)$ & $= x.app(y,z)$ &	\\
\hline
 & & &3&4& & & &8& & $rev([\;])$ & $= [\;]$ & $reverse$	\\
 & & &3&4& & & &8& & $rev(x.y)$ & $= app(rev(y),[x])$ &	\\
\hline
 & & & & & & & &8& & $rot([\;])$ & $= [\;]$ & $rotate$	\\
 & & & & & & & &8& & $rot(x.y)$ & $= app(y,[x])$ &	\\
\hline
 & & & & & & & &8& & $il(x,[\;])$ & $= x$ &	$interleave$	\\
 & & & & & & & &8& & $il([\;],x)$ & $= x$ &	\\
 & & & & & & & &8& & $il(x.y,z.w)$ & $= x.z.il(y,w)$ & \\
\hline
%  & & & & & & & & & & $val(nil)$ & $= 0$ & $bin\_value$	\\
%  & & & & & & & & & & $val(x::o)$ & $= dup(val(x))$ & \\
%  & & & & & & & & & & $val(x::i)$ & $= \suc(dup(val(x)))$&\\
% \hline
\end{tabular}
\end{center}
\caption{Background Equational Theories Used}
\label{Background Equational Theories Used}
\end{figure}

\begin{figure}
\begin{center}
\begin{tabular}[t]{@{}*{20}{l}@{}}
$s_0$ & $\sortdef$
	& $0$
	& $\mid if(s_p,s_0,s_n)$
	& $\mid if(s_0,s_n,s_0)$
	& $\mid ev(s_o)$
	& $\mid s_0+s_0$
	& $\mid s_0\1*s_n$
	& $\mid s_n\1*s_0$	\\
$s_1$ & $\sortdef$
	& $\suc(s_0)$
	& $\mid if(s_p,s_1,s_n)$
	& $\mid if(s_0,s_n,s_1)$
	& $\mid ev(s_e)$
	& $\mid s_1+s_0$
	& $\mid s_0\1+s_1$
	& $\mid s_1\1*s_1$	\\
$s_2$ & $\sortdef$
	& $\suc(s_1)$
	& $\mid if(s_p,s_2,s_n)$
	& $\mid if(s_0,s_n,s_2)$
	& $\mid s_2+s_0$
	& $\mid s_1+s_1$
	& $\mid s_0\1+s_2$
	& $\mid s_1\1*s_2$
	& $\mid s_2\1*s_1$	\\
$s_3$ & $\sortdef$
	& $\suc(s_2)$
	& $\mid if(s_p,s_3,s_n)$
	& $\mid if(s_0,s_n,s_3)$
	& $\mid s_3+s_0$
	& $\mid s_2+s_1$
	& $\mid s_1\1+s_2$
	& $\mid s_0\1+s_3$
	& $\mid s_1\1*s_3$
	& $\mid s_3\1*s_1$	\\
$s_4$ & $\sortdef$
	& $\suc(s_3)$
	& $\mid if(s_p,s_4,s_n)$
	& $\mid if(s_0,s_n,s_4)$
	& $\mid s_4+s_0$
	& $\mid s_3+s_1$
	& $\mid s_2\1+s_2$
	& $\mid s_1\1+s_3$
	& $\mid s_0\1+s_4$
	& $\mid s_1\1*s_4$
	& $\mid \! s_2\1*s_2$
	& $\mid \! s_4\1*s_1$	\\
$s_5$ & $\sortdef$
	& $\suc(s_4)$
	& $\mid if(s_p,s_5,s_n)$
	& $\mid if(s_0,s_n,s_5)$
	& $\mid s_5+s_0$
	& $\mid s_4+s_1$
	& $\mid s_3\1+s_2$
	& $\mid s_2\1+s_3$
	& $\mid s_1\1+s_4$
	& $\mid s_0\1+s_5$
	& $\mid \! s_1\1*s_5$
	& $\mid \! s_5\1*s_1$	\\
[0.2cm]
$s_p$ & $\sortdef$
	& $\suc(s_n)$
	& $\mid if(s_p,s_p,s_n)$
	& $\mid if(s_0,s_n,s_p)$
	& $\mid ev(s_e)$
	& $\mid s_p+s_n$
	& $\mid s_n\1+s_p$
	& $\mid s_p\1*s_p$	\\
$s_e$ & $\sortdef$
	& $0$
	& $\mid if(s_p,s_e,s_n)$
	& $\mid if(s_0,s_n,s_e)$
	& $\mid ev(s_o)$
	& $\mid \suc(s_o)$
	& $\mid s_e\1+s_e$
	& $\mid s_o\1+s_o$
	& $\mid s_e\1*s_n$
	& $\mid s_n\1*s_e$	\\
$s_o$ & $\sortdef$
	& $\suc(s_e)$
	& $\mid if(s_p,s_o,s_n)$
	& $\mid if(s_0,s_n,s_o)$
	& $\mid ev(s_e)$
	& $\mid s_o+s_e$
	& $\mid s_e\1+s_o$
	& $\mid s_o\1*s_o$	\\
$s_n$ & $\sortdef$
	& $0$
	& $\mid \suc(s_n)$
	& $\mid if(s_n,s_n,s_n)$
	& $\mid ev(s_n)$
	& $\mid s_n+s_n$
	& $\mid s_n\1*s_n$	\\
\end{tabular}
\end{center}
\caption{Equivalence Classes for Background Theory (5) from Fig.\
	\protect\ref{Background Equational Theories Used}}
\label{Equivalence Classes for Background Theory (5) from Fig.}
\end{figure}

\begin{figure}
\begin{center}
\begin{tabular}[t]{@{}*{10}{l}@{}}
$s_0$ & $\sortdef$
	& $0$
	& $\mid s_0+s_0$
	& $\mid s_0-s_0$
	& $\mid s_1-s_1$
	& $\mid s_2-s_2$
	& $\mid s_3-s_3$
	& $\mid s_4-s_4$
	& $\mid s_5-s_5$	\\
$s_1$ & $\sortdef$
	& $\suc(s_0)$
	& $\mid s_1+s_0$
	& $\mid s_0+s_1$
	& $\mid s_1-s_0$
	& $\mid s_2-s_1$
	& $\mid s_3-s_2$
	& $\mid s_4-s_3$
	& $\mid s_5-s_4$	\\
$s_2$ & $\sortdef$
	& $\suc(s_1)$
	& $\mid s_2+s_0$
	& $\mid s_1+s_1$
	& $\mid s_0+s_2$
	& $\mid s_2-s_0$
	& $\mid s_3-s_1$
	& $\mid s_4-s_2$
	& $\mid s_5-s_3$	\\
$s_3$ & $\sortdef$
	& $\suc(s_2)$
	& $\mid s_3+s_0$
	& $\mid s_2+s_1$
	& $\mid s_1+s_2$
	& $\mid s_0+s_3$
	& $\mid s_3-s_0$
	& $\mid s_4-s_1$
	& $\mid s_5-s_2$	\\
$s_4$ & $\sortdef$
	& $\suc(s_3)$
	& $\mid s_4+s_0$
	& $\mid s_3+s_1$
	& $\mid s_2+s_2$
	& $\mid s_1+s_3$
	& $\mid s_0+s_4$
	& $\mid s_4-s_0$
	& $\mid s_5-s_1$	\\
$s_5$ & $\sortdef$
	& $\suc(s_4)$
	& $\mid s_5+s_0$
	& $\mid s_4+s_1$
	& $\mid s_3+s_2$
	& $\mid s_2+s_3$
	& $\mid s_1+s_4$
	& $\mid s_0+s_5$
	& $\mid s_5-s_0$	\\
$s_n$ & $\sortdef$
	& $0$
	& $\mid \suc(s_n)$
	& $\mid s_n+s_n$
	& $\mid s_n-s_n$	\\
\end{tabular}
\end{center}
\caption{Equivalence Classes for Background Theory (6) from Fig.\
	\protect\ref{Background Equational Theories Used}}
\label{Equivalence Classes for Background Theory (6) from Fig.}
\end{figure}

Figure~\ref{Background Equational Theories Used} shows the equational
theories used in this paper; we will refer to them as background
theories.
For example, Theory (1) consists of all equations that have a ``1'' in
the leftmost column.
Figure~\ref{Sort Definition of $s_0,s_1$} refers to Theory (1) and
shows the sort definition of $s_0$ and $s_1$, representing
$\eqc{0}$ and $\eqc{\suc(0)}$, respectively.
Figure~\ref{Equivalence Classes for Background Theory (5) from Fig.}
shows the sort representation of $\eqc{0}$ to $\eqc{\suc^5(0)}$
wrt.\ Theory (5);
$s_p$, $s_e$, $s_o$, and $s_n$ denote the sets of positive
(i.e.\ $>0$), even, odd, and arbitrary natural numbers,
respectively.

In Theory (6), we have included the $(-)$ operator,
such that equivalence classes of terms are no longer
regular tree languages, and hence cannot be described by our sorts;
for example, the sort definition
$s_0 \sortdef 0 \mid \ldots
\mid s_0-s_0 \mid s_1-s_1 \mid \ldots \mid s_i-s_i \mid \ldots$
would become infinite.
We can, however, approximate the equivalence classes from below by
cutting off the sort definitions at a certain $n$, i.e., omitting all
greater numbers.
Figure~\ref{Equivalence Classes for Background Theory (6) from Fig.}
shows the approximations of $\eqc{0}$ to $\eqc{\suc^5(0)}$,
where we set the cut--off point to $n=5$, i.e.\ $s_6-s_6 \mid \ldots
\mid s_i-s_i \mid \ldots$ is missing;
hence, the equivalence classes do not contain terms that yield an
intermediate result greater than $5$ when evaluated to normal form.

In Sect.~\ref{Dynamic Sort Generation}, we discuss the use of
sort--definition schemes that allow several sort definitions to
be abbreviated by one scheme. Figure~\ref{Sort Definition Scheme
for Background Theory (0) from Fig.} shows a sort--definition
scheme for background Theory (0). Figure~\ref{Equivalence
Classes of Non--Ground Terms mod. $append,reverse$}, shows
sort--definition schemes of equivalence classes mod.\ $append$
and $reverse$. 

\cite{Emmelmann.1994} describes an algorithm for building the
sort definitions automatically from a given confluent and
noetherian term--rewriting system. However, there is no
implementation that fits our data structures. See also
Thm.~\eqr{6} and Cor.~\eqr{7} in Sect.~\ref{Equational Theories
of Finite Algebras}, which provide sufficient criteria for an
equational theory in order that the equivalence classes are
regular tree languages.

\subsection{Linear Generalizations}
\label{Linear Generalizations}

The following algorithm from \cite{Heinz.1995}
takes two sorts $s,s'$ and computes a sort $\theta$
containing all linear syntactical generalizations $t \sqcap t'$
of terms $t \in \L(s)$, $t' \in \L(s')$,
i.e.\

$$\theta \supset \{ t \sqcap t'
\mid t \in \L(s), t' \in \L(s'), t \sqcap t' \mbox{ linear } \}.$$

\ALGORITHM{ \eqd{rsg}
Let $s$ and $s'$ be sort names, and $\theta$ be a new sort name.
Let $V$ be an infinite set of new variables,
and $\phi: \S \times \S \ra V$ an injective mapping.
Define $rsg(s,s') := \theta$,
where a new sort definition is introduced for $\theta$:
\begin{enumerate}
\item If $rsg(s,s')$ has been called before,
	then $\theta$ is already defined.
\item If $s \sortdef s_1 \mid \ldots \mid s_n$, then define
	$\theta \sortdef rsg(s_1,s') \mid \ldots \mid rsg(s_n,s')$.
\item If $s' \sortdef s'_1 \mid \ldots \mid s'_n$, then define
	$\theta \sortdef rsg(s,s'_1) \mid \ldots \mid rsg(s,s'_n)$.
\item If $s \sortdef f(s_1,\ldots,s_n)$
	and $s' \sortdef f(s'_1,\ldots,s'_n)$,
	then define
	$\theta \sortdef f(rsg(s_1,s'_1),\ldots,rsg(s_n,s'_n))$.
\item If $s \sortdef f(s_1,\ldots,s_n)$
	and $s' \sortdef f'(s'_1,\ldots,s'_n)$
	with $f \neq f'$,
	then define
	$\theta \sortdef \phi(s,s')$.
\end{enumerate}
}

As shown in \cite{Heinz.1995},
$\L(rsg(s,s'))$ is a correct set of generalizations
which is complete for the linear ones,
i.e.,
$t \in \L(rsg(s_1,s_2))
\Ra \exists t_1 \in \L(s_1), \; t_2 \in \L(s_2) \;\;
t = t_1 \sqcap t_2$,
and
$t_1 \in \L(s_1) \wedge t_2 \in \L(s_2) \wedge
t := t_1 \sqcap t_2 \mbox{ linear}
\Ra  t \in \L(rsg(s_1,s_2))$.

Note that Case 1.\ requires us to maintain a set of argument
pairs for which $rsg$ has already been called, and to check the
argument pair of each new call against this set. This set is
called {\tt Occ} in the implementation; it was initially
implemented by a PROLOG list, then by a binary search tree, and
now by a balanced binary search tree. A membership test in a
balanced tree with 50 entries takes about 4 msec user time. See
also Sect.~\ref{E--Anti--Unification -- Runtimes} for the impact
of balancing on runtime. 

Assuming all sort definitions in head normal form, we can
slightly improve the above algorithm in order to produce less
variables. For example, given the sort definitions of $s_0,s_1$
from Fig.~\ref{Sort Definition of $s_0,s_1$}, the sort $s_{01}$
computed in Fig.~\ref{Computation of $hsg(s_0,s_1)$} would
comprise twelve different variables if computed by $rsg$, while
it actually --~computed by $hsg$ below~-- has only one variable. 

Since $\L(hsg(s_1,s_2)) \subset \L(rsg(s_1,s_2))$,
Alg.~\eqr{hsg} is still correct; since for each generalization
$t \in \L(rsg(s_1,s_2))$ we have $\sigma t \in \L(hsg(s_1,s_2))$
for some flat substitution $\sigma$, Alg.~\eqr{hsg} is also
complete for the linear generalizations.

\ALGORITHM{ \eqd{hsg}
Let $s$, $s'$, $V$, and $\phi$ be as in Alg.~\eqr{rsg}.
Define $hsg(s,s')$ as follows:
\begin{enumerate}
\item If $hsg(s,s')$ has been called earlier,
        its result is already defined.
\item If $s \sortdef \bigmid_{i=1}^m \; f_i(s_{i1},\ldots,s_{in_i})$,
	and
	$s' \sortdef \bigmid_{j=1}^{m'}
	\; f'_j(s'_{j1},\ldots,s'_{jn'_j})$,
	\\
	then let $\theta$ be a new sort name,
	\\
	define
	$\theta \sortdef \phi(s,s') \; \mid
	\; \bigmid_{i=1}^m \; \bigmid_{j=1}^{m'}
	\; hsg(f_i(s_{i1},\ldots,s_{in_i}),
	f'_j(s'_{j1},\ldots,s'_{jn'_j}))$;
	\\
	let $hsg(s,s') := \theta$.
\item Define $hsg(f(s_1,\ldots,s_n),f(s'_1,\ldots,s'_n))
	:= f(hsg(s_1,s'_1),\ldots,hsg(s_n,s'_n))$.
\item Define $hsg(f(s_1,\ldots,s_n),f'(s'_1,\ldots,s'_m)) := \bot$,
	if $f \neq f'$.
\end{enumerate}
}

Usually, most of the disjuncts in Case 2.\ start with different
function symbols, $f_i \neq f'_j$, and hence evaluate to $\bot$
by Case 4. In Sect.~\ref{Optimized Argument Selection}, we will
discuss an optimization that avoids these recursive calls of
$hsg$. 

Algorithms~\ref{rsg} and~\ref{hsg} can both be extended to
compute the linear generalizations of $N$ sorts simultaneously,
requiring $\phi: \S^N \ra V$. Our implementation is capable of
that, and we will tacitly use $rsg([s',s'',\ldots,s^{(N)}])$
with $N$ arguments where appropriate in this paper. Note that it
makes sense to have multiple occurrences of the same sort among
the input arguments. For example, using the sort definitions
from Fig.~\ref{Sort Definition of $s_0,s_1$}, we have $0*0
\sqcap (0+0)*(0+0) = v*v \in \L(hsg(s_0,s_0)) \setminus
\L(s_0)$. Moreover, it is important to maintain the order of
argument sorts during computation, since otherwise, for example 

$$
\begin{tabular}[t]{@{}l@{$\;$}ll@{}}
& $hsg(s_1,s_1) $	\\
$=$ & $\ldots \mid hsg(s_0,s_1)+hsg(s_1,s_0) \mid \ldots$	\\
$=$ & $\ldots \mid hsg(s_0,s_1)+hsg(s_0,s_1) \mid \ldots$	\\
$=$ & $\ldots \mid v_{01}+v_{01} \mid \ldots$	\\
\end{tabular}
$$

although $1$ is not equal to any instance of $v_{01}+v_{01}$.
For these reasons, we treat the argument of the extended $hsg$
as a list rather than as a set of sorts.

\subsection{Nonlinear Generalizations}
\label{Nonlinear Generalizations}

The result sorts of the above algorithm contains all linear
generalizations, but only some nonlinear ones. That is not
generally sufficient, as the example in Sect.~\ref{An
E--Anti--Unification Example} shows (see in particular the
difference between $s_{01}$ and $s'_{01}$ in
Fig.~\ref{Generalization of $s_0$ and $s_1$}). 

In order to obtain as well all nonlinear generalizations, which
are more specific, we need a second phase which introduces
common variables to related sorts. We use the abbreviation
$infs(V) := \{ \bigcap_{v \in V} \pi_i^N (\phi^{-1}(v)) \mid 1
\leq i \leq N \}$; note that $\phi^{-1}$ exists since $\phi$ is
injective. Below, we will assume $N=2$ for the sake of
simplicity. 

\cite{Heinz.1995} proceeds as follows:
for each set of variables $\{v_1,\ldots,v_n\}$ from the result
sorts, whenever $infs(\{v_1,\ldots,v_n\})$ does not contain the
empty sort, a new variable $v'$ is introduced which can be
thought of intuitively as generalizing the sorts in $infs(V)$,
and each occurrence of each $v_i$ is replaced by $v_i \mid v'$.
For example, using the definitions from Sect.~\ref{An
E--Anti--Unification Example}, we have 
$v_{01} = \phi(s_0,s_1)$,
$v_{0n} = \phi(s_0,s_n)$,
$v_{n1} = \phi(s_n,s_1)$,
and
$v_{nn} = \phi(s_n,s_n)$;
hence
$infs(\{v_{0n},v_{n1}\})
= \{ s_0 \cap s_n \; , \; s_n \cap s_1 \}
= \{s_0, s_1\}$
since $s_0 \cup s_1 \subset s_n$.
Thus, $v_{0n}$, $v_{0n}$ are replaced by $v_{0n} \1\mid v'$,
and $v_{0n} \1\mid v'$, respectively.
Similarly, for any subset of $\{ v_{01}, v_{0n}, v_{n1}, v_{nn} \}$
that contains more than one element, a new variable would be added.

We follow this approach, except that we only consider {\em
maximal} sets $\{v_1,\ldots,v_n\}$ of variables, i.e.\ where
$\{\} \in infs(\{v_1,\ldots,v_n,v_{n+1}\})$ for any $v_{n+1}$.
For each generalization term $t$ obtained by the
\cite{Heinz.1995} approach, we get a more specific
generalization term $\sigma t$ for some flat substitution
$\sigma$. By analogy with the remark in Sect.~\ref{Linear
Generalizations}, this shows the algorithm is still complete. In
the above example, we would add only one variable for the set
$\{ v_{01}, v_{0n}, v_{n1}, v_{nn} \}$, which is maximal. 

In the example in Sect.~\ref{An E--Anti--Unification Example} /
Fig.~\ref{Generalization of $s_0$ and $s_1$}, the second phase
leads to the modification of sort $s_{01}$ to $s'_{01}$, etc.
Unlike the former sort, the latter, for example, contains the
term $v'_{01}*v'_{01}$, meaning that $0$ and $1$ are both
quadratic numbers. 

We explain below how the $n$-tuples $v_1,\ldots,v_n$ of
variables with non--empty sort intersections
$\{\} \not \in infs(\{v_1,\ldots,v_n\})$
are found.
The replacement of each $v_i$ by $v_i \mid v'$ is then
straightforward.
As the computation of sort intersections is very time--consuming, we try
to minimize the number of such computations by applying some
---~correct~--- heuristics explained below.
Note, however, that when each defined sort name corresponds to an
equivalence class (unlike $s_n$), computing their intersections is
trivial, since different equivalence classes are always disjoint.

Let $V$ be the set of variables newly generated by the first phase.
Each variable $v \in V$ is assigned a pair
$\tpl{s_v,s'_v} := \phi^{-1}(v)$
of sorts such that $v$ represents all
generalizations of any term from $s_v$ with any term from $s'_v$.
Let $S := \{ s_v \mid v \in V \} \cup \{ s'_v \mid v \in V \}$ be
the set of all relevant sorts.
In the first step, we compute the set

$$I2 := \{ \tpl{s_1,s_2,s_{12}}
	\mid s_1,s_2 \in S, \{\} \neq s_{12} = s_1 \cap s_2,
	s_1 \prec s_2\}$$

of all non--empty intersections of sorts from $S$.
This is accomplished by the PROLOG predicate {\tt calc\_sort\_infs}.
The relation $(\prec)$ is an arbitrary irreflexive total ordering on
$\S$.
Without the condition $s_1 \prec s_2$, we had
$\tpl{s_2,s_1,s_{12}} \in I2$ whenever $\tpl{s_1,s_2,s_{12}} \in I2$;
the additional ordering condition helps to keep the memory
requirements small.
$I2$ is represented by a balanced binary tree.
Then, the set

$$L2 := \{ \tpl{v_1,v_2} \mid v_1,v_2 \in V,
	\{\} \not\in infs(\{v_1,v_2\}), v_1 \prec v_2 \}$$

of all pairs of variables with non--empty intersections
is computed by {\tt calc\_inf2s}.
The remarks on $s_1 \prec s_2$ apply similarly to the ordering
condition $v_1 \prec v_2$.
Each variable in the set $V$ is equipped with its number of occurrences
in $L2$, i.e.\
for each $v \in V$.
Let

$$i_v := \abs{ (\{ \tpl{v,v'} \mid \tpl{v,v'} \in L2 \}
	\cup \{ \tpl{v',v} \mid \tpl{v',v} \in L2\} )};$$

then the set
$V2 := \{ \tpl{v,i_v} \mid v \in V \}$
is computed using the predicates
{\tt calc\_varcnt} and {\tt collect\_varcnt}.
The predicate {\tt calc\_infs\_list} then takes $L2$ and $V2$ as input:
it calls {\tt calc\_infs} for each
$\tpl{v_1,v_2} \in L2$, which in turn
computes the set of all maximal
$V_{12}$ such that $v_1,v_2 \in V_{12}$,
and
$\{\} \not\in infs(V_{12})$.
The rules for {\tt calc\_infs} are given in Fig.~\ref{Rules for calc
infs}; 
the start configuration is

$$
\begin{tabular}[t]{@{}l@{\hspace*{0.5cm}}l@{\hspace*{0.5cm}}l@{}}
\hline
$\{v_1,v_2\}$ & $L2$ & $[\;]$	\\
\end{tabular}
$$

for each pair $\tpl{v_1,v_2} \in L2$.
All the
rules in this paper are priorized, i.e.\ the $n$-th rule is tried
only if the $1$-st, \ldots, $n-1$-th did not succeed.
$Out$ denotes the output so far.
The set $Susp$ contains ``suspended'' variable sets $V$ that are
subsumed by some earlier output, see rule \rulename{Suplist}.
The suspension is released as soon as a new variable can be joined to
the actual set $V$, see rule \rulename{Inf~Inh}.
The actual set $V$ is reported as a maximal one only if no suspended
variable sets exist, see rules \rulename{Output~Max},
\rulename{Max~Subs~Susp}.
Rule \rulename{Inf~Inh} has two successors, the output of both of which
has to be
joined; $\app$ denotes the appending of two lists.
Rule \rulename{Count} is justified since $v'$ can have non--empty
intersections with at most $i_{v'}$ different variables (``intersection
of variables'' $v_1,\ldots,v_n$ meaning $infs(\{v_1,\ldots,v_n\})$).
In order to allow fast lookup of $i_{v'}$ in rule \rulename{Count},
and fast testing of $\tpl{v',v'',\ldots} \not\in L2$
in rule \rulename{Sublist2}, both $L2$ and $V2$ are given additionally
in balanced--tree form to the algorithm.

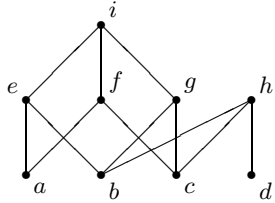
\begin{figure}
\begin{center}
\begin{picture}(2.7,3.7)
	%\put(0,0){\makebox(0,0){+}}
\put(0.300,0.300){\circle*{0.100}}
\put(1.300,0.300){\circle*{0.100}}
\put(2.300,0.300){\circle*{0.100}}
\put(3.300,0.300){\circle*{0.100}}
\put(0.300,1.300){\circle*{0.100}}
\put(1.300,1.300){\circle*{0.100}}
\put(2.300,1.300){\circle*{0.100}}
\put(3.300,1.300){\circle*{0.100}}
\put(1.300,2.300){\circle*{0.100}}
\put(0.400,0.200){\makebox(0.000,0.000)[tl]{$a$}}
\put(1.400,0.200){\makebox(0.000,0.000)[tl]{$b$}}
\put(2.400,0.200){\makebox(0.000,0.000)[tl]{$c$}}
\put(3.400,0.200){\makebox(0.000,0.000)[tl]{$d$}}
\put(0.200,1.400){\makebox(0.000,0.000)[br]{$e$}}
\put(1.400,1.400){\makebox(0.000,0.000)[bl]{$f$}}
\put(2.400,1.400){\makebox(0.000,0.000)[bl]{$g$}}
\put(3.400,1.400){\makebox(0.000,0.000)[bl]{$h$}}
\put(1.400,2.400){\makebox(0.000,0.000)[bl]{$i$}}
\put(0.300,0.300){\line(0,1){1.000}}
\put(0.300,0.300){\line(1,1){1.000}}
\put(1.300,0.300){\line(-1,1){1.000}}
%\put(1.300,0.300){\line(0,1){1.000}}
\put(1.300,0.300){\line(1,1){1.000}}
\put(2.300,0.300){\line(-1,1){1.000}}
\put(2.300,0.300){\line(0,1){1.000}}
\put(1.300,0.300){\line(2,1){2.000}}
\put(2.300,0.300){\line(1,1){1.000}}
\put(3.300,0.300){\line(0,1){1.000}}
\put(0.300,1.300){\line(1,1){1.000}}
\put(1.300,1.300){\line(0,1){1.000}}
\put(2.300,1.300){\line(-1,1){1.000}}
\end{picture}
\end{center}

\caption{Example: Intersection Hierarchy}
\label{Example: Intersection Hierarchy}
\end{figure}

Figure~\ref{Example: Intersection Hierarchy} shows an example
intersection hierarchy for the variables $a,b,c,d$;
a common upper bound of $v_1,\ldots,v_n$ indicates that
$\{\} \not \in infs(v_1,\ldots,v_n)$.
While setting up $I2$, the six intersection sets
$infs(\{a,b\}),
infs(\{a,c\}),
infs(\{a,d\}),
infs(\{b,c\}),
infs(\{b,d\}),
infs(\{c,d\})$
are computed and tested to determine whether they contain the
empty set. Note that these computations are unavoidable, since
any two sorts could intersect independently of the others.
During the run of {\tt calc\_infs\_list}, only two more
intersection sets need to be computed and checked; they are
flagged right to the rule \rulename{Inf~Inh} in
Figs.~\ref{Example: Intersection Computation Run}
and~\ref{Example: Intersection Computation Run (contd.)} which
show the corresponding example run. See also Sect.~\ref{An
E--Anti--Unification Example} for a second example.

\begin{figure}
\begin{center}
\begin{tabular}{@{}*{6}{l@{\hspace*{0.5cm}}}@{}}
$V$ & $[v' \mid List]$ & $Susp$
	&& \raisebox{-2ex}[0ex][0ex]{$v' \in V$}
	& \raisebox{-2ex}[0ex][0ex]{(Member)} \\
\cline{1-3}
$V$ & $List$ & $Susp$	\\
\mca{2}{} \\
$V$ & $[v' \mid List]$ & $Susp$
	&& \raisebox{-2ex}[0ex][0ex]{$i_{v'} < \abs{V}$}
	& \raisebox{-2ex}[0ex][0ex]{(Count)} \\
\cline{1-3}
$V$ & $List$ & $Susp$	\\
\mca{2}{} \\
$V$ & $[v' \mid List]$ & $Susp$
	&& $\exists v'' \in V \;\; \tpl{v',v'',\ldots} \not\in L2$
	& \raisebox{-2ex}[0ex][0ex]{(Sublist2)}	\\
\cline{1-3}
$V$ & $List$ & $Susp$
	&& $\wedge \tpl{v'',v',\ldots} \not\in L2$	\\
\mca{2}{} \\
$V$ & $[v' \mid List]$ & $Susp$
	&& \raisebox{-2ex}[0ex][0ex]
	{$\exists V' \in Out \;\; V \cup \{v'\} \subset V'$}
	& \raisebox{-2ex}[0ex][0ex]{(Suplist)}	\\
\cline{1-3}
$V$ & $List$ & $[v' \mid Susp]$	\\
\mca{2}{} \\
$V$ & $[v' \mid List]$ & $Susp$
	&& \raisebox{-2ex}[0ex][0ex]{$\{\} \not\in
		infs(V \cup \{v'\})$}
	& \raisebox{-2ex}[0ex][0ex]{(Inf Inh)}	\\
\cline{1-3}
$V \cup\{v'\}$ & $List \app Susp$ & $[\;]$	\\
$V$ & $List \app Susp$ & $[\;]$	\\
\mca{2}{} \\
$V$ & $[v' \mid List]$ & $Susp$
	&&& \raisebox{-2ex}[0ex][0ex]{(Inf Empty)}	\\
\cline{1-3}
$V$ & $List$ & $Susp$	\\
\mca{2}{} \\
$V$ & $[\;]$ & $[\;]$
	&& \raisebox{-2ex}[0ex][0ex]{$\exists V' \in Out \;\;
		V \subset V'$}
	& \raisebox{-2ex}[0ex][0ex]{(Max Subsumed)}	\\
\cline{1-3}
\\
\mca{2}{} \\
$V$ & $[\;]$ & $[\;]$
	&&& \raisebox{-2ex}[0ex][0ex]{(Output Max)}	\\
\cline{1-3}
\mca{2}{} & output $V$	\\
\mca{2}{} \\
$V$ & $[\;]$ & $[v' \mid Susp]$
	&&& \raisebox{-2ex}[0ex][0ex]{(Max Subs Susp)}	\\
\cline{1-3}
\\
\end{tabular}
\end{center}
\caption{Rules for {\tt calc\_infs}}
\label{Rules for calc infs}
\end{figure}

\begin{figure}
\begin{center}
\begin{tabular}[t]{@{}l@{$\;$}ll@{}}
$L2$ & $= \{ \tpl{a,b}, \tpl{a,c}, \tpl{b,c}, \tpl{b,d}, \tpl{c,d} \}$\\
$V2$ & $= \{ \tpl{a,2}, \tpl{b,3}, \tpl{c,3}, \tpl{d,2} \}$	\\
\end{tabular}

\begin{verbatim}
[a,b]   [a,b,c,d]   []
   (Member)
   [a,b]   [b,c,d]   []
      (Member)
      [a,b]   [c,d]   []
         (Inf Inh)            infs({a,b,c})
         [a,b,c]   [d]   []
            (Sublist2)
            [a,b,c]   []   []
               (Output Max [a,b,c])
         [a,b]   [d]   []
            (Sublist2)
            [a,b]   []   []
               (Max Subsumed)
[a,c]   [a,b,c,d]   []
   (Member)
   [a,c]   [b,c,d]   []
      (Suplist)
      [a,c]   [c,d]   [b]
         (Member)
         [a,c]   [d]   [b]
            (Sublist2)
            [a,c]   []   [b]
               (Max Subs Susp)
\end{verbatim}
(Contd.\ in Fig.~\ref{Example: Intersection Computation Run (contd.)})
\end{center}

\caption{Example: Intersection Computation Run}
\label{Example: Intersection Computation Run}
\end{figure}

\begin{figure}
\begin{center}
\begin{verbatim}
[b,c]   [a,b,c,d]   []
   (Suplist)
   [b,c]   [b,c,d]   [a]
      (Member)
      [b,c]   [c,d]   [a]
         (Member)
         [b,c]   [d]   [a]
            (Inf Inh)            infs({b,c,d})
            [b,c,d]   [a]   []
               (Sublist2)
                  [b,c,d]   []   []
                     (Output Max [b,c,d])
            [b,c]   [a]   []
               (Suplist)
               [b,c]   []   [a]
                  (Max Subs Susp)

[b,d]   [a,b,c,d]   []
   (Sublist2)
   [b,d]   [b,c,d]   []
      (Member)
      [b,d]   [c,d]   []
         (Suplist)
         [b,d]   [d]   [c]
            (Member)
            [b,d]   []   [c]
               (Max Subs Susp)

[c,d]   [a,b,c,d]   []
   (Sublist2)
   [c,d]   [b,c,d]   []
      (Suplist)
      [c,d]   [c,d]   [b]
         (Member)
         [c,d]   [d]   [b]
            (Member)
            [c,d]   []   [b]
               (Max subs susp)
\end{verbatim}
\end{center}

\caption{Example: Intersection Computation Run (contd.)}
\label{Example: Intersection Computation Run (contd.)}
\end{figure}

\clearpage
\subsection{Variable--Restricted E--Anti--Unification}
\label{Variable--Restricted E--Anti--Unification}

The following algorithm from \cite{Heinz.1995} takes two input
sorts $s,s'$ and a finite set $V$ of variables and computes the
set of all generalizations of terms from $s$ and $s'$ that
contain only variables from the set $V$. The call
$rsg_V(\{\},s,s')$ computes precisely the intersection of the
sorts $s$ and $s'$. Note that not only linear generalizations
are computed; hence, the computationally expensive second phase
described in Sect.~\ref{Nonlinear Generalizations} can be
omitted. See Sect.~\ref{E--Anti--Unification -- Runtimes} for
figures on the runtime improvement thus achieved. The remarks
from Sect.~\ref{Linear Generalizations} about extending the
algorithm to $N$ arguments also apply to $rsg_V$.

\ALGORITHM{ \label{vrsg}
Let $s$ and $s'$ be sorts and $\theta$ a new sort name.
Let $\phi: \S \times \S \ra \V$ be an injective mapping.
Let $V$ be a finite set of variables,
such that $TP := \phi^{-1}[V] \subset \T \times \T$
is a finite set of term pairs.
(Note that each term is also a sort expression.)
Define $rsg_V(V,s,s') := \theta$,
where a new sort definition for $\theta$
is introduced by the following cases.

\begin{enumerate}
\item If $rsg_V(V,s,s')$ has been called before,
	then $\theta$ is already defined.
\item If $s \sortdef s_1 \mid \ldots \mid s_n$, then define
	$\theta \sortdef
	rsg_V(V,s_1,s') \mid \ldots \mid rsg_V(V,s_n,s')$.
\item If $s' \sortdef s'_1 \mid \ldots \mid s'_n$, then define
	$\theta \sortdef
	rsg_V(V,s,s'_1) \mid \ldots \mid rsg_V(V,s,s'_n)$.
\item If $s \sortdef f(s_1,\ldots,s_n)$,
	and $s' \sortdef f(s'_1,\ldots,s'_n)$,
	\\
	then define
	$\theta \sortdef f(rsg_V(V,s_1,s'_1),\ldots,rsg_V(V,s_n,s'_n))$.
\item If $s \sortdef f(s_1,\ldots,s_n)$,
	and $s' \sortdef f'(s'_1,\ldots,s'_{n'})$,
	then define
	$\theta \sortdef
	\bigmid_{\tpl{t,t'} \in TP \cap (\L(s) \times \L(s'))}
	\; \phi(t,t')$.
\end{enumerate}
}

\LEMMA{ \eqd{vrsg1}
Given $\phi: \T^N \ra \V$ and $V$,
we may define
$\sigma_i \phi(t_1,\ldots,t_N) := t_i$
for all $\tpl{t_1,\ldots,t_N} \in \phi^{-1}[V]$
and $i=1,\ldots,N$.
We have $dom(\sigma_i) = V$ for all $i$.
We assume that for each $x \in V$, $x = \phi(t_1,\ldots,t_n)$
such that not all $t_1,\ldots,t_n$ start with the same function
symbol.

The result of Alg.~\eqr{vrsg} then satisfies the following correctness
property:

$$
t \in \L(rsg_V(V,[s_1,\ldots,s_N]))
\Lra vars(t) \subset V
\wedge \bigwedge_{i=1}^N \; \sigma_i t \in \L(s_i)$$

Hence, when the argument sorts of $rsg_V$ are equivalence classes of
terms, we have

$$t \in \L(rsg_V(V,[\eqc{t_1},\ldots,\eqc{t_N}]))
\Lra vars(t) \subset V \wedge \bigwedge_{i=1}^N \; \sigma_i t =_E t_i$$
}
\PROOF{
We show the lemma only for $N=2$,
using Thm.~\eqr{ind} with

$$
\begin{tabular}[t]{@{}ll@{\hspace*{0.5cm}}l@{}}
$p_\theta(t)$
	& $:\Lra \sigma_1 t \in \L(s) \wedge \sigma_2 t \in \L(s')
	\wedge vars(t) \subset V$
	& if $\theta \sortdef rsg_V(V,s,s')$	\\
$p_\theta(t)$
	& $:\Lra t = \phi(t',t'')$	
	& if $\theta \sortdef \phi(t',t'')$	\\
$p_\theta(t)$
	& $:\Lra t \in \L(\theta)$
	& else	\\
\end{tabular}
$$

In the latter case, we have nothing to show.
In the two former cases, we make a case distinction according to
the rule from Alg.~\eqr{vrsg} that defined $\theta$.
The second case has been introduced to put the result sort
definitions into normal form.
\be
\item In this case, no new sort definition is introduced.
\item We have
	\begin{tabular}[t]{@{}ll@{\hspace*{0.5cm}}l@{}}
	& $p_\theta(t)$	\\
	$\Lra$ & $\sigma_1 t \in \L(s) \wedge \sigma_2 t \in \L(s')
		\wedge vars(t) \subset V$
		& by Def.\ of $p_\theta$	\\
	$\Lra$ & $\exists i \;\; \sigma_1 t \in \L(s_i)
		\wedge \sigma_2 t \in \L(s')
		\wedge vars(t) \subset V$
		& by Def.\ of $\L(s)$	\\
	$\Lra$ & $\exists i \;\; p_{\theta_i}(t)$
		& by Def.\ of $p_{\theta_i}$	\\
	\end{tabular}
\item Similar to 2.
\item
	\begin{tabular}[t]{@{}ll@{\hspace*{0.5cm}}l@{}}
	& $p_\theta(t)$	\\
	$\Lra$ & $\sigma_1 t \in \L(s) \wedge \sigma_2 t \in \L(s')
		\wedge vars(t) \subset V$
		& by Def.\ of $p_\theta$	\\
	$\Lra$ & $\exists t_{11} \in \L(s_1),\ldots, t_{1n} \in \L(s_n)
		\;\; \exists t_{21} \in \L(s'_1),\ldots,
		t_{2n} \in \L(s'_n)$	\\
		& $\sigma_1 t = f(t_{11},\ldots,t_{1n}) \wedge
		\sigma_2 t = f(t_{21},\ldots,t_{2n}) \wedge
		vars(t) \subset V$
		& by Def.\ of $\L(s),\L(s')$	\\

	$\Lra$ & $\exists t_1,\ldots,t_n \;\;
		t=f(t_1,\ldots,t_n)$	\\
		& $\wedge
		\bigwedge_{i=1}^n
		\sigma_1 t_i \in \L(s_i) \wedge
		\sigma_2 t_i \in \L(s'_i) \wedge
		vars(t_i) \subset V$
		& $(*)$	\\

	$\Lra$ & $\exists t_1,\ldots,t_n \;\;
		t=f(t_1,\ldots,t_n) \wedge \bigwedge_{i=1}^n
		p_{\theta_i}(t_i)$
		& by Def.\ of $p_{\theta_i}$	\\
	\end{tabular}

	$(*)$:
	\begin{tabular}[t]{@{}ll@{\hspace*{0.5cm}}l@{}}
	``$\Ra$'': &
		since $\sigma_1 x$ and $\sigma_2 x$
		never start with the same function symbol $f$,	\\
		& we must have $t = f(t_1,\ldots,t_n)$	\\
		& with $\sigma_1 t_i = t_{1i} \in \L(s_i)$,	\\
		& the case is similar for $\sigma_2$	\\
	``$\La$'': &
		choose $t_{1i} := \sigma_1 t_i$,
		and $t_{2i} := \sigma_2 t_i$	\\
	\end{tabular}
\item
	\begin{tabular}[t]{@{}ll@{\hspace*{0.5cm}}l@{}}
	& $p_\theta(t)$	\\
	$\Lra$ & $\sigma_1 t \in \L(f(s_1,\ldots,s_n))$ and
		$\sigma_2 t \in \L(f'(s'_1,\ldots,s'_{n'}))$ and
		$vars(t) \subset V$
		& by Def.\ of $p_\theta$	\\
	$\Ra$ & $\sigma_1 t = f(\ldots)$ and $\sigma_2 t = f'(\ldots)$
		and $vars(t) \subset V$	\\
	$\Ra$ & $t \in \V$ and $vars(t) \subset V$	\\
	$\Ra$ & $t \in V$
		& by Def.\ of $vars$	\\
	$\Ra$ & $t = \phi(t',t'')$ for some $\tpl{t',t''} \in TP$
		& by Def.\ of $TP$	\\
	$\Ra$ & $\bigvee_{\tpl{t',t''}
		\in TP \cap (\L(s) \times \L(s'))}
		p_{\phi(t',t'')}(t)$
		& by Def.\ of $p_{\phi(t',t'')}$	\\
	\\
	\mca{2}{conversely:}	\\
	& $\bigvee_{\tpl{t',t''} \in TP \cap (\L(s) \times \L(s'))}
		p_{\phi(t',t'')}(t)$	\\
	$\Ra$ & $t = \phi(t',t'')$ for some
		$\tpl{t',t''} \in TP \cap (\L(s) \times \L(s'))$
		& by Def.\ of $p_{\phi(t',t'')}$	\\
	$\Ra$ & $\sigma_1 t = t' \in \L(s)$ and
		$\sigma_2 t = t'' \in \L(s')$ and $t \in V$
		& by Def.\ of $\sigma_1,\sigma_2$	\\
	$\Ra$ & $p_\theta(t)$
		& by Def.\ of $p_\theta$	\\
	\end{tabular}
\ee

}

In Case 5., all pairs in $\tpl{t,t'} \in TP$ are checked for whether
$t \in \L(s)$ and $t' \in \L(s')$.
For checking $t \in \L(s)$, a straightforward algorithm as shown in
Fig.~\ref{Simple Algorithm for Testing Sort Membership} is used.
The notation $t:s$ means that term $t$ is a member of the sort $s$;
intermediate results are cached if $s$ is a sort name.

\begin{figure}
\begin{center}
\begin{tabular}{@{}cl@{\hspace*{0.5cm}}l@{\hspace*{0.5cm}}l@{}}
$t \;:\; s$
	&& \raisebox{-2ex}[0ex][0ex]{$s \sortdef s'$}
	& \raisebox{-2ex}[0ex][0ex]{(Sort Def)} \\
\cline{1-1}
$t \;:\; s'$	\\
\\
$t \;:\; s_1 \1\mid s_2$
	&&& \raisebox{-2ex}[0ex][0ex]{(Disjunction L)}	\\
\cline{1-1}
$t \;:\; s_1$	\\
\\
$t \;:\; s_1 \1\mid s_2$
	&&& \raisebox{-2ex}[0ex][0ex]{(Disjunction R)}	\\
\cline{1-1}
$t \;:\; s_2$	\\
\\
$f(t_1,\ldots,t_n) \;:\; f(s_1,\ldots,s_n)$
	&&& \raisebox{-2ex}[0ex][0ex]{(Function)}	\\
\cline{1-1}
$t_1 \;:\; s_1$
	$\;\;\;\;\;\ldots\;\;\;\;\;$
	$t_n \;:\; s_n$	\\
\end{tabular}
\end{center}
\caption{Simple Algorithm for Testing Sort Membership}
\label{Simple Algorithm for Testing Sort Membership}
\end{figure}

\subsection{Optimized Argument Selection}
\label{Optimized Argument Selection}

We now discuss how to avoid recursive calls of $rsg_V$, and
analogously of $rsg$, that yield $\bot$.

Consider the sort definitions of $s_0$ and $s_1$ in
Fig.~\ref{Sort Definition of $s_0,s_1$}. In the naive
implementation, $hsg(s_0,s_1)$ amounts to sixteen recursive
calls, shown in lines 3 to 6 of Fig.~\ref{Computation of
$hsg(s_0,s_1)$}. However, by comparing the leading function
symbols in each call, we can see immediately that twelve of them
will return the empty sort, and can hence be ignored, while only
four of them start with identical function symbols and thus need
to be further evaluated: 

$$hsg(s_0+s_0,s_0+s_1) \mid hsg(s_0+s_0,s_1+s_0) \mid
hsg(s_0*s_n,s_1*s_1) \mid hsg(s_n*s_0,s_1*s_1).$$

The situation is slightly more complicated
when computing $rsg_V$: certain combinations
of different leading function symbols may lead to a variable in $V$ and
thus cannot be ignored;
e.g.\ if $v_1 = \phi(0,\suc(0))$
and $v_2 = \phi(0,0+0)$,
the calls $rsg_V(\{v_1,v_2\},0,\suc(0))$
and $rsg_V(\{v_1,v_2\},0,s_0+s_0)$
will result in $v_1$, and $v_2$, respectively.

\begin{figure}
\begin{center}
\begin{picture}(1.4,1.4)
	%\put(0,0){\makebox(0,0){+}}
\put(0.200,0.300){\circle*{0.100}}
	\put(0.200,0.200){\makebox(0.000,0.000)[t]{$v_1$}}
\put(1.200,0.300){\circle*{0.100}}
	\put(1.200,0.200){\makebox(0.000,0.000)[t]{$v_2$}}
\put(0.700,0.800){\circle*{0.100}}
\put(1.200,1.300){\circle*{0.100}}
\put(0.700,0.800){\vector(-1,-1){0.500}}
	\put(0.450,0.600){\makebox(0.000,0.000)[br]{$\suc$}}
\put(0.700,0.800){\vector(1,-1){0.500}}
	\put(0.950,0.600){\makebox(0.000,0.000)[bl]{$+$}}
\put(1.200,1.300){\vector(-1,-1){0.500}}
	\put(0.950,1.100){\makebox(0.000,0.000)[br]{$0$}}
\end{picture}
\end{center}
\caption{Example: Search Tree}
\label{Example: Search Tree}
\end{figure}
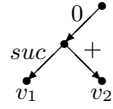

We present below an algorithm for selecting the argument pairs
for recursive $rsg_V$ calls.
In a first step, we use the set $V$ to build a search tree $VT^*$ of
pairs\footnote{$N$-tuples in the general case, where $N$ sorts are
simultaneously anti--unified.}
of different function symbols which need to be considered by $rsg_V$.
Building the search tree is a straightforward matter.
In the above example, we get the tree shown in Fig.~\ref{Example: Search
Tree}. 
In the general case, each path in the search tree has a length of $N$.
\\
The sort definitions are expected to be
in head normal form, i.e., of the general form

$$s \sortdef f_1(s_{11},\ldots,s_{1n_1}) \mid \ldots \mid
f_m(s_{m1},\ldots,s_{mn_m}).$$

We sort the disjuncts
$f_1(s_{11},\ldots,s_{1n_1})$,
\ldots, $f_m(s_{m1},\ldots,s_{mn_m})$
ascendingly by their leading function symbols,
assuming an arbitrary irreflexive total ordering $\prec$,
and group together all disjuncts with the same function symbol (called
``pre--grouping'' to distinguish it
from the kernel grouping algorithm).
For example, from the definitions of $s_0$ and $s_1$ in Fig.~\ref{Sort
Definition of $s_0,s_1$}, we get after sorting 

$$
\begin{tabular}[t]{@{}lllllll@{}}
$[$ & $0,$ & $s_0+s_0,$ & $s_0*s_n,s_n*s_0$ & $]$     \\
$[$ & $\suc(s_0),$ & $s_0+s_1,s_1+s_0,$ & $s_1*s_1$ & $]$, \\
\end{tabular}
$$

and after pre--grouping

$$
\begin{tabular}[t]{@{}llllll@{}}
$[$ & $[0],$ & $[s_0+s_0],$ & $[s_0*s_n,s_n*s_0]$ & $]$
	& $=: Ps_1$     \\
$[$ & $[\suc(s_0)],$ & $[s_0+s_1,s_1+s_0],$ & $[s_1*s_1]$ & $]$
	& $=: Ps_2$. \\
\end{tabular}
$$

In the general case, we get $N$ such lists of pre--groups.
These lists of pre--groups, together with the search tree,
are given to the kernel grouping algorithm shown in Fig.~\ref{Grouping
Rules}, 
which produces a list of $N$-tuples,
each serving as arguments for a
subsequent recursive $rsg_V$ call.

The kernel grouping algorithm takes four input arguments:
\bi
\item a flag $E$ which can take the values $eq$ or $ne$,
	the latter indicating that different function symbols
	have occurred in the current group
\item the search tree $VT^*$
\item the current group (a list of pre--groups)
\item the current list of lists of pre--groups.
\ei

A pre--group, like $[s_0+s_1,s_1+s_0]$, is denoted by $P$, or $P'$;
a list of pre--groups by $Ps$, $Ps'$, or $Group$.
$\deref P$ denotes the common head symbol of all terms in the
pre--group $P$.
For a search (sub)tree $VT$,
we denote by $VT_f$ the subtree of $VT$ at the
branch labeled $f$; if a branch labeled $f$ does not exist within
$VT$, we get $VT_f = nil$, which denotes the empty search tree.

The output accumulator is not shown in Fig.~\ref{Grouping Rules};
instead, in rule \rulename{Output}, the statement 
``output $reverse(Group)$'' occurs.
If this statement is reached, $Group = [P_N,\ldots,P_1]$ is a
list of length $N$, and a list
$[ \tpl{t_1,\ldots,t_N} \mid t_i \in P_i, i=1,\ldots,N]$
of all $N$-tuples, such
that the $i$-th component is a member of the $i$-th pre--group of
$reverse(Group)$,
is added to the output; the order of tuples within the list does not
matter.
The rule \rulename{Output} and the
rules named \rulename{Abort~\ldots} do not have a successor,
whereas rules \rulename{Join} and \rulename{Follow} have two.
The start configuration is

$$
\begin{tabular}[t]{@{}*{4}{l@{\hspace*{0.5cm}}}@{}}
\hline
$eq$ & $VT^*$ & $[\;]$ & $[Ps_1,\ldots,Ps_N]$	\\
\end{tabular}
$$

where $VT^*$ is the initial search tree
and $Ps_1,\ldots,Ps_N$ are the $N$ lists of pre--groups obtained after
pre--grouping.
Figure~\ref{Example: Argument Selection Run} shows a computation
example.
For the sake of readability, pre--groups are written in braces,
and the $Ps_i$ in the fourth argument are separated by semicolons.

The same algorithm can also be used for $rsg$ and $hsg$
instead of $rsg_V$
by providing an empty search tree.

\begin{figure}
\begin{center}
\begin{tabular}{@{}*{7}{l@{\hspace*{0.5cm}}}@{}}
$eq$ & $VT$ & $[P \mid Group]$
	& $[[P' \mid Ps'] \mid Rest]$
	&& \raisebox{-2ex}[0ex][0ex]{$f = \deref P = \deref P'$}
	& \raisebox{-2ex}[0ex][0ex]{(Join)}	\\
\cline{1-4}
$eq$ & $VT_f$ &
	$[P',P \mid Group]$
	& $Rest$	\\
$eq$ & $VT$ & $[P \mid Group]$
	& $[Ps' \mid Rest]$	\\
\\
\\
$E$ & $VT$ & $Group$ & $[\;]$
	&&& \raisebox{-2ex}[0ex][0ex]{(Output)}	\\
\cline{1-4}
&& \mca{2}{output $reverse(Group)$}	\\
\\
\\
$ne$ & $nil$ & $Group$ & $Any$
	&&& \raisebox{-2ex}[0ex][0ex]{(Abort Ne)}	\\
\cline{1-4}
\\	%&& \mca{2}{output $[\;]$}	\\
\\
\\
$eq$ & $nil$ & $[P \mid Group]$
	& $[[P' \mid Ps'] \mid Rest]$
	&& \raisebox{-2ex}[0ex][0ex]{$\deref P \prec \deref P'$}
	& \raisebox{-2ex}[0ex][0ex]{(Abort Eq)}	\\
\cline{1-4}
\\	%&& \mca{2}{output $[\;]$}	\\
\\
\\
$E$ & $VT$ & $Group$
	& $[[P' \mid Ps'] \mid Rest]$
	&& $f = \deref P'$
	& \raisebox{-2ex}[0ex][0ex]{(Follow)}	\\
\cline{1-4}
$ne$ & $VT_f$ & $[P' \mid Group]$ & $Rest$
	&& $Group \neq [\;]$	\\
$E$ & $VT$ & $Group$ & $[Ps' \mid Rest]$	\\
\\
\\
$E$ & $VT$ & $Group$ & $[[P' \mid Ps'] \mid Rest]$
	&& \raisebox{-2ex}[0ex][0ex]{$Group \neq [\;]$}
	& \raisebox{-2ex}[0ex][0ex]{(Skip)}	\\
\cline{1-4}
$E$ & $VT$ & $Group$ & $[Ps' \mid Rest]$	\\
\\
\\
$E$ & $VT$ & $[\;]$
	& $[[P' \mid Ps'] \mid Rest]$
	&& \raisebox{-2ex}[0ex][0ex]{$f = \deref P'$}
	& \raisebox{-2ex}[0ex][0ex]{(Init)}	\\
\cline{1-4}
$eq$ & $VT_f$ & $[P']$ & $Rest$	\\
$eq$ & $VT$ & $[\;]$ & $[Ps' \mid Rest]$	\\
\\
\\
$E$ & $VT$ & $Group$ & $[[\;] \mid Rest]$
	&&& \raisebox{-2ex}[0ex][0ex]{(Abort $[\;]$)}	\\
\cline{1-4}
\\	%&& \mca{2}{output $[\;]$}	\\
\end{tabular}
\end{center}
\caption{Grouping Rules}
\label{Grouping Rules}
\end{figure}

\begin{figure}
\begin{center}
\begin{tabular}[t]{@{}ll@{}l@{}}
Precedence: & {\tt (0)} $\prec$ {\tt (s)} $\prec$ {\tt (*)}
	$\prec$ {\tt (+)} \\
Input: & {\tt [[0,s0*sn,sn*s0,s0+s0],}
	& {\tt [\suc(s0),s1*s1,s0+s1,s1+s0]]}	\\
Pre--Grouped:
	& {\tt [[[0],[s0*sn,sn*s0],[s0+s0]],}
	& {\tt [[\suc(s0)],[s1*s1],[s0+s1,s1+s0]]]}	\\
Initial Search Tree $VT^*$:
	& $\stackrel{\tt 0}{\raa} \stackrel{\tt s}{\raa}$
	& ({\tt nil} denotes empty subtree)	\\
%Run: & (see below)	\\
Output: &
	\mca{2}{\tt [[0,\suc(s0)],[s0*sn,s1*s1],[sn*s0,s1*s1],%
	[s0+s0,s0+s1],[s0+s0,s1+s0]]}	\\
\end{tabular}

\begin{verbatim}
eq     0     []     [ [{0},{S0*Sn,Sn*S0},{S0+S0}] ; [{suc(S0)},{S1*S1},{S0+S1,S1+S0}] ]
   (Init)
   eq     s     [{0}]     [ [{suc(S0)},{S1*S1},{S0+S1,S1+S0}] ]
      (Follow s)
      ne     nil     [{suc(S0)},{0}]     [ ]
         (Output [0,suc(S0)])
      eq     s     [{0}]     [ [{S1*S1},{S0+S1,S1+S0}] ]
         (Skip)
         eq     s     [{0}]     [ [{S0+S1,S1+S0}] ]
            (Skip)
            eq     s     [{0}]     [ [] ]
               (Abort [])
   eq     0     []     [ [{S0*Sn,Sn*S0},{S0+S0}] ; [{suc(S0)},{S1*S1},{S0+S1,S1+S0}] ]
      (Init)
      eq     nil     [{S0*Sn,Sn*S0}]     [ [{suc(S0)},{S1*S1},{S0+S1,S1+S0}] ]
         (Skip)
         eq     nil     [{S0*Sn,Sn*S0}]     [ [{S1*S1},{S0+S1,S1+S0}] ]
            (Join *)
            eq     nil     [{S1*S1},{S0*Sn,Sn*S0}]     [ ]
               (Output [S0*Sn,S1*S1],[Sn*S0,S1*S1])
            eq     nil     [{S0*Sn,Sn*S0}]     [ [{S0+S1,S1+S0}] ]
               (Abort Eq)
      eq     0     []     [ [{S0+S0}] ; [{suc(S0)},{S1*S1},{S0+S1,S1+S0}] ]
         (Init)
         eq     nil     [{S0+S0}]     [ [{suc(S0)},{S1*S1},{S0+S1,S1+S0}] ]
            (Skip)
            eq     nil     [{S0+S0}]     [ [{S1*S1},{S0+S1,S1+S0}] ]
               (Skip)
               eq     nil     [{S0+S0}]     [ [{S0+S1,S1+S0}] ]
                  (Join +)
                  eq     nil     [{S0+S1,S1+S0},{S0+S0}]     [ ]
                     (Output [S0+S0,S0+S1],[S0+S0,S1+S0])
                  eq     nil     [{S0+S0}]     [ [] ]
                     (Abort [])
         eq     0     []     [ [] ; [{suc(S0)},{S1*S1},{S0+S1,S1+S0}] ]
            (Abort [])
\end{verbatim}
\end{center}

\caption{Example: Argument Selection Run}
\label{Example: Argument Selection Run}
\end{figure}

We give below an estimation of the time complexity of the
grouping algorithm.
Assume that we are given $N$ sorts whose definitions are in head normal
form and have each at most $m$ disjuncts on their right--hand side;
moreover, assume that the initial search tree $VT^*$ was built from $n$
variables.
Sorting the disjuncts by leading function symbols takes
$N \cdot m \cdot \log m$ time;
pre--grouping takes $N \cdot m$ time.
We obtain $N$ lists, each of which contains at most $m$ groups.
For the sake of brevity, we denote the third argument of the rules
in Fig.~\ref{Grouping Rules} by $G$, and the fourth by $H$.
Any 4-tuple $\tpl{E,VT,G,H}$ that occurs during the computation is
called a state; a state is called final if a rule without successors is
applied to it.
If $VT$ of a state is a subtree of $VT^*$, we call the list of
function
symbols along the way from the root of $VT^*$ to the root of $VT$
the path of that state. Proceeding from $VT$ to $VT_f$ means appending
$f$ to the actual path.
Observe the following properties of the rules:
\be
% \item \label{aufw1}
% 	The height of the actual search subtree $VT$ always equals
% 	$length(H)$.
\item \label{aufw3}
	To any state with $E = ne$ whose actual $VT$ is not a subtree of
	$VT^*$, the rule \rulename{Abort~Ne} will be applied,
	i.e.\ this state is final.
	Hence, each non--final state with $E = ne$ corresponds to a
	node in $VT^*$, viz.\ the root of the actual $VT$.
\item \label{aufw4}
	A non--final state with $E=ne$ and a path $[f_1,\ldots,f_k]$
	has $H = [Ps_{k+1},\ldots,Ps_N]$.
	Moreover, we have $f_i = \deref P_i$ for some $P_i$ in $Ps_i$
	for all $i=1,\ldots,k$.
\item \label{aufw5}
	Each successor of a state with $E=ne$
	corresponds to a different subtree
	$VT_f$ of $VT^*$ (possibly $VT_f = nil$),
	or has $head(H)$ shortened by one element.
	The latter makes it impossible to reach $VT_f$ again,
	since $f = \deref P' \prec P''$ for all later $P''$ in $Ps'$,
	using the terminology common to rules \rulename{Join},
	\rulename{Follow}, \rulename{Skip}, and \rulename{Init}.
\item \label{aufw6}
	At most $m$ different non--final states with $E = ne$ may
	correspond to the same node of $VT^*$.
	This follows from~\ref{aufw3} and~\ref{aufw5}.
\item \label{aufw2}
	$E$ has the value $eq$ iff there is a function symbol $f$
	such that $\deref P = f$ for all $P$ in $G$;
	in this case, we write $\deref G := f$.
\item \label{aufw7}
	At most $m$ different states with $E=eq$ and $G \neq [\;]$
	may have the same $\deref G$ and the same $length(G)$,
	since each rule either increases $length(G)$
	or shortens $head(H)$, the latter being possible at most $m$
	times.
	No rule decreases $length(G)$; and it always holds that
	$length(G) \leq N$.
\item \label{aufw8}
	Similar to~\ref{aufw7}, at most $m$ different states with $E=eq$
	and $G=[\;]$ can exist.
\item \label{aufw9}
	Each non--final state has at most two final states as
	successors.
\item \label{aufw10}
	The initial search tree has $N \cdot n + 1$ nodes,
	if no two variables share a part of their path except for the
	root node.
	In all other cases, the number of nodes is decreased by path
	sharing.
\ee

From~\ref{aufw10}, there are at most $N \cdot n + 1$ nodes in $VT^*$;
from~\ref{aufw3} and~\ref{aufw6}, there are at most
$m \cdot (N \cdot n + 1)$ non--final states with $E=ne$;
from~\ref{aufw7} and~\ref{aufw8}, there are at most
$m \cdot m \cdot N + m$ non--final states with $E=eq$.
Hence, there are at most
$m \cdot (N \cdot n + 1) + m \cdot m \cdot N + m
= m \cdot (N \cdot n + N \cdot m + 2)
\leq m \cdot N \cdot (m+n+1)$
non--final states,
and, by~\ref{aufw9},
at most
$3 \cdot m \cdot N \cdot (m+n+1)$ states in all.
A final state may result in an output of at most $g^N$ $N$-tuples,
where $g$ is the maximum number of disjuncts of a sort--definition's
right--hand side that start with the same function symbol.
Including sorting and pre--grouping,
our approach takes $\O(N \cdot m \cdot (m+n) \cdot g^N)$ time.
Since the naive approach ---~without grouping~--- 
takes $\O(m^N)$ time, grouping makes sense especially for $N > 2$,
i.e.\ when more than two input sorts are to be anti--unified.
Moreover, using the grouping algorithm makes the runtime more
insensitive to the number of different function symbols in the
background theory. 
Figure~\ref{Runtimes for Grouping} shows some
runtime measurements for grouping including pre--grouping. See
Sect.~\ref{E--Anti--Unification -- Runtimes} for more
measurements.

\begin{figure}
\begin{center}
\begin{tabular}[t]{@{}l@{\hspace*{0.5cm}}r@{\hspace*{0.5cm}}r@{}}
Sort Defs of & User Time in msec & No.\ of Output Groups	\\
\hline
$s_0,s_1$ & 13 & 4	\\
$s_0,s_1,s_2$ & 23 & 10	\\
$s_0,s_1,s_2,s_3$ & 39 & 32	\\
$s_0,s_1,s_2,s_3,s_4$ & 102 & 144	\\
\end{tabular}
\end{center}
\caption{Runtimes for Grouping}
\label{Runtimes for Grouping}
\end{figure}

%\clearpage
\subsection{Runtimes}
\label{E--Anti--Unification -- Runtimes}

In this section, we discuss some figures for runtime
measurements. All runtimes given in this paper refer to user
time of a SICStus PROLOG 2.1 \#8 implementation on a SUN 4/50GX
SPARC with 40 MHz clock and 64 MBytes main storage. 

\begin{figure}
\begin{center}
\begin{picture}(13.6,2.5)
	%\put(0,0){\makebox(0,0){+}}
\put(0.000,1.000){\circle*{0.100}}	% g
\put(2.640,1.000){\circle*{0.100}}	% v
\put(9.120,1.000){\circle*{0.100}}	% vgst
\put(11.760,1.000){\circle*{0.100}}	% vgsd
\put(12.720,1.000){\circle*{0.100}}	% vgsdb
\put(4.000,2.050){\circle*{0.100}}	% vst
\put(7.280,0.050){\circle*{0.100}}	% vgs
\put(0.000,1.200){\makebox(0.000,0.000)[b]{g}}
\put(2.640,1.200){\makebox(0.000,0.000)[b]{v}}
\put(9.120,1.200){\makebox(0.000,0.000)[b]{gstv}}
\put(11.760,1.200){\makebox(0.000,0.000)[br]{dgsv}}
\put(12.720,1.200){\makebox(0.000,0.000)[bl]{bdgsv}}
\put(4.000,2.250){\makebox(0.000,0.000)[b]{stv}}
\put(7.280,-0.150){\makebox(0.000,0.000)[t]{gsv}}
\put(1.320,1.100){\makebox(0.000,0.000)[b]{3}}
\put(5.600,1.100){\makebox(0.000,0.000)[b]{15}}
\put(10.440,1.100){\makebox(0.000,0.000)[b]{3}}
\put(12.240,1.100){\makebox(0.000,0.000)[b]{1.5}}
\put(6.560,1.600){\makebox(0.000,0.000)[bl]{8.3}}
\put(8.200,0.400){\makebox(0.000,0.000)[tl]{2.1}}
\put(0.000,1.000){\line(1,0){12.720}}
\put(9.120,1.000){\line(-5,1){5.120}}
\put(9.120,1.000){\line(-2,-1){1.840}}
\end{picture}
\end{center}
\caption{Graphical Representation of Improvement Factors}
\label{Graphical Representation of Improvement Factors}
\end{figure}
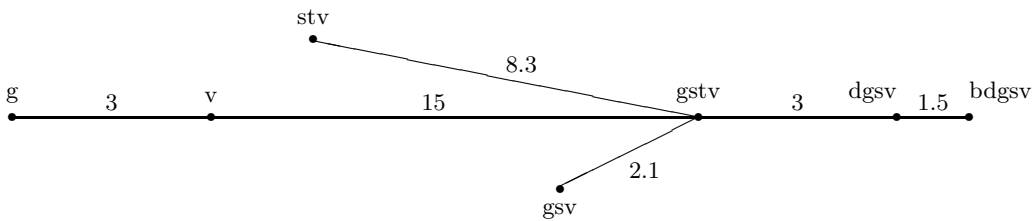

During development of the implementation prototype, we
experimented with several technical optimizations. The impact of
the successful ones (see Fig.~\ref{Technical Optimizations}) on
runtimes are discussed below.
Since runtime measurements have not been conducted
systematically at all stages of development, we tried to
estimate their impact from the available data shown in
Figs.~\ref{Lemma--Generation Runtimes} and~\ref{Series--Guessing
Runtimes}. These data are categorized and summarized in
Fig.~\ref{Measured Improvements of Technical Optimization
Groups}, yielding estimations of the improvement factors
obtained by applying several optimizations simultaneously.
Figure~\ref{Graphical Representation of Improvement Factors}
gives a graphical representation of these factors, the x--axis
being scaled logarithmically. In Fig.~\ref{Runtime Impact of
Single Technical Optimizations}, we have computed the impact of
each single technical improvement as far as possible, based on
the data from Fig.~\ref{Graphical Representation of Improvement
Factors}.

\begin{figure}
\begin{center}
\begin{tabular}{@{}|l|l|l|r|@{}}
\hline
Abbr. & Technical Optimization & Described in & Improvement	\\
&&& Factor	\\
\hline
b & use of balanced binary trees to implement {\tt Occ} sets \
	& Sect.~\ref{Linear Generalizations}
	& 1.5	\\
d & use of {\tt calc\_sort\_depths}
	& Sect.~\ref{Sort Enumeration} below	
	& 6.3	\\
g & use of {\tt group\_cr\_sorts}
	& Sect.~\ref{Optimized Argument Selection}
	& 8.3	\\
s & restricting the sort simplifier to {\tt simp1} \ldots {\tt simp4}
	& (obsolete) &	\\
t & without user trace && 2.1	\\
v & use of $rsg_V$
	& Sect.~\ref{Variable--Restricted E--Anti--Unification}
	& 21.4	\\
\hline
\end{tabular}
\end{center}
\caption{Technical Optimizations}
\label{Technical Optimizations}
\end{figure}

\begin{figure}
\begin{center}
\begin{tabular}[t]{@{}|l@{\hspace*{0.3cm}}l|r|r
	@{\hspace*{0.3cm}}r@{\hspace*{0.3cm}}r|@{}}
\hline
From  & To    & Factor & Line & vs.\ Line & in Figure	\\
\hline
\hline
{\bf g}     & {\bf v}     &  3.0 & \multicolumn{3}{|c|}{}	\\
\cline{3-3}
      &       &  1.9 &  5 &  6 & \ref{Lemma--Generation Runtimes}\\
      &       &  2.0 &  1 &  2 & \ref{Lemma--Generation Runtimes}\\
      &       &  2.0 &  3 &  4 & \ref{Lemma--Generation Runtimes}\\
      &       &  3.7 & 14 & 15 & \ref{Lemma--Generation Runtimes}\\
      &       &  3.7 & 16 & 17 & \ref{Lemma--Generation Runtimes}\\
      &       &  4.6 & 12 & 13 & \ref{Lemma--Generation Runtimes}\\
\hline
v     & {\bf gst}v  & 15.0 & \multicolumn{3}{|c|}{}	\\
\cline{3-3}
      &       & 12.0 & 17 & 18 & \ref{Lemma--Generation Runtimes}\\
      &       & 29.0 & 19 & 22 & \ref{Lemma--Generation Runtimes}\\
\hline
gs{\bf t}v  & {\bf d}gsv  &  3.0 & \multicolumn{3}{|c|}{}	\\
\cline{3-3}
      &       &   -- &  1 &  2 & \ref{Series--Guessing Runtimes} \\
      &       &  0.3 &  3 &  4 & \ref{Series--Guessing Runtimes} \\
      &       &  3.5 &  5 &  6 & \ref{Series--Guessing Runtimes} \\
      &       &  7.8 &  7 &  8 & \ref{Series--Guessing Runtimes} \\
\hline
dgsv  & {\bf b}dgsv & 1.5 & \multicolumn{3}{|c|}{}	\\
\cline{3-3}
      &       &  1.4 & 10 & 11 & \ref{Series--Guessing Runtimes} \\
      &       &  2.1 & 14 & 15 & \ref{Series--Guessing Runtimes} \\
\hline
stv   & {\bf g}stv & 8.3 & 20 & 22 & \ref{Lemma--Generation Runtimes}\\
\hline
gsv   & gs{\bf t}v & 2.1 & 21 & 22 & \ref{Lemma--Generation Runtimes}\\
\hline
\end{tabular}
\end{center}
\caption{Measured Improvements of Technical Optimization Groups}
\label{Measured Improvements of Technical Optimization Groups}
\end{figure}

Several measurements have been made in order to explore the
practical runtime behavior of the final version (with
optimizations $bdgsv$) of the algorithm wrt.\ number and size of
input sorts. 
Figure~\ref{E--Anti--Unification Runtime vs. Size of Sort
Definitions}(a) shows the E--anti--unification runtime vs.\ size
of sort definitions; figures denote runtime of
$rsg_V(\{\},Sorts)$ in seconds. $s_i$ denotes the equivalence
class of $i$ wrt.\ the respective background theory, cf.\
Fig.~\ref{Equivalence Classes for Background Theory (5) from
Fig.},~\ref{Equivalence Classes for Background Theory (6) from
Fig.}. ``$>n$'' means running out of memory after $n$ seconds. 

In the third column, we have included
the $(-)$ operator in the background
theory, such that equivalence classes of terms can no longer
be described by our sorts.
We therefore cut off the sort definitions at $20$ --- cf.\
Sect.~\ref{Modeling Equivalence Classes as Sorts}.
Since adding $(-)$ to the background theory led to an erratic
increase in runtime, we additionally measured the dependency of
runtime on the ``cut--off point''; the result is shown in
Fig.~\ref{E--Anti--Unification Runtime vs. Size of Sort
Definitions} (b), indicating the runtimes in seconds for
$rsg_V(\{\},s_0,s_2)$ with cut--off points $n$ from $2$ to $20$. 

Figure~\ref{E--Anti--Unification Runtime vs. Size and Number of
Sorts} shows the runtime for $rsg_V(\{\},s_X^i)$ wrt.\ the
background theory of $\{+\}$, $\{+,*\}$, and $\{+,*,/\}$,
respectively, where $s_X^i$ denotes the list that contains $i$
times the sort $s_X$.

\begin{figure}
\begin{center}
\begin{tabular}[t]{@{}|l|ll|rl|@{}}
\hline
Optimization & From & To & Factor &	\\
\hline
b  & dgsv & bdgsv &  1.5 &	\\
d  & gsv  & dgsv  &  6.3 & $= 2.1 * 3.0$	\\
g  & stv  & gstv  &  8.3 &	\\
t  & gsv  & gstv  &  2.1 &	\\
sv & g    & gsv   & 21.4 & $= 3.0 * 15.0 / 2.1$	\\
\hline
\end{tabular}
\end{center}
\caption{Runtime Impact of Single Technical Optimizations}
\label{Runtime Impact of Single Technical Optimizations}
\end{figure}

\begin{figure}
\begin{center}
\renewcommand{\arraystretch}{1.0}
\begin{tabular}[b]{@{}|l|r@{\hspace*{0.5cm}}r@{\hspace*{0.5cm}}r|@{}}
\hline
$Sorts$ & $+$ & $+,*$ & $+,-,*$	\\
\hline
 $[s_0,s_1,s_2]$ & 0 &	0 & $>266$	\\
 $[s_0,s_1,s_3]$ & 1 &	2 &		\\
 $[s_0,s_1,s_4]$ & 0 &	3 &		\\
 $[s_0,s_1,s_5]$ & 1 &	4 &		\\
 $[s_0,s_1,s_6]$ & 1 &	5 &		\\
 $[s_0,s_1,s_7]$ & 1 &	6 &		\\
 $[s_0,s_1,s_8]$ & 1 &	9 &		\\
 $[s_0,s_1,s_9]$ & 2 &	8 &		\\
$[s_0,s_1,s_{10}]$ & 2 & 10 &		\\
$[s_0,s_1,s_{11}]$ & 2 & 13 &		\\
$[s_0,s_1,s_{12}]$ & 2 & 15 &		\\
$[s_0,s_1,s_{13}]$ & 3 & 15 &		\\
$[s_0,s_1,s_{14}]$ & 3 & 17 &		\\
$[s_0,s_1,s_{15}]$ & 4 & 19 &		\\
$[s_0,s_1,s_{16}]$ & 6 & 21 &		\\
$[s_0,s_1,s_{17}]$ & 5 & 23 &		\\
$[s_0,s_1,s_{18}]$ & 7 & 24 &		\\
$[s_0,s_1,s_{19}]$ & 5 & 26 &		\\
$[s_0,s_1,s_{20}]$ & 7 & 29 &		\\
\hline
\multicolumn{1}{l}{}	\\
\multicolumn{4}{@{}c@{}}{(a)}	\\
\end{tabular}
\tab\tab\tab
\begin{tabular}[b]{@{}|l|r@{\hspace*{0.5cm}}r|@{}}
\hline
$n$	& $+,-$ & $+,-,*$	\\
\hline
$2$ & 0 & \\
$3$ & 1 & \\
$4$ & 3 & \\
$5$ & 7 & \\
$6$ & 15 &	\\
$7$ & 24 &	\\
$8$ & 41 &	\\
$9$ & 61 &	\\
$10$ & 96 &	\\
$11$ & 135 &	\\
$12$ & 194 &	\\
$13$ & 272 &	\\
$14$ & 363 &	\\
$15$ & 485 &	\\
$16$ & 640 &	\\
$17$ & 847 &	\\
$18$ & 1066 &	\\
$19$ & 1349 &	\\
$20$ & 1804   & 1912	\\
\hline
\multicolumn{1}{l}{}	\\
\multicolumn{3}{@{}c@{}}{(b)}	\\
\end{tabular}
\end{center}
\caption{E--Anti--Unification Runtime vs.\ Size of Sort Definitions}
\label{E--Anti--Unification Runtime vs. Size of Sort Definitions}
\end{figure}

\begin{figure}
{
\renewcommand{\arraystretch}{1.0}
\begin{picture}(15,11.5)
	%\put(0,0){\makebox(0,0){+}}
\put(0.000,2.000){\makebox(0.000,0.000)[bl]{%
	\begin{tabular}[b]{@{}|l|*{8}{r}|@{}}
	\hline
	$+$ & $i=2$    & $3$ & $4$ & $5$ & $6$ & $7$ & $8$ & $9$ \\
	\hline
	$s_0$	 & 0   & 1   & 1   & 1	 & 1   & 2   & 1   & 0	\\
	$s_1$	 & 0   & 0   & 1   & 6	 & 11  & 27  & 87  & 279 \\
	$s_2$	 & 1   & 2   & 16  & 84	 & 579 &$>3034$ &&	\\
	$s_3$	 & 1   & 13  & 106 & 1267&$>5313$	\\
	$s_4$	 & 2   & 31  & 576 &$>5009$	\\
	$s_5$	 & 4   & 92  & 2546	\\
	$s_6$	 & 7   & 221 &$>4518$	\\
	$s_7$	 & 10  & 492	\\
	$s_8$	 & 19  & 1080	\\
	$s_9$	 & 27  & 2092	\\
	$s_{10}$ & 39  &$>1874$ \\
	$s_{11}$ & 53	\\
	$s_{12}$ & 77	\\
	$s_{13}$ & 105	\\
	$s_{14}$ & 138	\\
	$s_{15}$ & 178	\\
	$s_{16}$ & 232	\\
	$s_{17}$ & 290	\\
	$s_{18}$ & 379	\\
	$s_{19}$ & 466	\\
	$s_{20}$ & 577	\\
	\cline{1-6}
	\end{tabular}
}}
\put(6.000,0.000){\makebox(0.000,0.000)[bl]{%
	\begin{tabular}[b]{@{}|l|*{8}{r}|@{}}
	\hline
	$+,*$ & $i=2$  & $3$ & $4$ & $5$ & $6$ & $7$ & $8$ & $9$ \\
	\hline
	$s_0$	 & 1   & 1   & 1   & 4	 & 15  & 38  & 113 & 344 \\
	$s_1$	 & 1   & 2   & 8   & 33	 & 138 & 577 & 2400&$>7464$   \\
	$s_2$	 & 2   & 9   & 49  & 318 & 2399&$>6466$	&&	\\
	$s_3$	 & 2   & 21  & 217 & 2632&$>5775$ &&&	\\
	$s_4$	 & 3   & 57  & 962 &$>5316$ &&&&	\\
	$s_5$	 & 7   & 133 &$>2511$ &&&&&	\\
	$s_6$	 & 10  & 309 &&&&&&	\\
	$s_7$	 & 17  & 643 &&&&&&	\\
	$s_8$	 & 24  & 1310&&&&&&	\\
	$s_9$	 & 34  & 2507&&&&&&	\\
	$s_{10}$ & 49  &$>2202$	 &&&&&&	\\
	$s_{11}$ & 67  &&&&&&&	\\
	$s_{12}$ & 96  &&&&&&&	\\
	$s_{13}$ & 123 &&&&&&&	\\
	$s_{14}$ & 160 &&&&&&&	\\
	$s_{15}$ & 209 &&&&&&&	\\
	$s_{16}$ & 261	\\
	$s_{17}$ & 335	\\
	$s_{18}$ & 423	\\
	$s_{19}$ & 525	\\
	$s_{20}$ & 632	\\
	\cline{1-5}
	\end{tabular}
}}
\put(11.000,0.000){\makebox(0.000,0.000)[bl]{%
	\begin{tabular}[b]{@{}|l|*{8}{r}|@{}}
	\hline
	$+,*,/$ & $i=2$& $3$ & $4$ & $5$ & $6$ & $7$ & $8$ & $9$ \\
	\hline
	$s_0$	 & 1   & 1   & 3   & 8	 & 17  & 44  &	   &	\\
	$s_1$	 & 794 &     &	   &	 &     &     &	   &	 \\
	$s_2$	 &     &     &	   &	 &     &     &	   &	\\
	\hline
	\end{tabular}
}}
\end{picture}
}

\caption{E--Anti--Unification Runtime vs.\ Size and Number of Sorts}
\label{E--Anti--Unification Runtime vs. Size and Number of Sorts}
\end{figure}

\clearpage
\section{Sort Enumeration}
\label{Sort Enumeration}

The algorithms $rsg$ and $rsg_V$ both return a sort as result.
In this section, we describe how the terms belonging to a given
sort are enumerated. 

Enumeration of all terms of a given sort is done depth first. It
is checked whether the sort is finite; in this case, a simple
depth--first traversion is sufficient. For an infinite sort, the
depth is limited to a fixed bound which is iteratively deepened.
The depth of a term is currently measured as the maximum number
of replacements of sort names by their definitions of all paths
from the term root to its leaves. However, several other depth
measures are possible; see the remarks on the $0,1,2,1,4,1,6$
series example in Sect.~\ref{Series Guessing -- Results and
Runtimes}. 

Cutting the depth at a fixed bound $n$ introduces many dead ends
in sort enumeration when disjunctions in sort definitions are
tried that can lead only to terms of a depth greater than $n$.
As this usually happens at each of the $n$ levels of recursive
descent, the number of dead ends is in the order of magnitude
$m^n$, where $m$ is the maximum number of disjuncts occurring in
a sort definition. For an example, see Figs.~\ref{Enumeration of
$s_{01}$} and~\ref{Enumeration of $s_{01}$ (contd.)} in
Sect.~\ref{An E--Anti--Unification Example}. An important
optimization therefore consists in recording for each sort $s$
the minimum depth $mindepth(s)$ of any of its terms; rule
\rulename{Sort~Def} is applicable only if this minimum depth is
smaller than the remaining depth bound. 

The rules of the optimized enumeration algorithm are given in
Fig.~\ref{Sort Enumeration Rules}. The notation $s : dp : t$
means that $t$ is a term of a depth $\leq dp$ in $\L(s)$, where
$s$ is an arbitrary sort expression. The resulting speedup of
6.3 compared to standard enumeration can be seen from line ``d''
in Fig.~\ref{Runtime Impact of Single Technical Optimizations}.
The values of the sort's minimum depths are computed using a
bottom--up algorithm acting on all (new) sort definitions
simultaneously, similar to the inhabitance algorithm from
\cite{Aiken.Murphy.1991b}, which is described in Sect.~\ref{Sort Depth
Computation}.

Originally, our implementation had the additional capability of
enumerating only terms in normal form wrt.\ a given set of
(linear) redices. However, line 9 in Fig.~\ref{Series--Guessing
Runtimes} shows that sort enumeration may still take
unacceptably long if only terms in normal form are to be
enumerated. The restriction to terms in normal form is therefore
achieved by intersecting the solution sort with the sort of all
terms in normal form (provided all redices are linear terms,
cf.\ Sect.~\ref{Filter Sorts}), rather than by restricting the
enumeration algorithm to terms in normal form. See line 10 in
Fig.~\ref{Series--Guessing Runtimes}.

\begin{figure}
\begin{center}
\begin{tabular}{@{}cl@{\hspace*{0.5cm}}l@{\hspace*{0.5cm}}l@{}}
$\bot \;:\; dp \;:\; t$
	&&& \raisebox{-2ex}[0ex][0ex]{(Bottom)} \\
\cline{1-1}
fail	\\
\\
$s \;:\; dp \;:\; t$
	&& $s \sortdef s'$,
	& \raisebox{-2ex}[0ex][0ex]{(Sort Def)}	\\
\cline{1-1}
$s' \;:\; dp\1-1 \;:\; t$
	&& $dp \geq mindepth(s)$	\\
\\
$s_1 \1\mid s_2 \;:\; dp \;:\; t$
	&&& \raisebox{-2ex}[0ex][0ex]{(Disjunction L)}	\\
\cline{1-1}
$s_1 \;:\; dp \;:\; t$	\\
\\
$s_1 \1\mid s_2 \;:\; dp \;:\; t$
	&&& \raisebox{-2ex}[0ex][0ex]{(Disjunction R)}	\\
\cline{1-1}
$s_2 \;:\; dp \;:\; t$	\\
\\
$f(s_1,\ldots,s_n) \;:\; dp \;:\; f(t_1,\ldots,t_n)$
	&&& \raisebox{-2ex}[0ex][0ex]{(Function)}	\\
\cline{1-1}
$s_1 \;:\; dp \;:\; t_1$
	$\;\;\;\;\;\ldots\;\;\;\;\;$
	$s_n \;:\; dp \;:\; t_n$	\\
\end{tabular}
\end{center}
\caption{Sort Enumeration Rules}
\label{Sort Enumeration Rules}
\end{figure}

\subsection{Sort Depth Computation}
\label{Sort Depth Computation}

Assume that all sort definitions are in head normal form.
To each sort name $s$,
associate a PROLOG variable $S$ that will hold the
minimal depth  of $s$ when the algorithm is finished.
For each sort definition
$s \sortdef
f_1(s_{11},\ldots,s_{1n_1}) \mid \ldots \mid
f_m(s_{m1},\ldots,s_{mn_m})$,
introduce an equation

$$S = min(max(S_{11},\ldots,S_{1n_1}), \ldots,
max(S_{m1},\ldots,S_{mn_m})) + 1$$

into the set of depth equations.
Since $max$ is idempotent, multiple occurrences among its arguments may
be replaced by one.
Since $max$ is monotonic, $max(S_{i1},\ldots,S_{in_i})$
can be omitted if $S$ itself occurs among $S_{i1},\ldots,S_{in_i}$.
If any nullary symbol is among $f_1,\ldots,f_m$,
the above depth equation simply reads $S=1$.
Building the set of depth equations is achieved by
the PROLOG predicate
{\tt calc\_sort\_depths1\_list},
which additionally distinguishes solved equations (i.e.\
ones
containing no uninstantiated PROLOG variables) from unsolved ones.

In a second phase, repeat the following solving algorithm until
no further progress is obtained, which will be the case after at
least $k$ cycles if there are $k$ sort definitions. The solving
algorithm checks each unsolved equation as to whether one or
more of its right--hand side $max(\ldots)$ expressions are
ground terms, meaning that all necessary data is available to
compute its value. These expressions are evaluated and
substituted by their result. If one of the results is minimal,
i.e.\ equals $i-1$ in the $i$-th cycle, the whole equation can
be solved: $S=i$. Fig.~\ref{Example: Sort Depth Computation}
gives an example of a computation of the depths of sorts from
Fig.~\ref{Equivalence Classes for Background Theory (5) from
Fig.}. 

All equations that remain unsolved after the final cycle
correspond to empty sorts, cf.\ \cite{Aiken.Murphy.1991b}. Their
sort definitions' right--hand sides are simply replaced by
$\bot$, and any occurrence of them in other sort definitions is
also replaced by $\bot$, which may give rise to some obvious
further simplifications. 

In order to supply a different depth measure, $max$ has to be
replaced accordingly, and the optimization based on its
idempotence property may no longer be applicable.

\begin{figure}
\begin{center}
\begin{tabular}[t]{@{}*{4}{l}@{}}
& $i=1$ & $2$ & $3$	\\
\hline
$S_0$ & $= 1$	\\
$S_1$ & $= min(S_0,S_e)+1$
	& $= 2$	\\
$S_2$ & $= min(S_1)+1$ & $= min(2)+1$
	& $= 3$	\\
$S_3$ & $= min(S_2,max(S_1,S_2))+1$
	& $= min(S_2,max(2,S_2))+1$
	& $= 4$	\\
$S_4$ & $= min(S_3,max(S_1,S_3),S_2)+1$
	& $= min(S_3,max(2,S_3),S_2)+1$
	& $= 4$	\\
$S_5$ & $= min(S_4,max(S_1,S_4),max(S_2,S_3))+1$	
	& $= min(S_4,max(2,S_4),max(3,S_3))+1$
	& $= 5$	\\
$S_p$ & $= min(S_n,S_e)+1$
	& $= 2$	\\
$S_e$ & $= 1$	\\
$S_o$ & $= min(S_e)+1$
	& $= 2$	\\
$S_n$ & $= 1$	\\
\end{tabular}
\end{center}
\caption{Example: Sort Depth Computation}
\label{Example: Sort Depth Computation}
\end{figure}

\clearpage
\section{An E--Anti--Unification Example}
\label{An E--Anti--Unification Example}

In this section, we give an elaborate example of how the
E--anti--unification algorithms work.
Consider the background equational Theory (1) from
Fig.~\ref{Background Equational Theories Used}, defining $(+)$
and $(*)$ over the $0$--$\suc$ algebra. The equivalence class of
$0$ and $\suc(0)$ can be defined as sorts $s_0$ and $s_1$,
respectively, as shown in Fig.~\ref{Sort Definition of
$s_0,s_1$}. $s_n$ is not an equivalence class, but the sort of
all terms. 
For example, the sort definition of $s_0$ can be read as follows:
there are four possibilities for forming a term of the value $0$:
first, the constant $0$ itself;
second, adding two terms of value $0$;
third, multiplying a term of value $0$ by any other term;
and fourth, multiplying them conversely.

The computation of $hsg(s_0,s_1)$, according to Alg.~\eqr{hsg},
is shown in Fig.~\ref{Computation of $hsg(s_0,s_1)$}. The
necessary auxiliary computations are shown in
Figs.~\ref{Computation of $hsg(s_0,s_0)$} to~\ref{Computation of
$hsg(s_n,s_n)$}. The resulting sort definitions are summarized
in the upper part of Fig.~\ref{Generalization of $s_0$ and
$s_1$}; They contain all the linear generalizations but only
some of the 
nonlinear ones.
Figures~\ref{Intersection Computation for $s_0,s_1$}
to~\ref{Intersection Computation for $s_0,s_1$ (3. Contin.)}
show the computation of variables with non--empty sort
intersections using the algorithm from Sect.~\ref{Nonlinear
Generalizations}, resulting in the new variable $v'_{00}$ for
$v_{00},v_{0n},v_{n0},v_{nn}$, and $v'_{01}$ for
$v_{01},v_{0n},v_{n1},v_{nn}$. The lower part of
Fig.~\ref{Generalization of $s_0$ and $s_1$} shows the
appropriately modified sort definitions. $\L(s'_{01})$ is the
set of all generalizations of $0$ and $\suc(0)$ modulo $+,*$. 

Figures~\ref{Enumeration of $s_{01}$} and~\ref{Enumeration of
$s_{01}$ (contd.)} show the enumeration of sort $s'_{01}$ with a
depth bound of 2. The actual depth bound is shown in the second
column, the $i$-th number referring to the $i$-th identifier
(variable or sort name) in the term. Terms flagged ``{\tt
<*****}'' do not contain any sort names and are output. In
Fig.~\ref{Enumeration of $s_{01}$ (contd.)}, most terms that do
not yield output are omitted.

\begin{figure}
\begin{center}
\begin{tabular}[t]{@{}ccccccccc@{}}
$s_0$ & $\sortdef$ & $0$ & $\mid$ & $s_0+s_0$ & $\mid$
	& $s_0*s_n$ & $\mid$ & $s_n*s_0$	\\
$s_1$ & $\sortdef$ & $\suc(s_0)$ & $\mid$ & $s_0+s_1$
	& $\mid$ & $s_1+s_0$ & $\mid$ & $s_1*s_1$ \\
$s_n$ & $\sortdef$ & $0$ & $\mid$ & $\suc(s_n)$ & $\mid$ & $s_n+s_n$
	& $\mid$ & $s_n*s_n$ \\
\end{tabular}
\end{center}
\caption{Sort Definition of $s_0,s_1$}
\label{Sort Definition of $s_0,s_1$}
\end{figure}

\begin{figure}
\begin{center}
\begin{tabular}[t]{@{}lllllllll@{}}
& $s_{01}$	\\
[0.2cm]
$\sortdef$ & $hsg(s_0,s_1)$	\\
[0.2cm]
$=$ & $v_{01} \mid hsg(0,\suc(s_0))$ & $\mid$ & $hsg(0,s_0+s_1)$
	& $\mid$ & $hsg(0,s_1+s_0)$
	& $\mid$ & $hsg(0,s_1*s_1)$ & $\mid$ \\
& $hsg(s_0+s_0,\suc(s_0))$ & $\mid$ & $hsg(s_0+s_0,s_0+s_1)$ & $\mid$
	& $hsg(s_0+s_0,s_1+s_0)$ & $\mid$
	& $hsg(s_0+s_0,s_1*s_1)$ & $\mid$ \\
& $hsg(s_0*s_n,\suc(s_0))$ & $\mid$ & $hsg(s_0*s_n,s_0+s_1)$ & $\mid$
	& $hsg(s_0*s_n,s_1+s_0)$ & $\mid$
	& $hsg(s_0*s_n,s_1*s_1)$ & $\mid$ \\
& $hsg(s_n*s_0,\suc(s_0))$ & $\mid$ & $hsg(s_n*s_0,s_0+s_1)$ & $\mid$
	& $hsg(s_n*s_0,s_1+s_0)$ & $\mid$ & $hsg(s_n*s_0,s_1*s_1)$ \\
[0.2cm]
$=$ & $v_{01} \mid \bot$ & $\mid$ & $\bot$ & $\mid$
	& $\bot$ & $\mid$ & $\bot$ & $\mid$ \\
& $\bot$ & $\mid$ & $hsg(s_0+s_0,s_0+s_1)$ & $\mid$
	& $hsg(s_0+s_0,s_1+s_0)$ & $\mid$
	& $\bot$ & $\mid$ \\
& $\bot$ & $\mid$ & $\bot$ & $\mid$
	& $\bot$ & $\mid$
	& $hsg(s_0*s_n,s_1*s_1)$ & $\mid$ \\
& $\bot$ & $\mid$ & $\bot$ & $\mid$
	& $\bot$ & $\mid$ & $hsg(s_n*s_0,s_1*s_1)$ \\
[0.2cm]
$=$ & $v_{01} \mid hsg(s_0,s_0)+hsg(s_0,s_1)$ & $\mid$
	& $hsg(s_0,s_1)+hsg(s_0,s_0)$ & $\mid$
	& $hsg(s_0,s_1)*hsg(s_n,s_1)$ & $\mid$
	& $hsg(s_n,s_1)*hsg(s_0*s_1)$ \\
[0.2cm]
$=$ & $v_{01} \mid s_{00}+s_{01}$ & $\mid$
	& $s_{01}+s_{00}$ & $\mid$
	& $s_{01}*s_{n1}$ & $\mid$
	& $s_{n1}*s_{01}$ \\
\end{tabular}
\end{center}
\caption{Computation of $hsg(s_0,s_1)$}
\label{Computation of $hsg(s_0,s_1)$}
\end{figure}

\begin{figure}
\begin{center}
\begin{tabular}[t]{@{}lllllllll@{}}
& $s_{00}$	\\
[0.2cm]
$\sortdef$ & $hsg(s_0,s_0)$	\\
[0.2cm]
$=$ & $v_{00} \mid hsg(0,0)$ & $\mid$ & $hsg(0,s_0+s_0)$ & $\mid$
	& $hsg(0,s_0*s_n)$ & $\mid$ & $hsg(0,s_n*s_0)$ & $\mid$ \\
& $hsg(s_0+s_0,0)$ & $\mid$ & $hsg(s_0+s_0,s_0+s_0)$ & $\mid$
	& $hsg(s_0+s_0,s_0*s_n)$ & $\mid$ & $hsg(s_0+s_0,s_n*s_0)$
	& $\mid$	\\
& $hsg(s_0*s_n,0)$ & $\mid$ & $hsg(s_0*s_n,s_0+s_0)$ & $\mid$
	& $hsg(s_0*s_n,s_0*s_n)$ & $\mid$ & $hsg(s_0*s_n,s_n*s_0)$
	& $\mid$	\\
& $hsg(s_n*s_0,0)$ & $\mid$ & $hsg(s_n*s_0,s_0+s_0)$ & $\mid$
	& $hsg(s_n*s_0,s_0*s_n)$ & $\mid$ & $hsg(s_n*s_0,s_n*s_0)$ \\
[0.2cm]
$=$ & $v_{00}$ & $\mid$ & $0$ & $\mid$
	& $hsg(s_0,s_0)+hsg(s_0,s_0)$ & $\mid$
	& $hsg(s_0,s_0)*hsg(s_n,s_n)$ & $\mid$	\\
	& $hsg(s_0,s_n)*hsg(s_n,s_0)$ & $\mid$
	& $hsg(s_n,s_0)*hsg(s_0,s_n)$ & $\mid$
	& $hsg(s_n,s_n)*hsg(s_0,s_0)$	\\
[0.2cm]
$=$ & $v_{00}$ & $\mid$ & $0$ & $\mid$
	& $s_{00}+s_{00}$ & $\mid$
	& $s_{00}*s_{nn}$ & $\mid$	\\
	& $s_{0n}*s_{n0}$ & $\mid$
	& $s_{n0}*s_{0n}$ & $\mid$
	& $s_{nn}*s_{00}$	\\
\end{tabular}
\end{center}
\caption{Computation of $hsg(s_0,s_0)$}
\label{Computation of $hsg(s_0,s_0)$}
\end{figure}

\begin{figure}
\begin{center}
\begin{tabular}[t]{@{}lllllllll@{}}
& $s_{0n}$	\\
[0.2cm]
$\sortdef$ & $hsg(s_0,s_n)$	\\
[0.2cm]
$=$ & $v_{0n} \mid hsg(0,0)$ & $\mid$ & $hsg(0,\suc(s_n))$ & $\mid$
	& $hsg(0,s_n+s_n)$ & $\mid$ & $hsg(0,s_n*s_n)$ & $\mid$ \\
& $hsg(s_0+s_0,0)$ & $\mid$ & $hsg(s_0+s_0,\suc(s_n))$ & $\mid$
	& $hsg(s_0+s_0,s_n+s_n)$ & $\mid$ & $hsg(s_0+s_0,s_n*s_n)$
	& $\mid$	\\
& $hsg(s_0*s_n,0)$ & $\mid$ & $hsg(s_0*s_n,\suc(s_n))$ & $\mid$
	& $hsg(s_0*s_n,s_n+s_n)$ & $\mid$ & $hsg(s_0*s_n,s_n*s_n)$
	& $\mid$	\\
& $hsg(s_n*s_0,0)$ & $\mid$ & $hsg(s_n*s_0,\suc(s_n))$ & $\mid$
	& $hsg(s_n*s_0,s_n+s_n)$ & $\mid$ & $hsg(s_n*s_0,s_n*s_n)$ \\
[0.2cm]
$=$ & $v_{0n} \mid 0$ & $\mid$
	& $hsg(s_0,s_n)+hsg(s_0,s_n)$ & $\mid$
	& $hsg(s_0,s_n)*hsg(s_n,s_n)$ & $\mid$
	& $hsg(s_n,s_n)*hsg(s_0,s_n)$	\\
[0.2cm]
$=$ & $v_{0n} \mid 0$ & $\mid$
	& $s_{0n}+s_{0n}$ & $\mid$
	& $s_{0n}*s_{nn}$ & $\mid$
	& $s_{nn}*s_{0n}$	\\
\end{tabular}
\end{center}
\caption{Computation of $hsg(s_0,s_n)$}
\label{Computation of $hsg(s_0,s_n)$}
\end{figure}

\begin{figure}
\begin{center}
\begin{tabular}[t]{@{}lllllllll@{}}
& $s_{n0}$	\\
[0.2cm]
$\sortdef$ & $hsg(s_n,s_n)$	\\
[0.2cm]
$=$ & $v_{n0} \mid hsg(0,0)$ & $\mid$ & $hsg(0,s_0+s_0)$ & $\mid$
	& $hsg(0,s_0*s_n)$ & $\mid$ & $hsg(0,s_n*s_0)$ & $\mid$ \\
& $hsg(\suc(s_n),0)$ & $\mid$ & $hsg(\suc(s_n),s_0+s_0)$ & $\mid$
	& $hsg(\suc(s_n),s_0*s_n)$ & $\mid$ & $hsg(\suc(s_n),s_n*s_0)$
	& $\mid$	\\
& $hsg(s_n+s_n,0)$ & $\mid$ & $hsg(s_n+s_n,s_0+s_0)$ & $\mid$
	& $hsg(s_n+s_n,s_0*s_n)$ & $\mid$ & $hsg(s_n+s_n,s_n*s_0)$
	& $\mid$	\\
& $hsg(s_n*s_n,0)$ & $\mid$ & $hsg(s_n*s_n,s_0+s_0)$ & $\mid$
	& $hsg(s_n*s_n,s_0*s_n)$ & $\mid$ & $hsg(s_n*s_n,s_n*s_0)$ \\
[0.2cm]
$=$ & $v_{n0} \mid 0$ & $\mid$
	& $hsg(s_n,s_0)+hsg(s_n,s_0)$ & $\mid$
	& $hsg(s_n,s_0)*hsg(s_n,s_n)$ & $\mid$
	& $hsg(s_n,s_n)*hsg(s_n,s_0)$	\\
[0.2cm]
$=$ & $v_{n0} \mid 0$ & $\mid$
	& $s_{n0}+s_{n0}$ & $\mid$
	& $s_{n0}*s_{nn}$ & $\mid$
	& $s_{nn}*s_{n0}$	\\
\end{tabular}
\end{center}
\caption{Computation of $hsg(s_n,s_0)$}
\label{Computation of $hsg(s_n,s_0)$}
\end{figure}

\begin{figure}
\begin{center}
\begin{tabular}[t]{@{}lllllllll@{}}
& $s_{n1}$	\\
[0.2cm]
$\sortdef$ & $hsg(s_n,s_1)$	\\
[0.2cm]
$=$ & $v_{n1} \mid hsg(0,\suc(s_0))$ & $\mid$ & $hsg(0,s_0+s_1)$
	& $\mid$ & $hsg(0,s_1+s_0)$
	& $\mid$ & $hsg(0,s_1*s_1)$ & $\mid$	\\
& $hsg(\suc(s_n),\suc(s_0))$ & $\mid$ & $hsg(\suc(s_n),s_0+s_1)$
	& $\mid$ & $hsg(\suc(s_n),s_1+s_0)$
	& $\mid$ & $hsg(\suc(s_n),s_1*s_1)$
	& $\mid$	\\
& $hsg(s_n+s_n,\suc(s_0))$ & $\mid$ & $hsg(s_n+s_n,s_0+s_1)$ & $\mid$
	& $hsg(s_n+s_n,s_1+s_0)$ & $\mid$ & $hsg(s_n+s_n,s_1*s_1)$
	& $\mid$	\\
& $hsg(s_n*s_n,\suc(s_0))$ & $\mid$ & $hsg(s_n*s_n,s_0+s_1)$ & $\mid$
	& $hsg(s_n*s_n,s_1+s_0)$ & $\mid$ & $hsg(s_n*s_n,s_1*s_1)$ \\
[0.2cm]
$=$ & $v_{n1} \mid \suc(hsg(s_n,s_0))$ & $\mid$
	& $hsg(s_n,s_0)+hsg(s_n,s_1)$ & $\mid$
	& $hsg(s_n,s_1)+hsg(s_n+s_0)$ & $\mid$
	& $hsg(s_n,s_1)*hsg(s_n*s_1)$ \\
[0.2cm]
$=$ & $v_{n1} \mid \suc(s_{n0})$ & $\mid$
	& $s_{n0}+s_{n1}$ & $\mid$
	& $s_{n1}+s_{n0}$ & $\mid$
	& $s_{n1}*s_{n1}$ \\
\end{tabular}
\end{center}
\caption{Computation of $hsg(s_n,s_1)$}
\label{Computation of $hsg(s_n,s_1)$}
\end{figure}

\begin{figure}
\begin{center}
\begin{tabular}[t]{@{}lllllllll@{}}
& $s_{nn}$	\\
[0.2cm]
$\sortdef$ & $hsg(s_n,s_n)$	\\
[0.2cm]
$=$ & $v_{nn} \mid hsg(0,0)$ & $\mid$ & $hsg(0,\suc(s_n))$ & $\mid$
	& $hsg(0,s_n+s_n)$ & $\mid$ & $hsg(0,s_n*s_n)$ & $\mid$ \\
& $hsg(\suc(s_n),0)$ & $\mid$ & $hsg(\suc(s_n),\suc(s_n))$ & $\mid$
	& $hsg(\suc(s_n),s_n+s_n)$ & $\mid$ & $hsg(\suc(s_n),s_n*s_n)$
	& $\mid$	\\
& $hsg(s_n+s_n,0)$ & $\mid$ & $hsg(s_n+s_n,\suc(s_n))$ & $\mid$
	& $hsg(s_n+s_n,s_n+s_n)$ & $\mid$ & $hsg(s_n+s_n,s_n*s_n)$
	& $\mid$	\\
& $hsg(s_n*s_n,0)$ & $\mid$ & $hsg(s_n*s_n,\suc(s_n))$ & $\mid$
	& $hsg(s_n*s_n,s_n+s_n)$ & $\mid$ & $hsg(s_n*s_n,s_n*s_n)$ \\
[0.2cm]
$=$ & $v_{nn} \mid 0$ & $\mid$
	& $\suc(hsg(s_n,s_n))$ & $\mid$
	& $hsg(s_n,s_n)+hsg(s_n,s_n)$ & $\mid$
	& $hsg(s_n,s_n)*hsg(s_n,s_n)$	\\
[0.2cm]
$=$ & $v_{nn} \mid 0$ & $\mid$
	& $\suc(s_{nn})$ & $\mid$
	& $s_{nn}+s_{nn}$ & $\mid$
	& $s_{nn}*s_{nn}$	\\
\end{tabular}
\end{center}
\caption{Computation of $hsg(s_n,s_n)$}
\label{Computation of $hsg(s_n,s_n)$}
\end{figure}

\begin{figure}
\begin{center}
\begin{tabular}[t]{@{}ccl*{12}{c}@{}}
\mca{15}{Linear Generalization:}	\\
$s_{00}$ & $\sortdef$ & $v_{00}$ & $\mid$ & $0$ & $\mid$
	& $s_{00}+s_{00}$ & $\mid$ & $s_{00}*s_{nn}$ & $\mid$
	& $s_{0n}*s_{n0}$ & $\mid$ & $s_{n0}*s_{0n}$ & $\mid$
	& $s_{nn}*s_{00}$	\\
$s_{01}$ & $\sortdef$ & $v_{01}$ & $\mid$ & $s_{00}+s_{01}$ & $\mid$
	& $s_{01}+s_{00}$ & $\mid$ & $s_{01}*s_{n1}$ & $\mid$
	& $s_{n1}*s_{01}$ \\
$s_{0n}$ & $\sortdef$ & $v_{0n}$ & $\mid$ & $0$ & $\mid$
	& $s_{0n}+s_{0n}$ & $\mid$ & $s_{0n}*s_{nn}$ & $\mid$
	& $s_{nn}*s_{0n}$	\\
$s_{n0}$ & $\sortdef$ & $v_{n0}$ & $\mid$ & $0$ & $\mid$
	& $s_{n0}+s_{n0}$ & $\mid$ & $s_{n0}*s_{nn}$ & $\mid$
	& $s_{nn}*s_{n0}$	\\
$s_{n1}$ & $\sortdef$ & $v_{n1}$ & $\mid$ & $\suc(s_{n0})$ & $\mid$
	& $s_{n0}+s_{n1}$ & $\mid$ & $s_{n1}+s_{n0}$ & $\mid$
	& $s_{n1}*s_{n1}$ \\
$s_{nn}$ & $\sortdef$ & $v_{nn}$ & $\mid$ & $0$ & $\mid$
	& $\suc(s_{nn})$ & $\mid$ & $s_{nn}+s_{nn}$ & $\mid$
	& $s_{nn}*s_{nn}$	\\
\\
\mca{15}{Complete Generalization:}	\\
$s'_{00}$ & $\sortdef$ & $v_{00} \mid v'_{00}$ & $\mid$ & $0$ & $\mid$
	& $s'_{00}+s'_{00}$ & $\mid$ & $s'_{00}*s'_{nn}$ & $\mid$
	& $s'_{0n}*s'_{n0}$ & $\mid$ & $s'_{n0}*s'_{0n}$ & $\mid$
	& $s'_{nn}*s'_{00}$	\\
$s'_{01}$ & $\sortdef$ & $v_{01} \mid v'_{01}$ & $\mid$
	& $s'_{00}+s'_{01}$ & $\mid$ & $s'_{01}+s'_{00}$ & $\mid$
	& $s'_{01}*s'_{n1}$ & $\mid$ & $s'_{n1}*s'_{01}$ \\
$s'_{0n}$ & $\sortdef$ & $v_{0n} \mid v'_{00} \mid v'_{01}$ & $\mid$
	& $0$ & $\mid$ & $s'_{0n}+s'_{0n}$ & $\mid$
	& $s'_{0n}*s'_{nn}$ & $\mid$ & $s'_{nn}*s'_{0n}$	\\
$s'_{n0}$ & $\sortdef$ & $v_{n0} \mid v'_{00}$ & $\mid$ & $0$ & $\mid$
	& $s'_{n0}+s'_{n0}$ & $\mid$ & $s'_{n0}*s'_{nn}$ & $\mid$
	& $s'_{nn}*s'_{n0}$	\\
$s'_{n1}$ & $\sortdef$ & $v_{n1} \mid v'_{01}$ & $\mid$
	& $\suc(s'_{n0})$ & $\mid$
	& $s'_{n0}+s'_{n1}$ & $\mid$ & $s'_{n1}+s'_{n0}$ & $\mid$
	& $s'_{n1}*s'_{n1}$ \\
$s'_{nn}$ & $\sortdef$ & $v_{nn} \mid v'_{00} \mid v'_{01}$ & $\mid$
	& $0$ & $\mid$ & $\suc(s'_{nn})$ & $\mid$
	& $s'_{nn}+s'_{nn}$ & $\mid$ & $s'_{nn}*s'_{nn}$	\\
\end{tabular}
\end{center}
\caption{Generalization of $s_0$ and $s_1$}
\label{Generalization of $s_0$ and $s_1$}
\end{figure}

\begin{figure}
\begin{center}
\begin{tabular}[t]{@{}lll@{}}
$S$ & $= \{$ & $s_0,s_1,s_n\}$	\\
$I2$ & $= \{$ & $ \tpl{s_0,s_n,s_0}, \tpl{s_1,s_n,s_1} \}$	\\
$L2$ & $= \{$ & $
	\tpl{v_{00},v_{0n}}, \tpl{v_{00},v_{n0}}, \tpl{v_{00},v_{nn}},
	\tpl{v_{01},v_{0n}}, \tpl{v_{01},v_{n1}}, \tpl{v_{01},v_{nn}},
	\tpl{v_{0n},v_{n0}}, \tpl{v_{0n},v_{n1}},
	\tpl{v_{0n},v_{nn}},$\\
	&& $\tpl{v_{n0},v_{nn}},
	\tpl{v_{n1},v_{nn}} \}$	\\
$V2$ & $= \{$ & $ \tpl{v_{00},3}, \tpl{v_{01},3}, \tpl{v_{0n},5},
	     \tpl{v_{n0},3}, \tpl{v_{n1},3}, \tpl{v_{nn},5} \}$	\\
\end{tabular}

\begin{verbatim}
[V00,V0n]      [V00,V01,V0n,Vn0,Vn1,Vnn]   []
   (Member)
   [V00,V0n]      [V01,V0n,Vn0,Vn1,Vnn]   []
      (Sublist2)
      [V00,V0n]      [V0n,Vn0,Vn1,Vnn]   []
         (Member)
         [V00,V0n]      [Vn0,Vn1,Vnn]   []
            (Inf Inh)                      infs({V00,V0n,Vn0})
            [V00,V0n,Vn0]      [Vn1,Vnn]   []
               (Sublist2)
               [V00,V0n,Vn0]      [Vnn]   []
                  (Inf Inh)                      infs({V00,V0n,Vn0,Vnn})
                  [V00,V0n,Vn0,Vnn]      []   []
                     (Output Max [V00,V0n,Vn0,Vnn])
                  [V00,V0n,Vn0]      []   []
                     (Max Subsumed)
            [V00,V0n]      [Vn1,Vnn]   []
               (Sublist2)
               [V00,V0n]      [Vnn]   []
                  (Suplist)
                  [V00,V0n]      []   [Vnn]
                     (Max Subs Susp)

[V00,Vn0]   [V00,V01,V0n,Vn0,Vn1,Vnn]   []
   (Member)
   [V00,Vn0]   [V01,V0n,Vn0,Vn1,Vnn]   []
      (Sublist2)
      [V00,Vn0]   [V0n,Vn0,Vn1,Vnn]   []
         (Suplist)
         [V00,Vn0]   [Vn0,Vn1,Vnn]   [V0n]
            (Member)
            [V00,Vn0]   [Vn1,Vnn]   [V0n]
               (Sublist2)
               [V00,Vn0]   [Vnn]   [V0n]
                  (Suplist)
                  [V00,Vn0]   []   [Vnn,V0n]
                     (Max Subs Susp)
\end{verbatim}
(Contd.\ in Fig.~\ref{Intersection Computation for $s_0,s_1$ (1. Contin.)})
\end{center}

\caption{Intersection Computation for $s_0,s_1$}
\label{Intersection Computation for $s_0,s_1$}
\end{figure}

\begin{figure}
\begin{center}
\begin{verbatim}
[V00,Vnn]   [V00,V01,V0n,Vn0,Vn1,Vnn]   []
   (Member)
   [V00,Vnn]   [V01,V0n,Vn0,Vn1,Vnn]   []
      (Sublist2)
      [V00,Vnn]   [V0n,Vn0,Vn1,Vnn]   []
         (Suplist)
         [V00,Vnn]   [Vn0,Vn1,Vnn]   [V0n]
            (Suplist)
            [V00,Vnn]   [Vn1,Vnn]   [Vn0,V0n]
               (Sublist2)
               [V00,Vnn]   [Vnn]   [Vn0,V0n]
                  (Member)
                  [V00,Vnn]   []   [Vn0,V0n]
                     (Max Subs Susp)

[V01,V0n]   [V00,V01,V0n,Vn0,Vn1,Vnn]   []
   (Sublist2)
   [V01,V0n]   [V01,V0n,Vn0,Vn1,Vnn]   []
      (Member)
      [V01,V0n]   [V0n,Vn0,Vn1,Vnn]   []
         (Member)
         [V01,V0n]   [Vn0,Vn1,Vnn]   []
            (Sublist2)
            [V01,V0n]   [Vn1,Vnn]   []
               (Inf Inh)                     infs({V01,V0n,Vn1})
               [V01,V0n,Vn1]   [Vnn]   []
                  (Inf Inh)                     infs({V01,V0n,Vn1,Vnn})
                  [V01,V0n,Vn1,Vnn]   []   []
                     (Output Max [V01,V0n,Vn1,Vnn])
                  [V01,V0n,Vn1]   []   []
                     (Max Subsumed)
            [V01,V0n]   [Vnn]   []
               (Suplist)
               [V01,V0n]   []   [Vnn]
                  (Max Subs Susp)

[V01,Vn1]   [V00,V01,V0n,Vn0,Vn1,Vnn]   []
   (Sublist2)
   [V01,Vn1]   [V01,V0n,Vn0,Vn1,Vnn]   []
      (Member)
      [V01,Vn1]   [V0n,Vn0,Vn1,Vnn]   []
         (Suplist)
         [V01,Vn1]   [Vn0,Vn1,Vnn]   [V0n]
            (Sublist2)
            [V01,Vn1]   [Vn1,Vnn]   [V0n]
               (Member)
               [V01,Vn1]   [Vnn]   [V0n]
                  (Suplist)
                  [V01,Vn1]   []   [Vnn,V0n]
                     (Max Subs Susp)
\end{verbatim}
(Contd.\ in Fig.~\ref{Intersection Computation for $s_0,s_1$ (2. Contin.)})
\end{center}

\caption{Intersection Computation for $s_0,s_1$ (1.\ Contin.)}
\label{Intersection Computation for $s_0,s_1$ (1. Contin.)}
\end{figure}

\begin{figure}
\begin{center}
\begin{verbatim}
[V01,Vnn]   [V00,V01,V0n,Vn0,Vn1,Vnn]   []
   (Sublist2)
   [V01,Vnn]   [V01,V0n,Vn0,Vn1,Vnn]   []
      (Member)
      [V01,Vnn]   [V0n,Vn0,Vn1,Vnn]   []
         (Suplist)
         [V01,Vnn]   [Vn0,Vn1,Vnn]   [V0n]
            (Sublist2)
            [V01,Vnn]   [Vn1,Vnn]   [V0n]
               (Suplist)
               [V01,Vnn]   [Vnn]   [Vn1,V0n]
                  (Member)
                  [V01,Vnn]   []   [Vn1,V0n]
                     (Max Subs Susp)

[V0n,Vn0]   [V00,V01,V0n,Vn0,Vn1,Vnn]   []
   (Suplist)
   [V0n,Vn0]   [V01,V0n,Vn0,Vn1,Vnn]   [V00]
      (Sublist2)
      [V0n,Vn0]   [V0n,Vn0,Vn1,Vnn]   [V00]
         (Member)
         [V0n,Vn0]   [Vn0,Vn1,Vnn]   [V00]
            (Member)
            [V0n,Vn0]   [Vn1,Vnn]   [V00]
               (Sublist2)
               [V0n,Vn0]   [Vnn]   [V00]
                  (Suplist)
                  [V0n,Vn0]   []   [Vnn,V00]
                     (Max Subs Susp)

[V0n,Vn1]   [V00,V01,V0n,Vn0,Vn1,Vnn]   []
   (Sublist2)
   [V0n,Vn1]   [V01,V0n,Vn0,Vn1,Vnn]   []
      (Suplist)
      [V0n,Vn1]   [V0n,Vn0,Vn1,Vnn]   [V01]
         (Member)
         [V0n,Vn1]   [Vn0,Vn1,Vnn]   [V01]
            (Sublist2)
            [V0n,Vn1]   [Vn1,Vnn]   [V01]
               (Member)
               [V0n,Vn1]   [Vnn]   [V01]
                  (Suplist)
                  [V0n,Vn1]   []   [Vnn,V01]
                     (Max Subs Susp)
\end{verbatim}
(Contd.\ in Fig.~\ref{Intersection Computation for $s_0,s_1$ (3. Contin.)})
\end{center}

\caption{Intersection Computation for $s_0,s_1$ (2.\ Contin.)}
\label{Intersection Computation for $s_0,s_1$ (2. Contin.)}
\end{figure}

\begin{figure}
\begin{center}
\begin{verbatim}
[V0n,Vnn]   [V00,V01,V0n,Vn0,Vn1,Vnn]   []
   (Suplist)
   [V0n,Vnn]   [V01,V0n,Vn0,Vn1,Vnn]   [V00]
      (Suplist)
      [V0n,Vnn]   [V0n,Vn0,Vn1,Vnn]   [V01,V00]
         (Member)
         [V0n,Vnn]   [Vn0,Vn1,Vnn]   [V01,V00]
            (Suplist)
            [V0n,Vnn]   [Vn1,Vnn]   [Vn0,V01,V00]
               (Suplist)
               [V0n,Vnn]   [Vnn]   [Vn1,Vn0,V01,V00]
                  (Member)
                  [V0n,Vnn]   []   [Vn1,Vn0,V01,V00]
                     (Max Subs Susp)

[Vn0,Vnn]   [V00,V01,V0n,Vn0,Vn1,Vnn]   []
   (Suplist)
   [Vn0,Vnn]   [V01,V0n,Vn0,Vn1,Vnn]   [V00]
      (Sublist2)
      [Vn0,Vnn]   [V0n,Vn0,Vn1,Vnn]   [V00]
         (Suplist)
         [Vn0,Vnn]   [Vn0,Vn1,Vnn]   [V0n,V00]
            (Member)
            [Vn0,Vnn]   [Vn1,Vnn]   [V0n,V00]
               (Sublist2)
               [Vn0,Vnn]   [Vnn]   [V0n,V00]
                  (Member)
                  [Vn0,Vnn]   []   [V0n,V00]
                     (Max Subs Susp)

[Vn1,Vnn]   [V00,V01,V0n,Vn0,Vn1,Vnn]   []
   (Sublist2)
   [Vn1,Vnn]   [V01,V0n,Vn0,Vn1,Vnn]   []
      (Suplist)
      [Vn1,Vnn]   [V0n,Vn0,Vn1,Vnn]   [V01]
         (Suplist)
         [Vn1,Vnn]   [Vn0,Vn1,Vnn]   [V0n,V01]
            (Sublist2)
            [Vn1,Vnn]   [Vn1,Vnn]   [V0n,V01]
               (Member)
               [Vn1,Vnn]   [Vnn]   [V0n,V01]
                  (Member)
                  [Vn1,Vnn]   []   [V0n,V01]
                     (Max Subs Susp)
\end{verbatim}
\end{center}

\caption{Intersection Computation for $s_0,s_1$ (3.\ Contin.)}
\label{Intersection Computation for $s_0,s_1$ (3. Contin.)}
\end{figure}

\begin{figure}
\begin{center}
\begin{verbatim}
S'01                              2
   V01                            1   <*****
   V'01                           1   <*****
   S'00+S'01                      1,1
      V00+V01                     0,0   <*****
      V00+V'01                    0,0   <*****
      V00+(S'00+S'01)             0,0
      V00+(S'01+S'00)             0,0
      V00+(S'01*S'n1)             0,0
      V00+(S'n1*S'01)             0,0
      V'00+V01                    0,0   <*****
      V'00+V'01                   0,0   <*****
      V'00+(S'00+S'01)            0,0
      V'00+(S'01+S'00)            0,0
      V'00+(S'01*S'n1)            0,0
      V'00+(S'n1*S'01)            0,0
      0+V01                       0,0   <*****
      0+V'01                      0,0   <*****
      0+(S'00+S'01)               0,0
      0+(S'01+S'00)               0,0
      0+(S'01*S'n1)               0,0
      0+(S'n1*S'01)               0,0
      (S'00+S'00)+V01             0,0
      (S'00+S'00)+V'01            0,0
      (S'00+S'00)+(S'00+S'01)     0,0
      (S'00+S'00)+(S'01+S'00)     0,0
      (S'00+S'00)+(S'01*S'n1)     0,0
      (S'00+S'00)+(S'n1*S'01)     0,0
      (S'00*S'nn)+V01             0,0
      (S'00*S'nn)+V'01            0,0
      (S'00*S'nn)+(S'00+S'01)     0,0
      (S'00*S'nn)+(S'01+S'00)     0,0
      (S'00*S'nn)+(S'01*S'n1)     0,0
      (S'00*S'nn)+(S'n1*S'01)     0,0
      (S'0n*S'n0)+V01             0,0
      (S'0n*S'n0)+V'01            0,0
      (S'0n*S'n0)+(S'00+S'01)     0,0
      (S'0n*S'n0)+(S'01+S'00)     0,0
      (S'0n*S'n0)+(S'01*S'n1)     0,0
      (S'0n*S'n0)+(S'n1*S'01)     0,0
      (S'n0*S'0n)+V01             0,0
      (S'n0*S'0n)+V'01            0,0
      (S'n0*S'0n)+(S'00+S'01)     0,0
      (S'n0*S'0n)+(S'01+S'00)     0,0
      (S'n0*S'0n)+(S'01*S'n1)     0,0
      (S'n0*S'0n)+(S'n1*S'01)     0,0
      (S'nn*S'00)+V01             0,0
      (S'nn*S'00)+V'01            0,0
      (S'nn*S'00)+(S'00+S'01)     0,0
      (S'nn*S'00)+(S'01+S'00)     0,0
      (S'nn*S'00)+(S'01*S'n1)     0,0
      (S'nn*S'00)+(S'n1*S'01)     0,0
\end{verbatim}
(Contd.\ in Fig.~\ref{Enumeration of $s_{01}$ (contd.)})
\end{center}
\caption{Enumeration of $s_{01}$}
\label{Enumeration of $s_{01}$}
\end{figure}

\begin{figure}
\begin{center}
\begin{verbatim}
   S'01+S'00                      1,1
      V01+V00                     0,0   <*****
      V01+V'00                    0,0   <*****
      V01+0                       0,0   <*****
      ...
      V'01+V00                    0,0   <*****
      V'01+V'00                   0,0   <*****
      V'01+0                      0,0   <*****
      ...
   S'01*S'n1                      1,1
      V01*Vn1                     0,0   <*****
      V01*V'01                    0,0   <*****
      ...
      V'01*Vn1                    0,0   <*****
      V'01*V'01                   0,0   <*****
      ...
   S'n1*S'01                      1,1
      Vn1*V01                     0,0   <*****
      Vn1*V'01                    0,0   <*****
      ...
      V'01*V01                    0,0   <*****
      V'01*V'01                   0,0   <*****
      ...
      (S'n1*S'n1)*(S'n1*S'01)     0,0
\end{verbatim}
\end{center}
\caption{Enumeration of $s_{01}$ (contd.)}
\label{Enumeration of $s_{01}$ (contd.)}
\end{figure}

\clearpage
\section{Lemma Generation}
\label{Lemma Generation}

\subsection{Implementation}
\label{Lemma Generation -- Implementation}

One application of E--anti--unification is for generating lemma
candidates in equational induction proofs. In this section, we
follow \cite{Heinz.1995} in showing how to generate candidates
that are applicable in a given blocked proof situation. The
algorithm presented here takes as input a term $t$ and a list
$[\sigma_1,\ldots,\sigma_n]$ of ground substitutions, and
generates as output the sort $s^r$ of all terms $t'$ such that
$\sigma_i t =_E \sigma_i t'$ for $i=1,\ldots,n$. In other words,
each generated $t'$ equals the given $t$ at least on all given
example instances. The appropriate selection of the $\sigma_i$
is discussed in Sect.~\ref{Selection of Ground
Instances} below. 

Note that for each universally valid equation $t =_E t'$, the
right--hand side $t'$ will be a member of $\L(s^r)$, and that for
each non--universally valid equation $t =_E t'$, $t'$ can be
excluded from $s^r$ by adding a ``counter--example instance''
$\sigma'$ with $\sigma' t \neq_E \sigma' t'$ to
$[\sigma_1,\ldots,\sigma_n]$. 

The algorithm first computes the simultaneous
syntactical anti--unification of all
example instances

$$t^l := \sigma_1 t \sqcap \ldots \sqcap \sigma_n t .$$

Usually, $t^l$ is a variant of $t$, cf.\ Sect.~\ref{Selection of
Ground Instances}. Next, we compute the sort of all
E--anti--unifiers of example instances 

$$s^r := rsg_V(vars(t^l),[ \eqc{\sigma_1 t},\ldots,\eqc{\sigma_n t}]).$$

For each $t^r \in \L(s^r)$, the equation $t^l = t^r$
is then a lemma candidate.
Note that using $rsg_V$ ensures that $vars(t^r) \subset vars(t^l)$,
thus excluding nonsense candidates like $(x_1+x_2)+x_3 = x_4+x_5$.
In order to additionally restrict $s^r$ to terms in normal form,
we may define the sort $s_{NF}$ of all terms in normal form
(see Sect.~\ref{Filter Sorts}
below) and intersect $s^r$ with $s_{NF}$. This will exclude
candidates such as $(x_1+x_2)+x_3 = x_1+(x_2+(x_3+0))$.

\subsection{Filter Sorts}
\label{Filter Sorts}

In order to restrict the output of lemma generation or other
applications to make it
satisfy certain additional criteria, we may intersect
the result sort of E--anti--unification with appropriate filter sorts.
The definition and construction
of these filter sorts is described below.

\bi
\item The top sort $s_{TOP}$ of all valid terms at all depends on the
        variables that have been introduced during E--anti--unification.
        Each such variable must be treated as a
	nullary function symbol. Let
        $V$ be the set of all variables occurring in any sort
        definition; then

	$$s_{TOP} \sortdef \bigmid_{x \in V} \; x
	\; \mid \; \bigmid_{f \in \F} \;
	f(s_{TOP},\ldots,s_{TOP}) .$$

\item Provided all term--rewrite rules of the given background 
	equational theory have linear left--hand sides,
	we can compute the sort $s_{RED}$
	of all reducible terms.
	Let $l_1,\ldots,l_n$ be the left--hand sides;
	let
	
	$$L_i := \{x \la s_{TOP} \mid x \in vars(l_i)\} \; (l_i)$$

	be the sorts obtained by replacing every variable by $s_{TOP}$.
	$L_i$ contains all terms where $l_i$ matches at the root.
	Define the set of all reducible terms by

	$$s_{RED} \sortdef \bigmid_{i=1}^n \; L_i
		\; \mid \; \bigmid_{f \in \F}
			\; f(s_{RED},s_{TOP},\ldots,s_{TOP})
			\mid \ldots \mid
			f(s_{TOP},\ldots,s_{TOP},s_{RED}) .$$

\item Since our sorts are closed wrt.\ relative complements,
	the sort of irreducible terms can simply be computed by
	$s_{NF} := s_{TOP} \setminus s_{RED}$.

\item If there are nonlinear redices, the set $\T_{RED}$ of
	reducible terms as well as the set $\T_{NF}$ of irreducible
	terms may no longer be a regular tree language
	and may thus not be
	representable by our sorts, cf.\ e.g.\
	\cite{Hofbauer.Huber.1994,Comon.1990}.
	However, by omitting all nonlinear left--hand sides, we can
	achieve
	$\L(s_{RED}) \subset \T_{RED}$,
	and $\T_{NF} \subset \L(s_{NF})$.
	Conversely, by replacing each nonlinear redex with a linear
	anti--instance, we get
	$\T_{RED} \subset \L(s_{RED})$
	and $\L(s_{NF}) \subset \T_{NF}$.

\item For every finite set of variables $V \subset W$,
	we can compute the sort $S_V^W$
	of all terms $t$ such that
	$V \subset vars(t) \subset W$.
	Let $A$ be the set of all occurring arities;
	for any $a \in A$, let
	
	$$parts(V,a) := \{ \tpl{P_1,\ldots,P_a} \mid
	P_1 \cup \ldots \cup P_a = V, \;\;
	\forall i,j \;\; i \neq j \ra P_i \cap P_j = \{\} \}$$

	denote the set of all partitions of the set $V$ into
	$a$ disjoint subsets;
	let $parts(V,0) = \{\}$.
	Define

	{
	$$
	\renewcommand{\arraystretch}{1.6}
	\begin{tabular}[t]{@{}ll@{}}
	$s_{\{\}}^W$
		& $\displaystyle\sortdef
		\bigmid_{x \in W} \;\; x \;\;\mid\;\;
		\bigmid_{f \in \F} \;\;
		f(s_{\{\}}^W,\ldots,s_{\{\}}^W)$,	\\
	$s_{\{v\}}^W$
		& $\displaystyle\sortdef v \;\;\mid\;\;
		\bigmid_{f \in \F, ar(f) \geq 1} \;\;
		f(s_{\{v\}}^W,s_{\{\}}^W,\ldots,s_{\{\}}^W)
		\mid \ldots \mid
		f(s_{\{\}}^W,\ldots,s_{\{\}}^W,s_{\{v\}}^W)$,	\\
	$s_V^W$
		& $\displaystyle\sortdef
		\bigmid_{a \in A} \;\;
		\bigmid_{f \in \F, ar(f) = a} \;\;
		\bigmid_{\tpl{P_1,\ldots,P_a} \in parts(V,a)} \;\;
		f(s_{P_1}^W,\ldots,s_{P_a}^W)$.	\\
	\end{tabular}
	$$
	}

	Fig.~\ref{Example: Variable Filter Sort}
	shows an example for $\F = \{0,s,+\}$,
	$V = \{v_1,v_2\}$ and $W = \{v_1,v_2,v_3,v_4\}$,
	using the abbreviations
	$s_{12}^{1234} := s_{\{v_1,v_2\}}^{\{v_1,v_2,v_3,v_4\}}$,
	$s_{\varepsilon}^{1234} := s_{\{\}}^{\{v_1,v_2,v_3,v_4\}}$,
	and so on.

	Note that filtering with $s_{\{\}}^V$ can be avoided by
	computing with $rsg_V$ instead of $rsg$, which is also
	much faster.
	However, for $V \neq \{\}$ the filter sorts $s^W_V$ provide a
	means of ensuring
	that each solution term contains at {\em least}
	all variables from a given set $V$.
\ei

\begin{figure}
\begin{center}
\begin{tabular}[t]{@{}ll@{}}
$s_{12}^{1234}$ & $\sortdef
	\suc(s_{12}^{1234}) \mid
	s_{12}^{1234}+s_{\varepsilon}^{1234} \mid
	s_{\varepsilon}^{1234}+s_{12}^{1234} \mid
	s_{1}^{1234}+s_{2}^{1234} \mid
	s_{2}^{1234}+s_{1}^{1234}$
	\\
$s_{1}^{1234}$ & $\sortdef
	v_1 \mid
	\suc(s_{1}^{1234}) \mid
	s_{\varepsilon}^{1234}+s_{1}^{1234} \mid
	s_{1}^{1234}+s_{\varepsilon}^{1234}$ \\
$s_{2}^{1234}$ & $\sortdef
	v_2 \mid
	\suc(s_{2}^{1234}) \mid
	s_{\varepsilon}^{1234}+s_{2}^{1234} \mid
	s_{2}^{1234}+s_{\varepsilon}^{1234}$ \\
$s_{\varepsilon}^{1234}$ & $\sortdef
	0 \mid v_1 \mid v_2 \mid v_3 \mid v_4 \mid
	\suc(s_{\varepsilon}^{1234}) \mid
	s_{\varepsilon}^{1234}+s_{\varepsilon}^{1234}$ \\
\end{tabular}
\end{center}

\caption{Example: Variable Filter Sort}
\label{Example: Variable Filter Sort}
\end{figure}

\subsection{Selection of Ground Instances}
\label{Selection of Ground Instances}

As shown in Sect.~\ref{Lemma Generation -- Implementation},
our lemma--generation algorithm needs some ground instances as
input.
Regarding their selection, we have the following requirements:
Let $\tpl{v_1,\ldots,v_n}$ be the variables occurring in the given
left--hand side term $t$.
A ground instance is defined by an $n$-tuple $\tpl{t_1,\ldots,t_n}$,
viz.\ $\{v_1 \la t_1,\ldots,v_n \la t_n\} \; (t)$.
Similarly, a matrix
$$G = \left(
\begin{array}{ccc}
t_{11} & \ldots & t_{1n}	\\
\vdots && \vdots	\\
t_{m1} & \ldots & t_{mn}	\\
\end{array}
\right)$$
defines $m$ ground instances of $t$, viz.\ one for each row.
We have the following requirements with respect to $G$:
\be
\item	In each column there occur two terms
	starting with different function symbols.
\item	No two columns are identical.

\item	In each column at least one ``non--trivial'' value occurs
	(e.g.\ $\not\in \{0,1\}$, $\not\in \{[\;], [a]\}$).

\item	There are not too many rows.
\item	No two rows are identical.
\item	The $t_{ij}$ are not too large values.
\ee

Requirements 1.\ and 2.\ ensure that the syntactical anti--unification
of all $m$ ground instances yields a variant of $t$;
3.\ enhances the quality of results;
4.\ to 6.\ are designed to obtain small runtimes.
Fig.~\ref{E--Anti--Unification Runtime vs. Size and Number of Sorts}
gives an impression of the size and number of ground
instances that can currently be handled by our implementation.
Note that by reimplementing in C and using up--to--date hardware, the
runtimes could be improved by about one order of magnitude.

By ``quality of results'' we mean --~informally~-- the ratio of
``desired''
terms to overall enumerated terms, weighted somehow by order of
appearance.
In other words,
the earlier a desired term appears, the higher the quality;
and the more desired terms appear, the higher the quality.
Which terms are considered to be ``desired'' depends on the
application; e.g.\ in lemma generation, $t^r$ is desired if the
equation $t^l = t^r$ is universally valid.
Since, in general, infinitely many terms are contained in the result
sort, a formal measure of quality can be defined e.g.\ by the expression
$q := \sum_{i=1}^\infty d_i \cdot \delta^i$,
where $d_i=1$ if the
$i$-th enumerated term belongs to the desired terms, $d_i=0$ else,
and $0 < \delta < 1$ is some real number.
We then always have $0 \leq q \leq \frac{\delta}{1-\delta}$.

Currently, the example ground instances still have to be provided
manually.
When automating this process, Requirements 1., 2., and 5.\ are easy to
satisfy as they are precisely defined.
Requirement 3.\ would call for special knowledge about the equational
theory which is not generally available. We propose dropping this
requirement which is vague anyway,
and enhancing the results' quality by
adding more instances (i.e.\ rows), maintaining, however, a
balance wrt.\ the contradicting Requirement 4.
On a similarly informal level, Requirement 3.\ usually contradicts
Requirement 6., since, for example,
the smallest terms are the nullary function
symbols
which mostly exhibit special behavior wrt.\ equationally defined
functions like $+,*,app$ and others.

The more the algorithms speeds can be improved, the less critical
the selection of appropriate ground examples becomes.

\subsection{Results and Runtimes}
\label{Lemma Generation -- Results and Runtimes}

Figure~\ref{Lemma--Generation Runtimes} shows some generated
lemmas together with the required runtimes --- cf.\ also
Sect.~\ref{E--Anti--Unification -- Runtimes}. The column
``T'' shows the background equational theory --- cf.\
Fig.~\ref{Background Equational Theories Used}. The left--hand
side of the equation in the column ``Law'' was input to the
algorithm; the right--hand side was generated; note, for
example, the difference between lines 23 and 24. Column ``Rhs''
indicates the normal forms of the left--hand side's ground
instances. This is considered a measure of the size of the input
sorts to be anti--unified. In lines 12 to 24, $\varepsilon$
stands for the empty list $[\;]$, while $ab$, for example,
stands for the list $[a,b]$. Column ``Nr'' indicates the number
of terms from the result sort that were enumerated before the
desired right--hand side appeared. Columns ``A'', ``I'', and
``E'' show the runtime for anti--unification, intersecting with
left--hand side's variables sort (left empty if $rsg_V$ is
used), and enumeration, respectively. Column ``$\Sigma$'' shows
the total runtime. Since the time resolution was 5 seconds,
shorter runtimes appear as ``0''. The rightmost column indicates
which technical optimizations have been switched on --- cf.\
Fig.~\ref{Technical Optimizations}. 

As an example, in line 1, the algorithm was provided with
the following instances
of the left--hand--side term $v1+(v2+v3)$:

$$
\begin{tabular}{ll}
$1+(0+0)$ & $= 1$,	\\
$0+(1+0)$ & $= 1$,	\\
$2+(0+1)$ & $= 3$.	\\
\end{tabular}
$$

It thus had to anti--unify the equivalence classes of $\suc(0)$,
$\suc(0)$, and $\suc(\suc(\suc(0)))$ wrt.\ the given background theory
consisting of the four equations defining $+$ and $*$.
Anti--Unification took 35 seconds; intersecting the sort of all
E--generalizations with the sort of all terms in variables
$v1,v2,v3$ took another 55 seconds, and enumerating the result
sort took 10 seconds until the desired right hand side $v1+(v2+v3)$
appeared.

As can be seen from a comparison of lines 7 and 8, for example, there
is a trade--off effect between runtime and the quality of the result
which can be controlled by the number and size of input sorts.

\begin{figure}
\begin{center}
\begin{tabular}{|r|l|r@{$\,=\,$}l|l|r|rrr|r|l@{}l@{}l@{}l|}
\hline
& T & \multicolumn{2}{|c|}{Law} & Rhs
	& Nr & A & I & E & $\Sigma$ &&&&	\\
\hline
\hline
1 & 1 & $v1+v2+v3$ & $v1+(v2+v3)$ & 1,1,3 & 6. & 35 & 55 & 10 & 100
	&g&&& \\
2 & 1 & $v1+v2+v3$ & $v1+(v2+v3)$ & 1,1,3 & 6. & 40 && 10 & 50 &&&& v \\
\cline{3-4}
3 & 1 & $v1*(v2+v3)$ & $v1*v2+v1*v3$ & 0,2,2 & 10. & 20 & 30 & 30 & 80
	&g&&&	\\
4 & 1 & $v1*(v2+v3)$ & $v1*v2+v1*v3$ & 0,2,2 & 10. & 30 && 10 & 40
	&&&& v \\
\cline{3-4}
5 & 1 & $v1*v1+v1*v2+v1*v2+v2*v2$ & $(v1+v2)*(v1+v2)$ & 4,1,4 & 4.
	& 285 & 25 & 70 & 380 &g&&&	\\
6 & 1 & $v1*v1+v1*v2+v1*v2+v2*v2$ & $(v1+v2)*(v1+v2)$ & 4,1,4 & 4.
	& 180 && 20 & 200 &&&& v	\\
\cline{3-4}
7 & 1 & $v1*v2$ & $v2*v1$ & 0,0 & 3. & 0 & 0 & 0 & 0  &g&&&\\
%\cline{3-4}
8 & 1 & $v1*v2*v3$ & $v1*(v2*v3)$ & 0,0,2 & 31. & 10 & 5 & 20 & 35
	&g&&& \\
\hline
\hline
9 & 2 & $dup(v1)+dup(v2)$ & $dup(v1+v2)$ & 2,4 & 2. & 5 & 20 & 5 & 30
	&g&&& \\
%\cline{3-4}
10 & 2 & $dup(v1)$ & $v1+v1$ & 0,4 & 4. & 0 & 5 & 0 & 5  &g&&&\\
%\cline{3-4}
11 & 2 & $v1*dup(v2)$ & $dup(v1*v2)$ & 0,0 & 13. & 0 & 0 & 5 & 5 &g&&&\\
\hline
\hline
12 & 3 & $app(rev(v1),rev(v2))$ & $rev(app(v2,v1))$ & ba,cd,cb & 1.
	& 185 & 5 & 230 & 420 &g&&&	\\
13 & 3 & $app(rev(v1),rev(v2))$ & $rev(app(v2,v1))$ & ba,cd,cb & 1.
	& 90 && 0 & 90 &&&& v	\\
14 & 3 & $app(rev(v1),rev(v2))$ & $rev(app(v2,v1))$
	& $\varepsilon$,bac,dcb & 1. & 110 & 5 & 150 & 265 &g&&& \\
15 & 3 & $app(rev(v1),rev(v2))$ & $rev(app(v2,v1))$
	& $\varepsilon$,bac,dcb & 1. & 70 && 0 & 70 &&&& v	\\
\cline{3-4}
16 & 3 & $app(v1,app(v2,v3))$ & $app(app(v1,v2),v3)$ & ab,cd,bd & 3.
	& 185 & 5 & 260 & 450 &g&&&	\\
17 & 3 & $app(v1,app(v2,v3))$ & $app(app(v1,v2),v3)$ & ab,cd,bd & 3.
	& 115 && 5 & 120 &&&& v	\\
18 & 3 & $app(v1,app(v2,v3))$ & $app(app(v1,v2),v3)$ & ab,cd,bd & 3.
	& ? && ? & 10 & g & s & t & v	\\
19 & 3 & $app(v1,app(v2,v3))$ & $app(app(v1,v2),v3)$ & abc,bde,cb & 3.
	& 855 && 15 & 870 &&&& v	\\
20 & 3 & $app(v1,app(v2,v3))$ & $app(app(v1,v2),v3)$ & abc,bde,cb & 3.
	& 240 && 10 & 250 && s & t & v	\\
21 & 3 & $app(v1,app(v2,v3))$ & $app(app(v1,v2),v3)$ & abc,bde,cb & 1.
	& 50 && 15 & 65 & g & s && v	\\
22 & 3 & $app(v1,app(v2,v3))$ & $app(app(v1,v2),v3)$ & abc,bde,cb & 1.
	& 20 && 10 & 30 & g & s & t & v	\\
\cline{3-4}
23 & 3 & $v1$ & $rev(rev(v1))$ & $\varepsilon$,ab & 4. &
	0 & 0 & 5 & 5 &g&&&	\\
24 & 3 & $rev(rev(v1))$ & $v1$ & $\varepsilon$,ab & 1. &
	0 & 0 & 5 & 5 &g&&&	\\
\hline
\hline
25 & 4 & $len(app(v1,v2))$ & $len(v1)+len(v2)$ & 1,2 & 4.
	& 5 & 10 & 5 & 20 &g&&&	\\
%\cline{3-4}
26 & 4 & $len(v1.app(v2,v3))$ & $\suc(len(v2)+len(v3))$ & 2,3 & 10.
	& 40 & 5 & 55 & 100 &g&&&	\\
\hline
\end{tabular}
\end{center}
\caption{Lemma--Generation Runtimes}
\label{Lemma--Generation Runtimes}
\end{figure}

\clearpage
\section{Series Guessing}
\label{Series Guessing}

\subsection{Implementation}
\label{Series Guessing -- Implementation}

A second application of E--anti--unification is the computation of
possible continuations of term sequences, wellknown from
intelligence tests.
In this section, we discuss a corresponding algorithm.
Given a list of terms $[t_n,\ldots,t_1]$
and an example count $k \leq n$,
the algorithm described below
generates the sort $s^r$ of all its possible formation laws wrt.\
a given equational background theory.
Each formation--law term $t' \in s^r$
computes each list member beyond $n-k$ from its predecessors,
using only functions of the equational background theory.
Note that for technical reasons we reverse the usual list order,
i.e.\ to compute formation laws of the quadratic numbers
$0,1,4,9,\ldots$, we start with the list
$[\suc^9(0),\suc^4(0),\suc(0),0]$.

We proceed in a similar way to lemma generation (cf.\
Sect.~\ref{Lemma Generation -- Implementation}). First, we
select the $k$ suffixes of length $n-k,\ldots,n-1$ and annotate
each one with its length plus one, i.e.\ with the rank of the
respective successor term within the series. We then
syntactically anti--unify these length--annotated suffixes: 

$$
\begin{tabular}[t]{@{}lll
c@{\hspace*{1.0cm}}c@{\hspace*{0.5cm}}c@{\hspace*{0.5cm}}
c@{\hspace*{0.5cm}}cc@{}}
&& $[$ & $\suc^{n-k+1}(0),$ & $t_{n-k},$ & $\ldots,$
	& $t_1$ & $]$	\\
&&&&& $\sqcap \ldots \sqcap$	\\
&& $[$ & $\suc^n(0),$ & $t_{n-1},$ & $\ldots,$
	& $t_k,$ & $t_{k-1},\ldots,t_1$ & $]$	\\
\cline{3-9}
$t^l$ & $=$ & $[$ & $\suc^{n-k+1}(v_{0 \ldots k-1}),$
	& $t'_{n-k \ldots n-1},$ & $\ldots,$
	& $t'_{1 \ldots k},$
	& $t''$ & $]$	\\
\end{tabular}
$$

where $t'_{i \ldots j} = t_i \sqcap \ldots \sqcap t_j$,
and $t'' = [\;] \sqcap [t_1] \sqcap \ldots \sqcap [t_{k-1},\ldots,t_1]$.
We compute the sort of all E--anti--unifiers
$s^r := rsg_V(vars(t^l),[\eqc{t_{n-k+1}},\ldots,\eqc{t_n}])$.
Each $t^r \in \L(s^r)$ can be interpreted as a series--formation law:
applying the term--rewriting rule
$t^l \leadsto t^r$ to a length--annotated suffix
$[\suc^i(0),t_i,\ldots,t_1]$ will result in the next series element
$t_{i+1}$, for $n-k \leq i < n$.
As in Sect.~\ref{Lemma Generation -- Implementation},
using $rsg_V$ ensures that $vars(t^r) \subset vars(t^l)$.
By intersecting $s^r$ with $s_{NF}$, we can additionally
restrict it to terms in normal form.
Fig.~\ref{Example: Series Guessing} gives an example for $n=4$
and $k=3$.

Note that $t^l$ is not usually a list of variables, but it
contains function symbols. Consequently, the terms from $s^r$
are more complicated than is intuitively expected; e.g.\ in the
example in Fig.~\ref{Example: Series Guessing}, the result term
considered is $\suc(v_1)*\suc(v_1)$, whereas the usual way would
be to 
give the series--construction law as $v*v$.
Alternatively, one might provide the $rsg_V$ call directly with
$V = \{ \phi(\suc^{n-k+1}(0),...,\suc^n(0)),
	\phi(t_{n-k},...,t_{n-1}), \ldots,
	\phi(t_1,...,t_k), \phi([\;],...,[t_{k-1},...,t_1]) \}$,
thus intentionally over--generalizing. This might require some
changes in the grouping algorithm from Sect.~\ref{Optimized
Argument Selection}. 
Moreover, Lemma~\ref{vrsg1} is no longer valid.

\begin{figure}
\begin{center}
\begin{tabular}[t]{@{}cc@{\hspace*{0.5cm}}ccc@{\hspace*{0.5cm}}lc
	@{\hspace*{0.5cm}}lc@{}}
\mca{7}{Given the series $0,1,4,9,\ldots$, and $k=3$}	\\
\\
$[$ & $\suc^3(0),$ & $\suc^4(0),$ & $\suc(0),0]$ & $\eqc{\suc^9(0)}$
	& $\ni$ & $\suc^3(0)*\suc^3(0)$	\\
$[$ & $\suc^2(0),$ & $\suc(0),$   & $0]$      & $\eqc{\suc^4(0)}$
	& $\ni$ & $\suc^2(0)*\suc^2(0)$	\\
$[$ & $\suc(0),$   & $0$       & $]$
	& \rule[-0.3cm]{0cm}{0cm}$\eqc{\suc(0)}$
	& $\ni$ & $\suc(0)*\suc(0)$	\\
\cline{1-7}
$[$ & $\suc(v_1),$ & $v_2 \mid$ & $v_3]$   & $\L(s_{1,4,9})$
	& $\ni$ & $\suc(v_1)*\suc(v_1)$	\\
\\
\mca{7}{For example, we have $\suc(v_1)*\suc(v_1) \in \L(s_{1,4,9})$;}\\
\mca{7}{the corresponding rewrite rule is:}	\\
$[$ & $\suc(v_1),$ & $v_2 \mid$ & $ v_3]$
	& $\leadsto \suc(v_1)*\suc(v_1)$ \\
\mca{7}{This rule rewrites:}	\\
$[$ & $\suc(0),$   & $0$       & $]$       & $\leadsto \suc(0)$	\\
$[$ & $\suc^2(0),$ & $\suc(0),$   & $0]$ & $\leadsto \suc^4(0)$	\\
$[$ & $\suc^3(0),$ & $\suc^4(0),$ & $\suc(0),0]$
	& $\leadsto \suc^9(0)$	\\
$[$ & $\suc^4(0),$ & $\suc^9(0),$ & $\suc^4(0),\suc(0),0]$
	& $\leadsto \suc^{16}(0)$ \\
\mca{7}{\ldots}	\\
\end{tabular}
\end{center}

\caption{Example: Series Guessing}
\label{Example: Series Guessing}
\end{figure}

\subsection{Results and Runtimes}
\label{Series Guessing -- Results and Runtimes}

Figure~\ref{Series--Guessing Runtimes} shows some series laws
that have been guessed together with the required runtime.
Columns ``T'', ``Nr'', ``A'', ``E'', ``$\Sigma$'' and the
rightmost one are as in Fig.~\ref{Lemma--Generation Runtimes}.
The column ``Series'' shows the series given to the algorithm;
the column ``L'' shows the number $k$ of example suffixes, i.e.\ the
number of anti--unified input sorts. The column ``Law'' shows the law
that is expected to be generated; $v_1$ denotes the previous member of
series, $v_2$ the pre--previous, and so on; $m$ denotes the rank
within the series.
Using the notions from 
Sect.~\ref{Series Guessing -- Results and Runtimes}, we have
$v_1 = t'_{n-k \ldots n-1}$,
$v_2 = t'_{n-k-1 \ldots n-2}$,
\ldots, and
$m = \suc^{n-k+1}(v_{0 \ldots k-1})$.
Columns ``D'' and ``N'' show the runtimes for computing sort
depths (cf.\ Sect.~\ref{Sort Enumeration}), and for intersecting
with the normal forms' sort (cf.\ Sect.~\ref{Filter Sorts}),
respectively. In lines 1 to 9, the enumeration algorithm was
restricted to output only terms in normal form. However, this
contradicted the technical optimization obtained by using sort
depths (cf.\ Sect.~\ref{Sort Enumeration}). Therefore, in lines
10 to 23, we explicitly intersected the sort of all
generalizations with the sort of all normal forms, dropping the
normal--form restriction from the enumeration algorithm. The
latter method makes sort enumeration dead--end--free, and is
thus faster for all but very small examples --- cf.\ lines 9 and
10. An $\infty$ in column ``Nr'' indicates that the desired law
has not been enumerated among, say, the first 100 terms. In this
case, the column ``$\Sigma$'' shows a ``$>$''. 

Since $t^l$ is not usually a list of variables, but contains
function symbols, $m,v_1,v_2,\ldots$
from Fig.~\ref{Series--Guessing Runtimes}
do, in fact, stand for proper terms.
For example, in line 23, we have $v_1 = \suc^2(x)$,
and the enumerated law, viz.\ $\suc(x)$,
has been abbreviated to $p(v_1)$,
the predecessor of $v_1$. The case is similar in line 21 ---
cf.\ also Fig.~\ref{Some Enumerated Series Laws}.

\begin{figure}
\begin{center}
\begin{tabular}{|r|l|l|l|l|r|rrrr|r|l@{}l@{}l@{}l@{}l@{}l|}
\hline
& T & Series & Law & L & Nr & A & D & N & E & $\Sigma$ &&&&&&\\
\hline
\hline
1 & 1 & $0,1,4$ & $m*m$ & 3? & 1. & 0 &&& 0 & 0	
	&&& g & s & t & v	\\
2 & 1 & $0,1,4$ & $m*m$ & 3 & 1. & 0 & 0 && 0 & 0
	&& d & g & s && v	\\
\cline{3-3}
3 & 1 & $0,2,4,6$ & $\suc(\suc(v_1))$ & 3? & 1. & 15 &&& 20 & 35
	&&& g & s & t & v	\\
4 & 1 & $0,2,4,6$ & $\suc(\suc(v_1))$ & 3 & 1. & 35 & 85 && 0 & 120
	&& d & g & s && v	\\
5 & 1 & $0,2,4,6$ & $m+m$ & 3? & 3. & 15 &&& 405 & 420
	&&& g & s & t & v	\\
6 & 1 & $0,2,4,6$ & $m+m$ & 3 & 3. & 35 & 85 && 0 & 120
	&& d & g & s && v	\\
\cline{3-3}
7 & 1 & $1,1,2,3,5$ & $v_1+v_2$. & 3? & 1. & 10 &&& 225 & 235
	&&& g & s & t & v	\\
8 & 1 & $1,1,2,3,5$ & $v_1+v_2$ & 3 & 1. & 15 & 15 && 0 & 30
	&& d & g & s && v	\\
\hline
\hline
9 & 5 & $1,1,2,3,5$ & $v_1+v_2$ & 3 & & 45 & 10 && $\infty$ & $\infty$
	&& d & g & s && v	\\
10 & 5 & $1,1,2,3,5$ & $v_1+v_2$ & 3 & 1. & 45 & 15 & 150 & 0 & 210	
	&& d & g & s && v	\\
11 & 5 & $1,1,2,3,5$ & $v_1+v_2$ & 3 & 1. & 45 & 5 & 100 & 0 & 150
	& b & d & g & s && v	\\
12 & 5 & $1,1,2,3,5$ & $v_1+v_2$ & 4 & $\infty$ & 200 & 55 & 610 & --
	& $>$865 & b & d & g & s && v	\\
\cline{3-3}
13 & 5 & $0,1,2,1,4,1,6$ & $if(ev(m),m,1)$ & 3 & $\infty$ & 60 & 20
	& 190 & & $>$270
	&& d & g & s && v	\\
14 & 5 & $0,1,2,1,4,1,6$ & $if(ev(m),m,1)$ & 4 & $\infty$ & 255
	& 285 & 160 & & $>$745
	&& d & g & s && v	\\
15 & 5 & $0,1,2,1,4,1,6$ & $if(ev(m),m,1)$ & 4 & $\infty$ & 200
	& 50 & 100 & & $>$350
	& b & d & g & s && v	\\
16 & 5 & $0,1,2,1,4,1$ & $if(ev(m),m,1)$ & 4 & 13. & 125
	& 25 & 170 & 0 & 320
	& b & d & g & s && v	\\
\cline{3-3}
17 & 5 & $0,0,1,1,0,0,1,1$ & $ev(v_2)$ & 5 & 1. & 500 & 155 & 1500
	& 0 & 2155
	&& d & g & s && v	\\
18 & 5 & $0,0,1,0,0,1$ & $ev(v_1+v_2)$ & 4 & 1. & 75 & 15 & 210 & 0
	& 300 && d & g & s && v	\\
19 & 5 & $0,1,3,7$ & $\suc(v_1+v_1)$ & 3 & 1. & 30 & 5 & 95 & 0 & 130
	&& d & g & s && v	\\
20 & 5 & $1,2,2,3,3,3,4,4,4,4$ & -- & 3 & $\infty$ & 60 & 20 & 205 & --
	& $>$285
	&& d & g & s && v	\\
21 & 5 & $2,3,5,9$ & $p(v_1+v_1)$ & 3 & 1. & 145 & 55 & 525 & 0 & 725
	&& d & g & s && v	\\
22 & 5 & $1,2,3,4,5,6$ & $\suc(m)$ & 3 & 1. & 125 & 40 & 425 & 0 & 590	
	&& d & g & s && v	\\
23 & 5 & $6,5,4,3,2,1$ & $p(v_1)$ & 3 & 1. & 15 & 0 & 40 & 0 & 55
	&& d & g & s && v	\\
\hline
\end{tabular}
\end{center}

\caption{Series--Guessing Runtimes}
\label{Series--Guessing Runtimes}
\end{figure}

\renewcommand{\a}{\tiny}

Some remarks on lines 13 to 16:
\\
The desired solution $if(ev(\suc(m)),\suc^3(m),\suc(0))$ has
the depth $5$.
Hence, all solutions of depth $4$ are enumerated before the desired
one,
the first being
$ev(x_1)*(x_3+x)+if(if(x,x_2,x_1),\suc(x_3),ev(x))$,
which is, in fact, considered a ``correct'' solution ---
see Fig.~\ref{Simplest Series Law for $0,1,2,1,4,1,6$}, where each column
corresponds to a subterm of the solution (the
marked column corresponds to
the whole term), irrelevant numbers are printed in small type.
The solution
$ev(x_1)*\suc(x_2)+\suc(ev(x_1))$
is shorter and of the same depth, but not in normal form.
The shortest solution seems to be
$\suc(ev(x_1)*\suc^2(x_2))$, which is deeper and not in normal form.
If the depth measure of the enumeration algorithm is changed such that
unfolding a sort definition is counted only if the head symbol of the
right--hand side is not in $\{0,s\}$, the desired solution is of
depth 2 and is thus enumerated first.

\begin{figure}
\begin{center}
\begin{tabular}[t]{@{}l@{$\;$}l@{}}
\mca{2}{We have}	\\
$m $ & $= \suc^3(x)$	\\
$v_1$ & $= \suc(x_1)$	\\
$v_2$ & $= \suc(x_2)$	\\
$v_3$ & $= x_3$ \\
\mca{2}{cf.\ line 16}	\\
\mca{2}{in Fig.~\ref{Some Enumerated Series Laws}}	\\
\end{tabular}
\tab
\begin{tabular}[t]{@{}*{6}{c}|c|*{9}{c}}
     &	       &     &	      &	    & \multicolumn{1}{c}{ }
& \multicolumn{1}{c}{$+$} \\
     &	       & $*$ &	      &	    & \multicolumn{2}{c}{ } &
$if$ \\
$ev$ &	       &     &	      & $+$ & \multicolumn{2}{c}{ } &
& $(if$ &
	&&& $s$ && $ev$ & \multicolumn{1}{r}{$)$}	\\
     & $(x_1)$ &     & $(x_3$ &	    & \multicolumn{1}{c}{$x)$}
& \multicolumn{1}{c}{ } & &	& $(x,$
	& $x_2$ & $,x_1),$ &	 & $(x_3),$ &	   & $(x)\;\;$ \\
%$ev$ & $(x_1)$ & $*$ & $(x_3$ & $+$ & $x)$ & $+$ & $if$ & $(if$ & $(x,$
%	& $x_2$ & $,x_1),$ & $s$ & $(x_3),$ & $ev$ & $(x))$	\\
\hline
0&1 &0 &\a 0&\a 0 &\a 0 &1 &1 &1 &0 &\a 0 & 1 &	  1 &	0 &\a 1 &\a 0 \\
1&0 &2 &   1&	2 &   1 &4 &2 &1 &1 & 1 &\a 0 &	  2 &	1 &\a 0 &\a 1 \\
0&3 &0 &\a 2&\a 4 &\a 2 &1 &1 &0 &2 & 0 &\a 3 &\a 3 &\a 2 &   1 &   2 \\
1&0 &4 &   1&	4 &   3 &6 &2 &3 &3 & 3 &\a 0 &	  2 &	1 &\a 0 &\a 3 \\
\hline
0&5 &0 &\a 4&\a 8 &\a 4 &1 &1 &0 &4 & 0 &\a 5 &\a 5 &\a 4 &   1 &   4 \\
1&0 &6 &   1&	6 &   5 &8 &2 &5 &5 & 5 &\a 0 &	  2 &	1 &\a 0 &\a 5 \\
0&7 &0 &\a 6&\a 12&\a 6 &1 &1 &0 &6 & 0 &\a 7 &\a 7 &\a 6 &   1 &   6 \\
1&0 &8 &   1&	8 &   7 &10&2 &7 &7 & 7 &\a 0 &	  2 &	1 &\a 0 &\a 7 \\
\end{tabular}
\end{center}

\caption{Simplest Series Law for $0,1,2,1,4,1,6,\ldots$}
\label{Simplest Series Law for $0,1,2,1,4,1,6$}
\end{figure}

Some remarks on line 18:
\\
While the function $ev(\cdot)$ allows one to distinguish series
members with even and odd rank, no function is
available in background theory (5) to distinguish ranks
$0$, $1$, and $2$ mod.\ $3$. For this reason, the
series in line 18 was expected by the authors to have no
solution at all when given to the algorithm. Note that
the trivial solution $v_3$ was prevented by setting $L = 4$,
thus forcing the algorithm to compute the first $1$ from the two
preceding
$0$'s. However, the algorithm found the solution
$ev(v_1+v_2)$, which is ``correct''. This seems to indicate a
possible strength of the algorithm: building series laws from a
well--defined limited set of functions rather than predicting the
next series member(s).

\begin{figure}
\begin{center}
\begin{tabular}[t]{@{}|r|l|l|r|r@{$:$}l|*{2}{l}|@{}}
\hline
& T & Series & D & $m$ & $v_1$,$v_2$,\ldots & Laws & \\
\hline
 1 & 1 & ;0,1,4 & $\infty$ & $x_1$ &
	& $x_1*x_1$ &	\\
 2 & 1 & 1,2,3;4,5,6 & $\infty$ & $\suc^3(x_1)$ & $\suc^3(x_1),\ldots$
	& $\suc^4(x_1)$ & \\
 3 & 1 & 0;2,4,6 & $\infty$ & $\suc(x_1)$ & $x_2$
	& $\suc^2(x_2)$
	& $\suc^2(x_1+x_1)$	\\
 4 & 1 & 0,1,2,1;4,1,6 & $\infty$ & $\suc^4(x_1)$ & $\suc(x_2)\ldots$
	&& \\
\hline
 5 & 1 & 1,1;2,3,5 & 2 & $\suc^2(x_1)$ & $\suc(x_1),\suc(x_2)$
	& $\suc^2(x_2+x_1)$
	& $\suc^2(x_2*x_2+x_1)$	\\
 6 & 1 & 1,1,2;3,5,8 & 2 & $\suc^3(x_1)$
	& $\suc^2(x_2),\suc(x_1),\suc(x_3)$
	& $\suc^3(x_3\1+(x_1\1+x_1))$
	& $\suc^3(x_2+x_1)$	\\
 7 & 1 & 0;1,3,7 & 5 & $\suc(x_1)$ & $x_2$
	& $\suc(x_1+x_1*x_1)$
	& $\suc(x_2+x_2)$	\\
 8 & 1 & 2,3;5,9 & 2 & $\suc^2(x_1)$ & $\suc^3(x_2),\suc^2(x_1)$
	& $\suc^5(x_2*x_2)$
	& $\suc^5(x_2+x_2)$	\\
\hline
 9 & 6 & 0;1,4 & 4 & $\suc(x_1)$ & $x_1$
	& $\suc(x_1+x_1+x_1)$
	& $x_1\1+x_1\1+x_1\1+(\suc(x_1)\1-x_1)$	\\
10 & 7 & ;0,1,4 & 5 & $x_1$ & $x_2$
	%(1418 s.)
	& $x_1*x_1$
	& $3*x_1^3-x_1^2-x_1^4$ \\
11 & 7 & 0,0,1;0,0,1 & 3 & $\suc^3(x_1)$ & $x_2,x_3,x_4$
	& $x_4$
	& $(x_4+x_3)*(x_4-x_2)$ \\
\hline
12 & 5 & ;0,1,4 & 3 & $x_1$
	&& $x_1*x_1$
	& $x_1*if(x_1,x_1,0)$	\\
13 & 5 & 1,1;2,3,5 & 4 & $\suc^2(x_1)$ & $\suc(x_1),\suc(x_2)$
	& $\suc(\suc(x_1+x_2))$
	& $\suc(if(x_1,x_1\1+x_1,\suc(0)))$	\\
14 & 5 & 0,0,1,1;0,0,1,1 & 3 & $\suc^4(x_1)$ & $x_2,x_3,x_4,x_5$
	& $x_5$
	& $ev(x_3)$	\\
15 & 5 & 0,0;1,0,0,1 & 3 & $\suc^2(x_1)$ & $x_2,x_3$
	& $ev(x_2+x_3)$
	& $ev(if(x_2,x_1,x_3))$	\\
16 & 5 & 0,1,2,1;4,1,6 & 4 & $\suc^4(x_1)$
	& $\suc(x_2),\suc(x_3),\suc(x_4),x_1$
	& \multicolumn{2}{l|}
		{$(ev(x_1)+if(x_1,x_4,x_3))*(ev(x_1)+(x_3+x_1))$} \\
\hline
17 & 8 & abc;bca,cab & $\infty$ & $\suc(x_1)$ & $[x_2,x_3,x_4]$
	& $[x_4,x_2,x_3]$ &	\\
18 & 8 & abab,aba;ab,a & $\infty$ & $\suc^2(x_1)$
	& $[a,b|x_2],[a,b,a|x_3]$ && \\
19 & 8 & baba,aba;ba,a & 2 & $\suc^2(x_1)$ & $[x_2,x_3|x_4],\ldots$
	& $il([x_3],x_4)$
	& $[x_3|rev(x_4)]$	\\
20 & 8 & a,ba;aba,baba & 2 & $\suc^2(x_1)$ & $[x_2,x_3|x_4],[x_3|x_4]$
	& $[x_3,x_2,x_3|x_4]$
	& $il([x_3\1,x_3],app(x_4\1,[x_2]))$	\\
\hline
\end{tabular}
\end{center}
\caption{Some Enumerated Series Laws}
\label{Some Enumerated Series Laws}
\end{figure}

Figure~\ref{Some Enumerated Series Laws} gives some enumerated series
laws.
The column ``T'' indicates the background theory used.
In the column ``Series'', the semicolon indicates the number of example
suffixes: one for each series member to the right of the semicolon.
Column ``D'' shows the depth up to which the shown laws appeared. An
``$\infty$'' means that the result sort is finite and is given
completely
in the column ``Laws''; in all other cases, an arbitrary selection of
the enumerated laws is shown.
The column ``$m:v_1,v_2,\ldots$'' shows the result of syntactical
anti--unification of examples.

Note that in lines 1 to 4, and 17 to 18,
the result sort is finite,
and precise statements can be made about the series and their possible
continuations.
For example,
from line 3 we can conclude that there are only two ways to
build the series $0,2,4,6$ from functions $+$ and $*$.
From line 4 we can conclude that there is no way to build
$0,1,2,1,4,1,6$ from these functions.

\clearpage
\section{Other Potential Applications}
\label{Other Potential Applications}

In this section, we look at some other potential applications of
E--anti--unification which we have considered but not implemented.

\subsection{Divergence Handling in Knuth--Bendix Completion}
\label{Divergence Handling in Knuth--Bendix Completion}

The Knuth--Bendix completion procedure is an important tool in
equational reasoning. However, its applicability is limited by
the possibility of its generating an infinite sequence of
term--rewriting rules. Kirchner
\cite{Kirchner.1987,Kirchner.1989} proposes schematizing such a
sequence using meta variables; Avenhaus \cite{Avenhaus.1989}
proposes consistently introducing new function symbols and new
equations to achieve termination of the Knuth--Bendix
completion.

% To automate such a process it is necessary to recognize
% non--termination of a sequence of new generated equations. If a
% sequence seems to be non--terminating, then a finite number
% of new equations generated so far can be used to be
% anti--unified. The result is one or more generalized schemes of
% equations. E.g.\ see the sequence of equations generated by
% Knuth--Bendix completion and the equational theory listed in
% Fig.~\ref{avenbsp}. There, anti--unification of the diverging
% new equations leads to the schemata
% $f(\suc^n(x),\suc^n(0))=f(x,\suc^n(0))$. This schemata can be
% used as a new equation to replace the diverging sequence. 

% \begin{figure}
% Equational Theory
% \\
% \begin{tabular}{lll}
% $\suc^0(x)$ &$=$ & $x$ \\
% $\suc^{\suc(n)}(x)$ &$=$& $\suc(\suc^n(x))$
% \end{tabular}
% \\
% Diverging Sequence
% \\
% \begin{tabular}{lll}
% $f(x,0)$ &$=$&$f(x,0)$ \\
% $f(\suc(x),\suc(0))$ &$=$& $f(x,\suc(0))$ \\
% $f(\suc(\suc(x)),\suc(\suc(0)))$ &$=$& $g(x,\suc(\suc(0)))$ \\
% $\vdots$ \\
% \end{tabular}
% \caption{Completion with Diverging Sequence} \label{avenbsp}
% \end{figure}

% \cite{Thomas.Watson.1993}

In any approach, it is necessary to recognize schemata in rule or
term sequences.
Kirchner proposes using generalization modulo an equivalence
relation to find useful schemata candidates, but has not developed
any approaches to
anti--unification modulo equational theories.
Nevertheless, useful equational candidates could be suggested by
using the mechanisms for anti--unification presented in this
paper. 

Below, we sketch a possible approach for finding useful
schemas using Exm.\ 2.1 from \cite{Kirchner.1989}. Consider the
rewrite rules shown in Fig.~\ref{Input Rewrite Rules from
Kirchner's Example 2.1} axiomatizing the ``signed binary tree
theory''. From these rules, the completion procedure generates
two infinite families of rules shown in Fig.~\ref{Completion of
the Theory from Fig. Input Rewrite Rules from Kirchner's Example
2.1}. The rightmost column shows the origin of each rule.
For example,
Rule~$5$ arose from superimposing Rule~$4$ at position~$1$ on
Rule~$2$ at the root, denoted by $\varepsilon$. Kirchner states in 
Example~2.3 that the first and second family of rules can be schematized
by the meta rule $\forall X,Y \;\; f(-X,f(X,Y)) \leadsto Y$ and
$\forall X,Y \;\; f(f(Y,X),-X) \leadsto Y$, respectively, modulo
the equational theory consisting of Rules~1 and~2 from
Fig.~\ref{Input Rewrite Rules from Kirchner's Example 2.1}.

\begin{figure}
\begin{center}
\begin{tabular}[t]{@{}r@{\hspace*{0.5cm}}rl@{\hspace*{0.5cm}}l@{}}
$1$ & $- -x$ & $\leadsto x$	\\
$2$ & $-f(x,y)$ & $\leadsto f(-y,-x)$	\\
$3$ & $f(f(y,x),-x)$ & $\leadsto y$	\\
$4$ & $f(-x,f(x,y))$ & $\leadsto y$	\\
\end{tabular}
\end{center}
\caption{Input Rewrite Rules from Kirchner's Example 2.1}
\label{Input Rewrite Rules from Kirchner's Example 2.1}
\end{figure}

\begin{figure}
\begin{center}
\begin{tabular}[t]{@{}r@{\hspace*{0.5cm}}rl@{\hspace*{0.5cm}}l@{}}
$5$ & $f(f(-x_2,-x_1),f(f(x_1,x_2),y))$ & $\leadsto y$
	& $o(4,\varepsilon.1,2,\varepsilon)$	\\
$6$ & $f(f(-x_2,x_1),f(f(-x_1,x_2),y))$ & $\leadsto y$
	& $o(5,\varepsilon.1.2,1,\varepsilon)$	\\
$7$ & $f(f(x_2,-x_1),f(f(x_1,-x_2),y))$ & $\leadsto y$
	& $o(5,\varepsilon.1.1,1,\varepsilon)$	\\
$8$ & $f(f(x_2,x_1),f(f(-x_1,-x_2),y))$ & $\leadsto y$
	& $o(6,\varepsilon.1.1,1,\varepsilon)$ 
	or $o(7,\varepsilon.1.2,1,\varepsilon)$	\\
$9$ & $f(f(f(-x_3,-x_2),-x_1),f(f(x_1,f(x_2,x_3)),y))$ & $\leadsto y$
	& $o(5,\varepsilon.1.1,2,\varepsilon)$	\\
& \ldots	\\
$10$ & $f(f(f(f(-x_4,-x_3),-x_2),-x_1),f(f(x_1,f(x_2,f(x_3,x_4))),y))$
	& $\leadsto y$
	& $o(9,\varepsilon.1.1.1,2,\varepsilon)$	\\
& \ldots	\\
\\
$11$ & $f(f(y,f(x_1,x_2)),f(-x_2,-x_1))$ & $\leadsto y$
	& $o(3,\varepsilon.1.2,2,\varepsilon)$	\\
$12$ & $f(f(y,f(-x_1,x_2)),f(-x_2,x_1))$ & $\leadsto y$
	& $o(11,\varepsilon.1.2.1,1,\varepsilon)$	\\
$13$ & $f(f(y,f(x_1,-x_2)),f(x_2,-x_1))$ & $\leadsto y$
	& $o(11,\varepsilon.1.2.2,1,\varepsilon)$	\\
$14$ & $f(f(y,f(-x_1,-x_2)),f(x_2,x_1))$ & $\leadsto y$
	& $o(12,\varepsilon.1.2.2,1,\varepsilon)$ 
	or $o(13,\ldots,1,\varepsilon)$ \\
$15$ & $f(f(y,f(x_1,f(x_2,x_3))),f(f(-x_3,-x_2),-x_1))$
	& $\leadsto y$
	& $o(11,\varepsilon.1.2.2,2,\varepsilon)$	\\
& \ldots	\\
\end{tabular}
\end{center}
\caption{Completion of the Theory from Fig.\ 
	\protect\ref{Input Rewrite Rules from Kirchner's Example 2.1}}
\label{Completion of the Theory from Fig.
	Input Rewrite Rules from Kirchner's Example 2.1}
\end{figure}

We now focus on the first family and discuss the
problems that arise when we try to reduce the detection of the
first meta rule to a series--guessing task like the ones
described in Sect.~\ref{Series Guessing -- Implementation}.

\paragraph{Finding appropriate subsequences.}
A fair completion strategy will not produce the rules in the
order shown in Fig.~\ref{Completion of the Theory from Fig.
Input Rewrite Rules from Kirchner's Example 2.1}; rather, it will
produce a sequence of rules interleaving both rule
families. Thus we are left with the task of detecting subsequences
that may obey common schemes. To achieve this, we propose applying
the
series--guessing algorithm for the empty equational theory to
the origin information.

Each rule generated by the completion process can be assigned an
``origin term'' constructed from the numbers of the input rules as
constants and a $4$--ary operator $o$ (for ``overlay''). For
example, Rules~$4$, $5$, $9$, and~$10$ from Figs.~\ref{Input
Rewrite Rules from Kirchner's Example 2.1} and~\ref{Completion of the
Theory from Fig. Input Rewrite Rules from Kirchner's Example 2.1} are
represented by the origin terms 
$4$, 
$o(4,\varepsilon.1,2,\varepsilon)$,
$o(o(4,\varepsilon.1,2,\varepsilon),\varepsilon.1.1,2,\varepsilon)$,
and
$o(o(o(4,\varepsilon.1,2,\varepsilon),\varepsilon.1.1,2,\varepsilon),
\varepsilon.1.1.1,2,\varepsilon)$,
respectively. 
Note that the path concatenation operator ``$.$'' associates to the
left.
Applying the series--guessing algorithm to this
term sequence, and setting, say, the example count $k=2$, we
obtain one single law, viz.\
$[\suc(\suc(v_1)),o(v_2,v_3.1,2,\varepsilon),v_2|v_4] \leadsto
o(o(v_2,v_3.1,2,\varepsilon),v_3.1.1,2,\varepsilon)$. If,
by way of a
negative example, we apply the series--guessing algorithm to
Rules~$4$, $5$, $11$, $15$, we get no law at all, thus
indicating that this sequence does not belong to a rule family. 

Since series guessing wrt.\ the empty theory is quite fast, one
might check all possible ascending rule sequences whether their
origin terms reveal a common law or not. Some optimization may
be achieved using the fact that a sequence has no law if any of its
prefixes has no law. Moreover, efficient data structures like
substitution tree indexes \cite{Graf.1994} might help to find
all rule subsequences that have a law.

\paragraph{Applying series guessing to a selected rule subsequence.}
Once we have selected one or more rule sequences as described
above, we may apply the series--guessing algorithm to the rules.
Each series law thus obtained will correspond to a meta rule
that generalizes all rules of the subsequence modulo the chosen
equational background theory. For example, in a first attempt, we
applied series guessing to the rules shown in
Fig.~\ref{Series--Guessing Approach (A) Applied to the Rule
Sequence $4,5,9,10$}, however without success.
Figure~\ref{Series--Guessing Approach (B) Applied to the Rule
Sequence $4,5,9,10$} documents a successful attempt at series
guessing for both $k=2$ and $k=3$, taking 17 and 28 seconds user time,
respectively.

\begin{figure}
\begin{center}
\begin{tabular}[t]{@{}ll@{}}
Input Series:
\\[-0.2cm]
\begin{tabular}[t]{@{}|lllllllllllllll|@{}}
\hline
$[$
& $f($ & $f($ & $f($ & $f(-x_4,-x_3)$ & $,-x_2),$ & $-x_1)$
	& $,f($ & $f($ & $x_1,$ & $f(x_2,$ & $f(x_3,x_4)))$ & $,y))$
	& $\leadsto y$ & $,$	\\
& $f($ & $f($ & $f($ & $-x_3$ & $,-x_2),$ & $-x_1)$
	& $,f($ & $f($ & $x_1,$ & $f(x_2,$ & $x_3))$ & $,y))$ 
	& $\leadsto y$ & $,$ \\
& $f($ & $f($ & \mc{3}{$-x_2,$} & $-x_1)$
	& $,f($ & $f($ & $x_1,$ & \mc{2}{$x_2)$} & $,y))$
	& $\leadsto y$ & $,$	\\
& $f($ & \mc{5}{$-x_1$}
	& $,f($ & \mc{4}{$x_1$} & $,y))$
	& $\leadsto y$
	& $]$	\\
\hline
\end{tabular}
\\
\\
Variables ($k=2$):
\\[-0.2cm]
\begin{tabular}[t]{@{}|llrl|@{}}
\hline
$v_1$ & $=$ & $\phi(0,$ & $\suc(0))$	\\
$v_2$ & $=$ & $\phi(-x_2,$ & $f(-x_3,-x_2))$	\\
$v_3$ & $=$ & $\phi(x_2,$ & $f(x_2,x_3))$	\\
$v_4$ & $=$ & $\phi(-x_1,$ & $f(-x_2,-x_1))$	\\
$v_5$ & $=$ & $\phi(x_1,$ & $(fx_1,x_2))$	\\
$v_6$ & $=$ & $\phi([\;],$ & $[f(-x_1,f(x_1,y)) \leadsto y])$ \\
\hline
\end{tabular}
\\
\\
No laws found ($k=2$, $k=3$)%
&
%.
\\
\end{tabular}
\end{center}
\caption{Series--Guessing Approach (A) Applied to the Rule Sequence
	$4,5,9,10$}
\label{Series--Guessing Approach (A) Applied to the Rule Sequence
	$4,5,9,10$}
\end{figure}

\begin{figure}
\begin{center}
\begin{tabular}[t]{@{}l@{\hspace*{1.5cm}}l@{}}
\mca{2}{Input Series:}
\\[-0.2cm]
\mca{2}{%
\begin{tabular}[t]{@{}|lllllllllllllll|@{}}
\hline
$[$
& $f($ & $f($ & $f($ & $f(-x(1),-x(2))$ 
	& $,-x(3)),$ & $-x(4))$
	& $,f($ & $f($ & $x(4),$ & $f(x(3),$ 
	& $f(x(2),x(1))))$ & $,y))$
	& $\leadsto y$ & $,$	\\
& $f($ & $f($ & $f($ & $-x(2)$ & $,-x(3)),$ 
	& $-x(4))$
	& $,f($ & $f($ & $x(4),$ & $f(x(3),$ 
	& $x(2)))$ & $,y))$ 
	& $\leadsto y$ & $,$ \\
& $f($ & $f($ & \mc{3}{$-x(3),$} & $-x(4))$
	& $,f($ & $f($ & $x(4),$ & \mc{2}{$x(3))$} 
	& $,y))$
	& $\leadsto y$ & $,$	\\
& $f($ & \mc{5}{$-x(4)$}
	& $,f($ & \mc{4}{$x(4)$} & $,y))$
	& $\leadsto y$ & $]$	\\
\hline
\end{tabular}%
}
\\
\\
Variables ($k=2$):
&
Variables ($k=3$):
\\[-0.2cm]
\begin{tabular}[t]{@{}|llrl|@{}}
\hline
$v_1$ & $=$ & $\phi(0,$ & $1)$	\\
$v_2$ & $=$ & $\phi(-x(1),$ & $f(-x(1),-x(2)))$ \\
$v_3$ & $=$ & $\phi(x(1),$ & $f(x(2),x(1)))$ \\
$v_4$ & $=$ & $\phi([\;],$ & $[\ldots])$ \\
\hline
\end{tabular}
&
\begin{tabular}[t]{@{}|llrcl|@{}}
\hline
$v_1$ & $=$ & $\phi(0,$ & $1,$ & $2)$	\\
$v_2$ & $=$ & $\phi(-x(1),$ & $f(-x(1),-x(2)),$
	& $f(f(-x(1),-x(2)),-x(3)))$ \\
$v_3$ & $=$ & $\phi(x(1),$ & $f(x(2),x(1)),$
	& $f(x(3),f(x(2),x(1))))$ \\
$v_4$ & $=$ & $\phi([\;],$ & $[\ldots],$ & $[\ldots])$ \\
\hline
\end{tabular}
\\
\\
\mca{2}{Laws found ($k=2$):}
\\[-0.2cm]
\mca{2}{%
\begin{tabular}[t]{@{}|lll|@{}}
\hline
$f(f(f(v_2,-x(\suc^2(v_1))),-x(\suc^3(v_1))),$
	& $f(f(x(\suc^3(v_1)),f(x(\suc^2(v_1)),v3)),y))$
	& $\leadsto y$	\\
$f(f(f(v_2,-x(\suc^2(v_1))),-x(\suc^3(v_1))),$
	& $f(f(x(\suc^3(v_1)),f(x(\suc^2(v_1)),-v_2)),y))$
	& $\leadsto y$	\\
$f(f(f(-v_3,-x(\suc^2(v_1))),-x(\suc^3(v_1))),$
	& $f(f(x(\suc^3(v_1)),f(x(\suc^2(v_1)),v_3)),y))$
	& $\leadsto y$	\\
$f(f(f(-v_3,-x(\suc^2(v_1))),-x(\suc^3(v_1))),$
	& $f(f(x(\suc^3(v_1)),f(x(\suc^2(v_1)),-v_2)),y))$
	& $\leadsto y$	\\
\hline
\end{tabular}%
}
\\
\\
\mca{2}{Laws found ($k=3$):}
\\[-0.2cm]
\mca{2}{%
\begin{tabular}[t]{@{}|lll|@{}}
\hline
$f(f(v_2,-x(\suc^2(v_1))),$ & $f(f(x(\suc^2(v_1)),v3),y))$
	& $\leadsto y$	\\
$f(f(v_2,-x(\suc^2(v_1))),$ & $f(f(x(\suc^2(v_1)),-v_2),y))$
	& $\leadsto y$	\\
$f(f(-v_3,-x(\suc^2(v_1))),$ & $f(f(x(\suc^2(v_1)),v_3),y))$
	& $\leadsto y$	\\
$f(f(-v_3,-x(\suc^2(v_1))),$ & $f(f(x(\suc^2(v_1)),-v_2),y))$
	& $\leadsto y$	\\
\hline
\end{tabular}%
}
\end{tabular}
\end{center}
\caption{Series--Guessing Approach (B) Applied to the Rule Sequence
	$4,5,9,10$}
\label{Series--Guessing Approach (B) Applied to the Rule Sequence
	$4,5,9,10$}
\end{figure}

\paragraph{Dealing with variable names.}
Since the E--anti--unification approach presented here is based
on regular tree grammars, we treated variables as constants.
This is no longer sufficient for the present completion example:
if the variables are named as in Figs.~\ref{Input Rewrite Rules
from Kirchner's Example 2.1} and~\ref{Completion of the Theory from
Fig. Input Rewrite Rules from Kirchner's Example 2.1}, the
syntactical anti--unification of length--annotated suffixes will
yield the variables shown in Fig.~\ref{Series--Guessing Approach
(A) Applied to the Rule Sequence $4,5,9,10$}. Note that $x_4$
does not occur in any argument of $\phi$; in particular, a
variable $v = \phi(-x_3,f(-x_4,-x_3))$ is missing. This is the
reason for the non--existence of a series law in
Fig.~\ref{Series--Guessing Approach (A) Applied to the Rule
Sequence $4,5,9,10$}. If we number the variables $x_1$ to $x_4$
in reverse order, we have a similar problem, since then
$\phi(x_3,x_4)$ is missing. 

In a second attempt, we tried to circumvent this problem by
modeling variables as functions of natural numbers, i.e.\
$x(i)$ for $x_i$, where $i$ denotes $\suc^i(0)$ for the sake of
brevity. The rule sequence and the corresponding $\phi$ is
shown in Fig.~\ref{Series--Guessing Approach (B) Applied to the
Rule Sequence $4,5,9,10$}. Note that we also had to
reverse the numbering of variables, since otherwise
$\phi(x(3),f(x(3),x(4)))$ would again be missing. 

What is needed, though, is an approach to E--anti--unification
that is
able to deal with variables rather than just constants.

\begin{figure}
\begin{center}
\begin{tabular}[t]{@{}lll@{}}
$\eqc{X \leadsto Y}$ & $\sortdef$ & $\eqc{X} \leadsto \eqc{Y}$	\\
$\eqc{f(X,Y)}$ & $\sortdef$ 
	& $f(\eqc{X},\eqc{Y}) \mid - -\eqc{f(X,Y)}
	\mid -f(\eqc{-Y},\eqc{-X})$	\\
$\eqc{-f(X,Y)}$ & $\sortdef$ & $\eqc{f(-Y,-X)}$	\\
$\eqc{- -X}$ & $\sortdef$ & $\eqc{X}$	\\
$\eqc{-X}$ & $\sortdef$ & $-\eqc{X} \mid - -\eqc{-X}$	\\
$\eqc{X}$ & $\sortdef$ & $X \mid - -\eqc{X}$	\\
\end{tabular}
\end{center}
\caption{Tree--Grammar Scheme for Equivalence Classes in Kirchner's
	Example 2.1}
\label{Tree--Grammar Scheme for Equivalence Classes in Kirchner's
	Example 2.1}
\end{figure}

\paragraph{Setting up the equational background theory's tree grammar.}
The main problem here consists in determining an appropriate
equational background theory in the first place. In the above example,
we simply took the theory given in \cite{Kirchner.1989},
consisting of Rules~1 and~2 from Fig.~\ref{Input Rewrite
Rules from Kirchner's Example 2.1}. In general, though, it is not at
all clear which background theory is appropriate. In Kirchner's
examples, it is always taken from the input rules. In many other
examples, it simply defines the repeated application of certain
functions; e.g., for the completion example\footnote{This
example was provided by J\"org Denzinger, University of
Kaiserslautern.} in Fig.~\ref{Divergent Completion Schematizable
using omega Terms}, it is sufficient to define $b^i(x)$ by
$b^0(x) = x$ and $b^{s(i)}(x) = b(b^i(x))$, and to define
$d^i(x)$ similarly. This is a special case of $\omega$ terms
which have been investigated to deal with infinite sets of terms
occurring in Knuth--Bendix completion or elsewhere
\cite{Comon.1995}. It seems to be a good default measure to use
a theory of appropriate $\omega$ terms as the equational background
theory. 

Once the background theory itself is fixed, the next problem is
how to find an appropriate regular tree grammar for the
equivalence classes modulo this theory. In the above example,
this was the most time--consuming and error--prone task.
Figure~\ref{Tree--Grammar Scheme for Equivalence Classes in
Kirchner's Example 2.1} shows the tree--grammar scheme (cf.\
Sect.~\ref{Dynamic Sort Generation}) we eventually came up with;
the rules are priorized PROLOG facts. Note that the approach of
\cite{Emmelmann.1994} mentioned in Sect.~\ref{Modeling
Equivalence Classes as Sorts} is not applicable, since it
requires a confluent and noetherian term--rewriting system, which
is precisely what the completion process itself is about to build.
Except for the cases where the background theory is based on
$\omega$ terms, a satisfactory approach to automatically obtaining
a tree grammar for the equivalence classes has yet to be found.

\begin{figure}
\begin{center}
\begin{tabular}[t]{@{}r@{\hspace{0.5cm}}rl@{}}
\mca{3}{Ordering LPO: $c \succ b \succ a \succ d \succ f \succ e$} \\
\mca{3}{Input Equations:} \\
1 & $a(c(x))$ & $\leadsto x$	\\
2 & $c(f(x))$ & $\leadsto e(x)$	\\
3 & $a(e(x))$ & $\leadsto f(x)$	\\
[0.1cm]
\mca{3}{Completion:}	\\
4 & $c(d(x))$ & $\leadsto b(c(x))$	\\
5 & $a(b(c(x)))$ & $\leadsto d(x)$	\\
6 & $a(b(e(x)))$ & $\leadsto d(f(x))$	\\
7 & $a(b(b(c(x))))$ & $\leadsto d(d(x))$	\\
8 & $a(b(b(e(x))))$ & $\leadsto d(d(f(x)))$	\\
9 & $a(b(b(b(c(x)))))$ & $\leadsto d(d(d(x)))$	\\
10 & $a(b(b(b(e(x)))))$ & $\leadsto d(d(d(f(x))))$	\\
11 & $a(b(b(b(b(c(x))))))$ & $\leadsto d(d(d(d(x))))$	\\
12 & $a(b(b(b(b(e(x))))))$ & $\leadsto d(d(d(d(f(x)))))$	\\
13 & $a(b(b(b(b(b(c(x)))))))$ & $\leadsto d(d(d(d(d(x)))))$	\\
14 & $a(b(b(b(b(b(e(x)))))))$ & $\leadsto d(d(d(d(d(f(x))))))$	\\
15 & $a(b(b(b(b(b(b(c(x))))))))$ & $\leadsto d(d(d(d(d(d(x))))))$ \\
16 & $a(b(b(b(b(b(b(e(x))))))))$ & $\leadsto d(d(d(d(d(d(f(x)))))))$ \\
\mca{3}{\ldots}	\\
\end{tabular}
\end{center}
\caption{Divergent Completion Schematizable using $\omega$ Terms}
\label{Divergent Completion Schematizable using omega Terms}
\end{figure}

\paragraph{Considering equations in anti--unifying length--annotated
suffixes.}
A closer look at the above completion example reveals a flaw in
our approach to series guessing as presented in
Sect.~\ref{Series Guessing -- Implementation}.
Consider the variables $v_2$ and
$v_3$ in Fig.~\ref{Series--Guessing Approach (B) Applied to the
Rule Sequence $4,5,9,10$}. Their pre--images under $\phi$ are
inverse to each other modulo the background equational theory.
Hence, it was desirable to establish the relation $v_2 = -v_3$,
and to delete $v_3$ from the set $V$ of variables that may
appear in a series law, thus avoiding certain redundant series laws.
In fact, the set of enumerated laws for
$k=2$ collapses into the singleton set

$$\{f(f(f(v_2,-x(\suc^2(v_1))),-x(\suc^3(v_1))),
f(f(x(\suc^3(v_1)),f(x(\suc^2(v_1)),-v_2)),y)) \leadsto y\}$$

modulo the above relation; the case is similar for $k=3$. The reason for
the appearance of both $v_2$ and $v_3$ is that we did not consider the
equational background theory when anti--unifying the
length--annotated suffixes. 
However, by simply E--anti--unifying them using the algorithm from
Sect.~\ref{Linear Generalizations}~/~\ref{Nonlinear Generalizations},
we would still get both $v_2$ and $v_3$ in the variables set $V$.

%\clearpage
\subsection{Hoare Invariants}
\label{Hoare Invariants}

When verifying imperative programs, a still unresolved problem is
how to generate loop invariants automatically.
We show below that E--anti--unification could be used for
this purpose.

Consider the following imperative program:

\begin{center}
\vbox{\tt
\begin{tabbing}
x := 0;	\\
for \=i:=0 by 1 to n	\\
do \>x:=x+2*i	\\
done	\\
\end{tabbing}
}
\end{center}

By anti--unifying sample values of $\tpl{{\tt i},{\tt x}}$
for a couple of iterations,
e.g.\ $\tpl{0,0}, \tpl{1,1}, \tpl{2,4}, \tpl{3,9}$
modulo an appropriate theory
(which should define at least {\tt +} and {\tt *}),
we obtain the candidate {\tt x = i*i} for the loop invariant.
Note that {\tt 0 <= i <= n} cannot be obtained since it is not an
equation.

%\clearpage
\subsection{Reengineering of Functional Programs}
\label{Reengineering of Functional Programs}

When programming software, it is often difficult to attain an
adequate level of abstraction at the first try.
Consequently, one often has different procedures that do
almost ---~but not quite~--- the same thing. E--anti--unification
could potentially help in reengineering such programs by
computing a most specific generalization. 

By way of an example, consider a functional program where two almost
identical procedures $f_1(\cdot)$ and $f_2(\cdot)$ are defined: 

$$
\begin{tabular}[t]{@{}rl@{}}
$f_1(x)$ & $:=F[ \; x+1 \; ]$	\\
$f_2(x)$ & $:=F[ \; 2*x \; ]$	\\
$g(x)$ & $:= G[ \; f_1(x_1) \; , \; f_2(x_2) \; ]$	\\
\end{tabular}
$$

where $F[\cdot]$ denotes the possibly complex body expression
common to both procedures; the third line indicates calls to
$f_1$ and $f_2$ in the body $G[\cdot]$ of another procedure $g$.
Purely syntactic anti--unification of the procedure bodies of
$f_1$ and $f_2$ yields $F[y]$, leading to the following
reengineered program: 

$$
\begin{tabular}[t]{@{}rl@{}}
$f(x,y)$ & $:=F[ \; y \; ]$	\\
$g(x)$ & $:= G[ \; f(x_1,x_1+1) \; , \; f(x_2,2*x_2) \; ]$	\\
\end{tabular}
$$

Note that the calls to $+$ and $*$ have been moved outward since they
are considered to have nothing in common.
Using E--anti--unification modulo the theory of $+$ and $*$,
we get a more specific generalization of the bodies, viz.\
$F[a*x+b]$, leading to a better factorization of the commonalities of
$f_1$ and $f_2$:

$$
\begin{tabular}[t]{@{}rl@{}}
$f(x,a,b)$ & $:=F[ \; a*x+b \; ]$	\\
$g(x)$ & $:= G[ \; f(x_1,1,1) \; , \; f(x_2,2,0) \; ]$	\\
\end{tabular}
$$

In practice, the literally identical part $F[\cdot]$ of two
procedure bodies will usually be small, or even non--existent.
It is thus of great importance to be able to deal with
commonalities that are revealed only by applying a background
theory about programming language semantics. To the extent that
this theory is an equational one and admits regular equivalence
classes, our E--anti--unification approach can be employed to
detect the commonalities.

%\clearpage
\subsection{Strengthening of Induction Hypotheses}
\label{Strengthening of Induction Hypotheses}

It is well known that sometimes a formula $F$ cannot be proven
by induction, while a stronger formula $G$ such that $G \Ra F$ is
provable. The reason is that a stronger formula also provides a
stronger induction hypothesis. In inductive theorem proving, it
is therefore a common technique to strengthen $F$ if its original
proof attempt failed. This is usually done by replacing some
constants or terms by variables, or by decoupling different
occurrences of the same variable; in other words, by building
anti--instances of $F$. The main problem is not to
overgeneralize $F$ such that it is no longer valid. 

There are several heuristics known from the literature, e.g.\
\cite{Boyer.Moore.1979,Hummel.1990,Bundy.Harmelen.Ireland.1990}.
Manna and
Waldinger \cite{Manna.Waldinger.1980} propose using
anti--unification, based on the data acquired during the failed
original proof attempt, to find the least strengthening of $F$
that has a chance of being proven inductively. However, they failed to
notice that their example does in fact need anti--unification
modulo an equational theory.

\begin{figure}
\begin{center}
\begin{tabular}[t]{@{}rrl@{\hspace*{0.5cm}}l@{}}
1. & $app([\;],l)$ & $= l$	\\
2. & $app([h \mid t],l)$ & $= [h \mid app(t,l)]$	\\
3. & $rev([\;])$ & $= [\;]$	\\
4. & $rev([h \mid t])$ & $= app(rev(t),[h])$	\\
\cline{1-3}
5. & $rev\_a([\;],l)$ & $= l$	\\
6. & $rev\_a([h \mid t],l)$ & $= rev\_a(t,[h] \mid l)$	\\
[0.2cm]
\mca{4}{Equations 1.\ to 4.\ are given as background theory,
	defining $append$ and $reverse$;}	\\
\mca{4}{equations 5.\ and 6.\ are the usual optimized version
	of $reverse$ using an accumulator.}	\\
\mca{4}{The goal is to show the correctness of $rev\_a$,
	i.e.\ to show}	\\
[0.2cm]
7. & $rev\_a(l,[\;])$ & $= rev(l)$	\\
[0.2cm]
\mca{4}{We try a proof by induction on $l$; $l = [\;]$ is trivial;} \\
\mca{4}{in case $l = [h \mid t]$, we get the following goal sequence:}\\
[0.2cm]
8. & $rev\_a([h \mid t],[\;])$ & $= rev([h \mid t])$	\\
9. & $rev\_a(t,[h])$ & $= app(rev(t),[h])$ & by 6.\ and 4.\	\\
10. & $rev\_a(t,[h])$ & $= app(rev\_a(t,[\;]),[h])$ & by I.H.\ (7.) \\
\end{tabular}
\end{center}
\caption{Failed Induction--Proof Attempt}
\label{Failed Induction--Proof Attempt}
\end{figure}

Consider the equational theory and the
induction proof attempt shown in Fig.~\ref{Failed Induction--Proof
Attempt}.
Manna and Waldinger suggest anti--unifying the actual goal
with the induction hypothesis if simple rewriting with it does not
lead to success.
This way, overgeneralization could be avoided, since anti--unification
computes the most specific anti--instance of the induction hypothesis
that is just general enough to make it applicable.
Backtracking to 9., we obtain by E--anti--unification
with the induction hypothesis:

$$
\begin{tabular}[t]{@{}lll@{\hspace*{0.5cm}}l@{}}
$rev\_a(t,[h])$ && $= app(rev(t),[h])$ & 9.	\\
$rev\_a(l,[\;])$ & $= rev(l)$ & $= app(rev(l),[\;])$
	& I.H.\ (7.), and $l=app(l,[\;])$	\\
\hline
$rev\_a(t',l')$ && $= app(rev(t'),l')$ & strengthened I.H.	\\
\end{tabular}
$$

It is well known that the proof succeeds using the strengthened
induction hypothesis.

In order to do the E--anti--unification shown above, we need to
define equivalence classes of non--ground terms as regular tree
languages. In our example, we get the grammar scheme shown in
Fig.~\ref{Equivalence Classes of Non--Ground Terms mod.
$append,reverse$}.

\begin{figure}
\begin{center}
\begin{tabular}[t]{@{}ll@{\hspace*{-0.5cm}}l@{}}
$\eqc{rev(A)}$ & $\sortdef rev(\eqc{A}) 
	\mid app(\eqc{[\;]},\eqc{rev(A)})
	\mid app(\eqc{rev(A)},\eqc{[\;]})$	\\
$\eqc{app(A,B)}$ & $\sortdef app(\eqc{A},\eqc{B})
	 \mid app(\eqc{app(A,B)},\eqc{[\;]})
	 \mid app(\eqc{[\;]},\eqc{app(A,B)})
	 \mid rev(\eqc{rev(app(A,B))})$	\\
$\eqc{A}$ & $\sortdef A
	\mid app(\eqc{[\;]},\eqc{A})
	\mid app(\eqc{A},\eqc{[\;]})
	\mid rev(\eqc{rev(A)})$
	& if $app(\ldots) \neq A \neq rev(\ldots)$	\\
\end{tabular}
\end{center}
\caption{Equivalence Classes of Non--Ground Terms mod.\ $append,reverse$}
\label{Equivalence Classes of Non--Ground Terms mod. $append,reverse$}
\end{figure}

In the example, it is sufficient to syntactically anti--unify the 
left--hand sides,
and to provide the variable set $\{t',l'\}$ where $t' = \phi(t,l)$
and $l' = \phi([h],[\;])$ to $rsg_V$ of the right--hand sides, viz.\
$rsg_V(\{t',l'\},\eqc{app(rev(t),[h])},\eqc{rev(l)})$.

The desired solution is the one enumerated first, and it is found
in 1 second user time.
We list all enumerated terms
at depth level 3, which is the minimum level
of the result sort:

$$app(rev(t'),l'), \;\; app(rev(t'),rev(l')),
\;\; app(rev(t'),app(l',[\;])), \;\; app(rev(t'),app([\;],l')) .$$

Note that the second term is not a correct solution, the reason
being that instances of $l'$, viz.\ $[h]$ and $[\;]$, are both
``trivial'' values in the sense of requirement 3.\ from
Sect.~\ref{Selection of Ground Instances}, i.e.\ they satisfy
$rev(x)=x$. The third and fourth solution are correct, but not in
normal form; they would have been filtered out by a normal form
filter.

% The PROLOG goal sequence is thus ({\tt c(A)} denoting the equivalence
% class of {\tt A}):
%
% \begin{verbatim}
% [rar].
%
% assert(c(rev(A)) sortdef  rev(c(A))
%                         !app(c([]),c(rev(A)))
%                         !app(c(rev(A)),c([]))).
% assert(c(app(A,B)) sortdef        app(c(A),c(B))
%                         !app(c(app(A,B)),c([]))
%                         !app(c([]),c(app(A,B)))
%                         !rev(c(rev(app(A,B))))).
% assert(c(A) sortdef      A
% 			!app(c([]),c(A))
%                         !app(c(A),c([]))
%                         !rev(c(rev(A)))).
% assert((A sortdef B :- A sortdf B)).
%
% aus([rev\_a(vl,[]),rev\_a(vt,[vh])],A),
% setof((Ss,var,V),origin_v(Ss,V),OccL),
% auv(inf,[c(rev(vl)),c(app(rev(vt),[vh]))],OccL,B),
% assert(redices([])),
% find_rhs(B,[],R).
% \end{verbatim}

%\clearpage
\section{Dynamic Sort Generation}
\label{Dynamic Sort Generation}

For many background equational theories, some of the equivalence
classes follow a common schema. For example, consider Theory (3)
from Fig.~\ref{Background Equational Theories Used}, defining
$append$ and $reverse$ of lists. The equivalence class of any
list $[A,B]$ can be defined by 

$$\eqc{[A,B]}
\sortdef [\eqc{A},\eqc{B}]
\mid app(\eqc{[A,B]},\eqc{[\;]})
\mid app(\eqc{[A]},[\eqc{B}])
\mid app(\eqc{[\;]},\eqc{[A,B}])
\mid rev(\eqc{[B,A]}),$$

irrespective of the value of $A$ and $B$. We thus get a schema
of sort definitions abbreviating any of its instances. This can
be implemented in PROLOG by a corresponding fact containing $A$
and $B$ as PROLOG variables.

\begin{figure}
\begin{center}
\begin{tabular}[t]{@{}l@{$\;$}ll@{}}
$(\eqc{0}$ & $\sortdef 0 \mid \eqc{0}+\eqc{0}).$	\\
$(\eqc{\suc(N)}$ & $\sortdef \suc(\eqc{N}) \mid Sums)$
	& $\la all\_sums(0,N,Sums).$	\\
[0.2cm]
\mca{2}{$all\_sums(I,0,\eqc{I}+\eqc{0}).$}	\\
\mca{2}{$all\_sums(I,\suc(J),\eqc{I}+\eqc{\suc(J)} \mid Sums)$}
	& $\la all\_sums(\suc(I),J,Sums).$	\\
\end{tabular}
\end{center}
\caption{Sort Definition Scheme for Background Theory (0) from Fig.\
	\protect\ref{Background Equational Theories Used}}
\label{Sort Definition Scheme for Background Theory (0) from Fig.}
\end{figure}

Moreover, we may even use PROLOG rules to dynamically generate
sort definitions as needed. For example, the schema shown in
Fig.~\ref{Sort Definition Scheme for Background Theory (0) from
Fig.} will generate the equivalence class of each natural number
modulo the theory of $(+)$. The predicate $all\_sums(I,J,Sums)$
computes a sort disjunction 

$$Sums \;\;\;=\;\;\;
\eqc{I}+\eqc{J} \mid \eqc{I\1+1}+\eqc{J\1-1}
\mid \ldots \mid \eqc{I\1+J}+\eqc{0}.$$

Similarly, the theory of $(+)$ and $(*)$ has been
schematized, requiring
a predicate that computes all factors of a given number.

%\clearpage
\section{Depth--Bounded E--Anti--Unification}
\label{Depth--Bounded E--Anti--Unification}

In most applications of E--anti--unification, one is interested only in
the first few enumerated terms, which are the simplest wrt.\ the
employed depth measure.
In such cases, it may be sufficient to cut off the $rsg_V$ algorithm
each time it reaches a given depth, thus avoiding the computation of
larger solutions that will subsequently be ignored anyway.
This way, even large terms can be E--anti--unified.

For example, consider the series
$2,4,16,256$.
The series--prediction algorithm
from Sect.~\ref{Series Guessing -- Implementation}
with example count $k=2$ will call
$rsg_V(\{v_1,v_2,v_3,v_4\},s_{16},s_{256})$,
where
$v_1 = \phi(16-4,0)$,
$v_2 = \phi(4-2,0)$,
$v_3 = \phi([2],[\;])$,
and
$v_4 = \phi(4-3,0)$.

The first enumerated solution should be
$\suc^4(v_1)*\suc^4(v_1)$, which has depth 5, or depth 1 if
$\suc$ is not counted. In any case, it is not necessary to
descend to depth 256 to obtain this solution. 

In this example, however, owing to the growing length of sort
definitions caused by $(+)$, the search--tree breadth is too
large to get reasonable runtimes, even for depth 2.

\clearpage
\section{Equational Theories of Finite Algebras}
\label{Equational Theories of Finite Algebras}

In this section,
we show that for each finite algebra we can always generate a closed
representation of all its quantifier--free and variable--bounded
theorems.

We first
give some sufficient criteria for an equational
theory so that the equivalence classes are regular tree languages
(Sect.~\ref{Criteria for Regular Equivalence Classes}).
We then show that
the set $\jTH_n(\A)$
of formal equations in up to $n$ variables, which are
universally valid over a given finite algebra $\A$,
is always a regular tree language.
This implies a constructive way of describing $\jTH_n(\A)$
by a finite term--rewriting system (Sect.~\ref{Equational Theories in
$n$ Variables}).
By extending the approach to typed algebras and including a
type $Bool$, we are able to construct an axiom set
describing the set of all quantifier--free and variable--bounded
theorems of a given finite algebra with arbitrary functions,
predicates, and junctors.
See Fig.~\ref{Theory of N mod 3 with +, <, =, and Bool with wedge
and vee}
for an axiom system that describes
$\jTH_2(\tpl{(\N \;mod\; 3),(+),(<),(=),(\wedge),(\vee)})$.

\newcommand{\sn}[3]{s_{#1 #2 \ldots #3}}

\subsection{Criteria for Regular Equivalence Classes}
\label{Criteria for Regular Equivalence Classes}

\LEMMA{ \eqd{6}
(Sufficient Criterion for Regular Equivalence Classes)	\\
If for each $t \in \NF$ and each $f \in \F$ only finitely many
$\tpl{t_1,\ldots,t_n} \in \NF^n$ exist
\\
such that $f(t_1,\ldots,t_n) =_E t$,
\\
then the equivalence class $\eqc{t}$
of each term $t \in \T$
can be represented as a regular sort.
}
\PROOF{ $\;$	\\
For each $t \in \NF$, we may define the sorts
\\
\begin{tabular}[t]{@{}rl@{\hspace*{0.5cm}}l@{}}
$\eqc{t}$ & $\sortdef
	\bigmid_{f \in \F,
		t_1,\ldots,t_n \in \NF,
		f(t_1,\ldots,t_n) =_E t} \;
	\sn{f}{t_1}{t_n}$
	& and	\\
$\sn{f}{t_1}{t_n}$ & $\sortdef f(\eqc{t_1},\ldots,\eqc{t_n})$
	& for each $t_1,\ldots,t_n \in \NF$.	\\
\end{tabular}

We use the induction principle (Thm.~\eqr{ind} from
Sect.~\ref{Definitions and Notations}) with 
\\
\begin{tabular}[t]{@{}rl@{\hspace*{0.5cm}}l@{}}
$P_{\eqc{t}}(t')$ & $:\Lra t' =_E t$ & and 	\\
$P_{\sn{f}{t_1}{t_n}}(t')$
	& $:\Lra t' =_E f(t_1,\ldots,t_n)$.	\\
\end{tabular}

\bi
\item	Let $t' = f(t'_1,\ldots,t'_n)$
	with $t'_i \in \T$,
	then
	\\
	\begin{tabular}[t]{@{}ll@{\hspace*{0.5cm}}l@{}}
	& $P_{\eqc{t}}(t')$	\\
	$\Ra$ & $t' =_E t$ & Def.\ $P$	\\
	$\Ra$ & $f(t'_1,\ldots,t'_n) =_E t$ & Def.\ $t'$	\\
	$\Ra$ & $t' =_E f(t_1,\ldots,t_n)$
		& where $t_i := nf(t'_i)$	\\
	$\Ra$ & $P_{\sn{f}{t_1}{t_n}}(t')$
		for some $f \in \F$, $t_1,\ldots,t_n \in \NF$
		with $f(t_1,\ldots,t_n) =_E t$	\\
	\mca{2}{conversely:}	\\
	& $P_{\sn{f}{t_1}{t_n}}(t')$
		for some $f(t_1,\ldots,t_n) =_E t$	\\
	$\Ra$ & $t' =_E f(t_1,\ldots,t_n) =_E t$ & Def.\ $P$	\\
	$\Ra$ & $P_{\eqc{t}}(t')$ & Def.\ $P$	\\
	\end{tabular}
\item	\begin{tabular}[t]{@{}ll@{\hspace*{0.5cm}}l@{}}
	& $P_{\sn{f}{t_1}{t_n}}(t')$	\\
	$\Lra$ & $t' =_E f(t_1,\ldots,t_n)$ & Def.\ $P$	\\
	$\Lra$ & $\exists t'_1,\ldots,t'_n \in \T \;\;
		t' = f(t'_1,\ldots,t'_n)
		\wedge t'_1 =_E t_1 \wedge \ldots \wedge t'_n =_E t_1$
		& $(*)$	\\
	$\Lra$ & $\exists t'_1,\ldots,t'_n \in \T \;\;
		t' = f(t'_1,\ldots,t'_n) \wedge
		P_{\eqc{t_1}}(t'_1) \wedge \ldots \wedge
		P_{\eqc{t_n}}(t'_n)$
	\end{tabular}
	\\
	\begin{tabular}[t]{@{}lll@{}}
	$(*)$: & ``$\Ra$'': & choose $t'_i := t_i$	\\
	& ``$\La$'': & $t' =_E f(t'_1,\ldots,t'_n)
		=_E f(t_1,\ldots,t_n)$	\\
	\end{tabular}
\ei
Hence, the sort $\eqc{t}$ contains all term that are
equivalent mod.\ $E$ to $t$.
}

\COROLLARY{ \eqd{7}
(Sufficient Criterion for Regular Equivalence Classes)	\\
Let $\prec$ be a well--founded ordering on $\NF$
with a finite branching degree,
and let $\preceq$ be its reflexive closure.
\\
If $t_i \preceq nf(f(t_1,\ldots,t_n))$
for all $f \in \F$, $t_1,\ldots,t_n \in \NF$
and all $i \in \{1,\ldots,n\}$,
\\
then $\eqc{t}$ is representable as a regular sort
for all $t \in \T$.
}
\PROOF{ $\;$	\\
Under the above assumptions, we have for fixed $t \in \NF$
and $f \in \F$:
\\
$t_1,\ldots,t_n \in \NF \wedge f(t_1,\ldots,t_n) =_E t
\Ra t_i \preceq nf(f(t_1,\ldots,t_n)) = t$,
hence
\\
$\{\tpl{t_1,\ldots,t_n} \in \NF^n \mid f(t_1,\ldots,t_n) =_E t\}
\subset \{\tpl{t_1,\ldots,t_n} \in \NF^n
	\mid t_1 \preceq t, \ldots, t_n \preceq t\}$,
\\
where the latter set is finite.
The conclusion follows by Lemma \eqr{6}.
}

\COROLLARY{ \eqd{7.5}
$\;$	\\
If $\T /_E$ is finite,
each equivalence class can be represented as a regular tree language.
}
\PROOF{
Trivially, $\NF$ has as many elements as $\T /_E$.
The conclusion follows by Lemma \eqr{6}.
}

\subsection{Equational Theories in One Variable}
\label{Equational Theories in One Variable}

\DEFINITION{ \eqd{x0.4}
$\;$	\\
Let $(:) \not\in \F$ be a symbol of arity 2;
we call a term of the form $t_1:t_2$ a (formal) equation
if $t_1,t_2 \in \T$.
An equation $t_1:t_2$ is called universally valid
if $\sigma t_1 =_E \sigma t_2$ for every substitution $\sigma$.
Define $\jTH_n(\T /_E)$ as the set of all formal equations in up
to $n$ variables that are universally valid over $\T /_E$.
}

\DEFINITION{ \eqd{x0.5}
$\;$	\\
We will assume below that $\T/_E$ is finite,
and hence we have only a finite number $N$ of normal forms.
Define
$B := \{ \tpl{b_1,\ldots,b_N}
\mid \forall t \in \T \; \exists i \in \{1,\ldots,N\} \;\;
t =_E b_i \}$
the set of all $N$-tuples that represent all $N$ normal forms.
Let $\eqc{\vec{b}} := \tpl{\eqc{b_1},\ldots,\eqc{b_N}}$
denote the equivalence class of $\vec{b}$.
In this section,
let $x := \phi(\vec{b})$ for some $\vec{b} \in B$,
and let
$A := \{ \eqc{\vec{a}} \in (\T /_E)^N \mid
\L(rsg_V(\{ x \},\eqc{\vec{a}})) \neq \{\} \}$.
}

\LEMMA{ \eqd{x1}
Let $t_1,t_2 \in \T$
with $vars(t_1) \cup vars(t_2) \subset \{x\}$.
The equation $t_1 : t_2$ is universally valid
iff there exists an $\eqc{\vec{a}} \in A$ such that
$t_1,t_2 \in \L(rsg_V(\{ x \},\eqc{\vec{a}}))$.
}
\PROOF{
$\;$	\\
``$\Ra$'':
\\
Consider the $\sigma_i$ from Lemma \eqr{vrsg1} in
Sect.~\ref{Variable--Restricted E--Anti--Unification}. 
Define $a_i := \sigma_i t_1$;
then $t_1 \in \L(rsg_V(\{ x \},\eqc{\vec{a}}))$
by Lemma \eqr{vrsg1};
similarly, $\sigma_i t_2 =_E \sigma_i t_1 = a_i$ for all $i$ implies
$t_2 \in \L(rsg_V(\{ x \},\eqc{\vec{a}}))$.

``$\La$'':
\\
Let $t_1,t_2 \in \L(rsg_V(\{ x \},\eqc{\vec{a}}))$;
consider an arbitrary substitution $\{x \la t\}$.
We have

\begin{tabular}[t]{@{}ll@{\hspace*{0.5cm}}l@{}}
& $\{x \la t\} \; (t_1)$	\\
$=_E$ & $\{x \la b_i\} \; (t_1)$
	& since $t =_E b_i$ for some $i \in \{1,\ldots,N\}$
	by Def.~\eqr{x0.5}	\\
$=$ & $\sigma_i t_1$
	& by Def.\ of $\sigma_i$ and since $vars(t_1) \subset \{x\}$ \\
$=_E$ & $a_i$
	& by Lemma \eqr{vrsg1}
	since $t_1 \in \L(rsg_V(\{ x \},\eqc{\vec{a}}))$\\
$=_E$ & $\{x \la t\} \; (t_2)$ & similarly	\\
\end{tabular}

Hence, the equation $t_1 : t_2$ is universally valid.
}

\THEOREM{ \eqd{x2}
If $\T /_E$ is finite,
the set of all universally valid equations in one variable $x$ is a
regular tree language.
}
\PROOF{
Let $\phi$ be such that $\phi(\vec{b}) = x$ for some $\vec{b} \in B$.
By Cor.~\eqr{7.5}, $\eqc{a_i}$ is a regular tree language
for each $a \in \T$.
Define
$s_{valid} \sortdef \bigmid_{\eqc{\vec{a}} \in A} \;
rsg_V(\{ x \},\eqc{\vec{a}}):rsg_V(\{ x \},\eqc{\vec{a}})$.
Let $t_1,t_2 \in \T$ with $vars(t_1) \cup vars(t_2) \subset \{x\}$.
Then,

\begin{tabular}[t]{@{}ll@{\hspace*{0.5cm}}l@{}}
& $t_1 : t_2$ is universally valid	\\
$\Lra$ & exists $\vec{a} \in \T^N$
	such that $t_1,t_2 \in \L(rsg_V(\{ x \},\eqc{\vec{a}}))$
	& by Lemma \eqr{x1}	\\
$\Lra$ & exists $\eqc{\vec{a}} \in A$ such that
	$(t_1:t_2) \in
	\L(rsg_V(\{x\},\eqc{\vec{a}}):rsg_V(\{x\},\eqc{\vec{a}}))$ \\
$\Lra$ & $(t_1:t_2) \in \L(s_{valid})$	\\
\end{tabular}
}

\EXAMPLE{ \eqd{0.6}
$\;$	\\
Consider $(\N \;mod\; 2)$ with $+$.
We have, letting $v_{01} = \phi(0,1)$,
and abbreviating $s_{0} := \eqc{0}$, $s_{1} := \eqc{1}$,
and $s_{ij} := rsg_V(\{v_{01}\},s_i,s_j)$:
\\
\begin{tabular}[t]{@{}*{11}{c}@{}}
$s_{0}$ & $\sortdef$ & $0$
	& $\mid$ & $s_{0}+s_{0}$ & $\mid$ & $s_{1}+s_{1}$ \\
$s_{1}$ & $\sortdef$ & $1$
	& $\mid$ & $s_{0}+s_{1}$ & $\mid$ & $s_{1}+s_{0}$ \\
[0.1cm]
$s_{00}$ & $\sortdef$ & $0$
	& $\mid$ & $s_{00}+s_{00}$ & $\mid$ & $s_{01}+s_{01}$
	& $\mid$ & $s_{10}+s_{10}$ & $\mid$ & $s_{11}+s_{11}$ \\
$s_{01}$ & $\sortdef$ & $v_{01}$
	& $\mid$ & $s_{00}+s_{01}$ & $\mid$ & $s_{01}+s_{00}$
	& $\mid$ & $s_{10}+s_{11}$ & $\mid$ & $s_{11}+s_{10}$ \\
$s_{10}$ & $\sortdef$ &&& $s_{00}+s_{10}$ & $\mid$ & $s_{01}+s_{11}$
	& $\mid$ & $s_{10}+s_{00}$ & $\mid$ & $s_{11}+s_{01}$ \\
$s_{11}$ & $\sortdef$ & $1$
	& $\mid$ & $s_{00}+s_{11}$ & $\mid$ & $s_{01}+s_{10} $
	& $\mid$ & $s_{10}+s_{01}$ & $\mid$ & $s_{11}+s_{00}$ \\
[0.1cm]
$s_{valid}$ & $\sortdef$ &&& $s_{00}:s_{00}$ & $\mid$ & $s_{01}:s_{01}$
	& $\mid$ & $s_{10}:s_{10}$ & $\mid$ & $s_{11}:s_{11}$	\\
\end{tabular}
}

\subsection{Equational Theories in $n$ Variables}
\label{Equational Theories in $n$ Variables}

Theorem \eqr{x2} can be generalized to $n$ variables,
which we will do in this section.
The definitions given
below generalize Defs.~\eqr{x0.4} and \eqr{x0.5} to $n$
variables.

\DEFINITION{ \eqd{y0.5}
Modifying Def.~\eqr{x0.5}, define

$$B :=
\left\{
\left(
\begin{array}{ccc}
b_{1,1} & \ldots & b_{1,N^n}	\\
\vdots && \vdots	\\
b_{n,1} & \ldots & b_{n,N^n}	\\
\end{array}
\right)
\raisebox{-4.5ex}{\rule{0.03cm}{10ex}} \;\;
\forall t_1,\ldots,t_n \in \T \;\;
\exists j \in \{1,\ldots,N^n\} \;\;
\forall i \in \{1,\ldots,n\} \;\;
b_{i,j} =_E t_i
\right\}$$

$B$ is non--empty since $(\T /_E)^n$ is finite, viz.\ of
cardinality $N^n$.
Intuitively, for each matrix $\vec{b} \in B$
there is a one--to--one correspondence between its
column vectors and the elements of $(\T /_E)^n$.

In this section,
let $x_1 := \phi(b_{1,1},\ldots,b_{1,N^n})$, \ldots,
$x_n := \phi(b_{n,1},\ldots,b_{n,N^n})$
for some $\vec{b} \in B$,
and let $V := \{ x_1,\ldots,x_n \}$.
Let
$A := \{ \eqc{\vec{a}} \in (\T /_E)^{(N^n)} \mid
\L(rsg_V(\{ x_1,\ldots,x_n \},\eqc{\vec{a}})) \neq \{\} \}$.
}

\LEMMA{ \eqd{y1}
Let $t_1,t_2 \in \T$
with $vars(t_1) \cup vars(t_2) \subset V$.
The equation $t_1 : t_2$ is universally valid
iff there exists an $\vec{a} \in A$ such that
$t_1,t_2 \in \L(rsg_V(\{ x_1,\ldots,x_n \},\eqc{\vec{a}}))$.
}
\PROOF{	$\;$	\\
``$\Ra$'':
\\
Consider the $\sigma_i$ from Lemma \eqr{vrsg1} in
Sect.~\ref{Variable--Restricted E--Anti--Unification}. 
Define $a_i := \sigma_i t_1$ for $i=1,\ldots,N^n$;
\\
then $t_1 \in \L(rsg_V(\{ x_1,\ldots,x_n \},\eqc{\vec{a}}))$
by Lemma \eqr{vrsg1};
\\
similarly, $\sigma_i t_2 =_E \sigma_i t_1 = a_i$ for all $i$ implies
$t_2 \in \L(rsg_V(\{ x_1,\ldots,x_n \},\eqc{\vec{a}}))$.

``$\La$'':
\\
Let $t_1,t_2 \in \L(rsg_V(\{ x_1,\ldots,x_n \},\eqc{\vec{a}}))$;
consider an arbitrary substitution
$\{x_1 \la t'_1,\ldots,x_n \la t'_n\}$.
We have
\\
\begin{tabular}[t]{@{}ll@{\hspace*{0.5cm}}l@{}}
& $\sigma_j x_i$	\\
$=$ & $\sigma_j \phi(b_{i,1},\ldots,b_{i,N^n})$ & Def.\ $x_i$	\\
$=$ & $b_{i,j}$ & Def.\ $\sigma_j$	\\
$=$ & $\{x_1 \la b_{1,j},\ldots,x_n \la b_{n,j}\} \; (x_i)$	\\
\end{tabular}
\\
By induction on $t$, it follows that
$\sigma_j t = \{x_1 \la b_{1,j},\ldots,x_n \la b_{n,j}\} \; (t)$
for all $t \in \T$ with $vars(t) \subset \{x_1,\ldots,x_n\}$.
Thus:
\\
\begin{tabular}[t]{@{}ll@{\hspace*{0.5cm}}l@{}}
& $\{x_1 \la t'_1,\ldots,x_n \la t'_n\} \; (t_1)$	\\
$=_E$ & $\{x_1 \la b_{1,j},\ldots,x_n \la b_{n,j}\} \; (t_1)$
	& since
	$\exists j \in \{1,\ldots,N^n\} \;
	\forall i \in \{1,\ldots,n\} \;\; b_{i,j} =_E t'_i$
	by Def.~\eqr{y0.5}	\\
$=$ & $\sigma_j t_1$ & as shown above	\\
$=_E$ & $a_j$
	& by Lemma \eqr{vrsg1},
	since $t_1 \in \L(rsg_V(\{ x_1,\ldots,x_n \},\eqc{\vec{a}}))$\\
$=_E$ & $\{x_1 \la t'_1,\ldots,x_n \la t'_n\} \; (t_2)$ & similarly \\
\end{tabular}
\\
Hence, the equation $t_1 : t_2$ is universally valid.
}

\THEOREM{ \eqd{y2}
If $\T /_E$ is finite,
the set $\jTH_n(\T/_E)$
of all universally valid equations in $n$ variables is a
regular tree language for each $n$.
}
\PROOF{
Let $\phi$ be
such that $\phi(b_{1,1},\ldots,b_{1,N^n}) = x_1$, \ldots,
$\phi(b_{n,1},\ldots,b_{n,N^n}) = x_n$ for some $\vec{b} \in B$;
let $V := \{ x_1,\ldots,x_n \}$.
By Cor.~\eqr{7.5}, $\eqc{a_i}$ is a regular tree language
for each $a \in \T$.
Define
$s_{valid} \sortdef \bigmid_{\eqc{\vec{a}} \in A}
\; rsg_V(\{ x_1,\ldots,x_n \},\eqc{\vec{a}}):rsg_V(\{ x_1,\ldots,x_n
\},\eqc{\vec{a}})$.
Let $t_1,t_2 \in \T$ with $vars(t_1) \cup vars(t_2) \subset V$.
Then
\\
\begin{tabular}[t]{@{}ll@{\hspace*{0.45cm}}l@{}}
& $t_1 : t_2$ universally valid \\
$\Lra$ & exists $\vec{a} \in A$
	such that
	$t_1,t_2 \in \L(rsg_V(\{ x_1,\ldots,x_n \},\eqc{\vec{a}}))$
	& by Lemma \eqr{y1}	\\
$\Lra$ & exists $\eqc{\vec{a}} \in A$ such that
	$(t_1:t_2) \in \L(rsg_V(\{x_1,\ldots,x_n\},\eqc{\vec{a}})
	:rsg_V(\{x_1,\ldots,x_n\},\eqc{\vec{a}}))$ \\
$\Lra$ & $(t_1:t_2) \in \L(s_{valid})$	\\
\end{tabular}
}

\EXAMPLE{	\eqd{y0.6}
Consider again $(\N \;mod\; 2)$ with $+$.
Let
$v_{0011} = \phi(0,0,1,1)$,
$v_{0101} = \phi(0,1,0,1)$,
$V = \{v_{0011}, v_{0101}\}$,
and abbreviate $s_{0} := \eqc{0}$, $s_{1} := \eqc{1}$,
and $s_{ijkl} := rsg_V(\{v_{0011}, v_{0101}\},s_i,s_j,s_k,s_l)$.
The sort definitions are shown in Fig.~\ref{Sort Definition in 
Exm. \eqR{y0.6}}; 
the sorts $s_{0001}$, $s_{0010}$, $s_{0100}$, $s_{0111}$, $s_{1000}$,
$s_{1011}$, $s_{1101}$, and $s_{1110}$
are all empty and have been omitted
in the definition of $s_{valid}$.
For example, the commutativity law
$v_{0011}+v_{0101}:v_{0101}+v_{0011}$
is contained in
$\L(s_{0110}:s_{0110})$.
}

\begin{figure}
\begin{center}
\begin{tabular}[t]{@{}*{19}{c}@{}}
$s_{0000}$ & $\sortdef$ & $0$
	& $\mid$ & $s_{0000}+s_{0000}$ & $\mid$ & $s_{0011}+s_{0011}$
	& $\mid$ & $s_{0101}+s_{0101}$ & $\mid$ & $s_{0110}+s_{0110}$ \\
      &&& $\mid$ & $s_{1001}+s_{1001}$ & $\mid$ & $s_{1010}+s_{1010}$
	& $\mid$ & $s_{1100}+s_{1100}$ & $\mid$ & $s_{1111}+s_{1111}$ \\
$s_{0011}$ & $\sortdef$ & $v_{0011}$
	& $\mid$ & $s_{0000}+s_{0011}$ & $\mid$ & $s_{0011}+s_{0000}$
	& $\mid$ & $s_{0101}+s_{0110}$ & $\mid$ & $s_{0110}+s_{0101}$ \\
      &&& $\mid$ & $s_{1001}+s_{1010}$ & $\mid$ & $s_{1010}+s_{1001}$
	& $\mid$ & $s_{1100}+s_{1111}$ & $\mid$ & $s_{1111}+s_{1100}$ \\
$s_{0101}$ & $\sortdef$ & $v_{0101}$
	& $\mid$ & $s_{0000}+s_{0101}$ & $\mid$ & $s_{0011}+s_{0110}$
	& $\mid$ & $s_{0101}+s_{0000}$ & $\mid$ & $s_{0110}+s_{0011}$ \\
      &&& $\mid$ & $s_{1001}+s_{1100}$ & $\mid$ & $s_{1010}+s_{1111}$
	& $\mid$ & $s_{1100}+s_{1001}$ & $\mid$ & $s_{1111}+s_{1010}$ \\
$s_{0110}$ & $\sortdef$ &
	&        & $s_{0000}+s_{0110}$ & $\mid$ & $s_{0011}+s_{0101}$
	& $\mid$ & $s_{0101}+s_{0011}$ & $\mid$ & $s_{0110}+s_{0000}$ \\
      &&& $\mid$ & $s_{1001}+s_{1111}$ & $\mid$ & $s_{1010}+s_{1100}$
	& $\mid$ & $s_{1100}+s_{1010}$ & $\mid$ & $s_{1111}+s_{1001}$ \\
$s_{1001}$ & $\sortdef$ &
	&        & $s_{0000}+s_{1001}$ & $\mid$ & $s_{0011}+s_{1010}$
	& $\mid$ & $s_{0101}+s_{1100}$ & $\mid$ & $s_{0110}+s_{1111}$ \\
      &&& $\mid$ & $s_{1001}+s_{0000}$ & $\mid$ & $s_{1010}+s_{0011}$
	& $\mid$ & $s_{1100}+s_{0101}$ & $\mid$ & $s_{1111}+s_{0110}$ \\
$s_{1010}$ & $\sortdef$ &
	&        & $s_{0000}+s_{1010}$ & $\mid$ & $s_{0011}+s_{1001}$
	& $\mid$ & $s_{0101}+s_{1111}$ & $\mid$ & $s_{0110}+s_{1100}$ \\
      &&& $\mid$ & $s_{1001}+s_{0011}$ & $\mid$ & $s_{1010}+s_{0000}$
	& $\mid$ & $s_{1100}+s_{0110}$ & $\mid$ & $s_{1111}+s_{0101}$ \\
$s_{1100}$ & $\sortdef$ &
	&        & $s_{0000}+s_{1100}$ & $\mid$ & $s_{0011}+s_{1111}$
	& $\mid$ & $s_{0101}+s_{1001}$ & $\mid$ & $s_{0110}+s_{1010}$ \\
      &&& $\mid$ & $s_{1001}+s_{0101}$ & $\mid$ & $s_{1010}+s_{0110}$
	& $\mid$ & $s_{1100}+s_{0000}$ & $\mid$ & $s_{1111}+s_{0011}$ \\
$s_{1111}$ & $\sortdef$ & $1$
	& $\mid$ & $s_{0000}+s_{1111}$ & $\mid$ & $s_{0011}+s_{1100}$
	& $\mid$ & $s_{0101}+s_{1010}$ & $\mid$ & $s_{0110}+s_{1001}$ \\
      &&& $\mid$ & $s_{1001}+s_{0110}$ & $\mid$ & $s_{1010}+s_{0101}$
	& $\mid$ & $s_{1100}+s_{0011}$ & $\mid$ & $s_{1111}+s_{0000}$ \\
$s_{valid}$ & $\sortdef$ &
	&        & $s_{0000}:s_{0000}$ & $\mid$ & $s_{0011}:s_{0011}$
	& $\mid$ & $s_{0101}:s_{0101}$ & $\mid$ & $s_{0110}:s_{0110}$ \\
      &&& $\mid$ & $s_{1001}:s_{1001}$ & $\mid$ & $s_{1010}:s_{1010}$
	& $\mid$ & $s_{1100}:s_{1100}$ & $\mid$ & $s_{1111}:s_{1111}$ \\
\end{tabular}
\end{center}
\caption{Sort Definitions in Exm.\ \protect\eqr{y0.6}}
\label{Sort Definition in Exm. \eqR{y0.6}}
\end{figure}

\begin{figure}
\begin{center}
\begin{tabular}[t]{@{}cr*{15}{c}@{}}
$s_{ffff}$ & $\sortdef$
	& $f$
	& $\mid$ & $s_{ffff} \vee s_{ffff}$
	& $\mid$ & $s_{fttt} \wedge s_{ffff}$
	& $\mid$ & $s_{tttt} \wedge s_{ffff}$
	& $\mid$ & $s_{ffff} \wedge s_{fttt}$
	& $\mid$ & $s_{ffff} \wedge s_{tttt}$
	& $\mid$ & $s_{ffff} \wedge s_{ffff}$	\\
	&&& $\mid$ & $s_{ffff} \wedge s_{ffft}$
	& $\mid$ & $s_{ffff} \wedge s_{fftt}$
	& $\mid$ & $s_{ffff} \wedge s_{ftft}$
	& $\mid$ & $s_{ffft} \wedge s_{ffff}$
	& $\mid$ & $s_{fftt} \wedge s_{ffff}$
	& $\mid$ & $s_{ftft} \wedge s_{ffff}$	\\
$s_{ffft}$ & $\sortdef$
	&&& $s_{ffff} \vee s_{ffft}$
	& $\mid$ & $s_{ffft} \vee s_{ffff}$
	& $\mid$ & $s_{ffft} \vee s_{ffft}$
	& $\mid$ & $s_{fttt} \wedge s_{ffft}$
	& $\mid$ & $s_{tttt} \wedge s_{ffft}$
	& $\mid$ & $s_{ffft} \wedge s_{fttt}$
	& $\mid$ & $s_{ffft} \wedge s_{tttt}$	\\
	&&& $\mid$ & $s_{ffft} \wedge s_{ffft}$
	& $\mid$ & $s_{ffft} \wedge s_{fftt}$
	& $\mid$ & $s_{ffft} \wedge s_{ftft}$
	& $\mid$ & $s_{fftt} \wedge s_{ffft}$
	& $\mid$ & $s_{fftt} \wedge s_{ftft}$
	& $\mid$ & $s_{ftft} \wedge s_{ffft}$
	& $\mid$ & $s_{ftft} \wedge s_{fftt}$	\\
$s_{fftt}$ & $\sortdef$
	& $v_{fftt}$
	& $\mid$ & $s_{ffff} \vee s_{fftt}$
	& $\mid$ & $s_{ffft} \vee s_{fftt}$
	& $\mid$ & $s_{fftt} \vee s_{ffff}$
	& $\mid$ & $s_{fftt} \vee s_{ffft}$
	& $\mid$ & $s_{fftt} \vee s_{fftt}$	\\
	&&& $\mid$ & $s_{fttt} \wedge s_{fftt}$
	& $\mid$ & $s_{tttt} \wedge s_{fftt}$
	& $\mid$ & $s_{fftt} \wedge s_{fttt}$
	& $\mid$ & $s_{fftt} \wedge s_{tttt}$
	& $\mid$ & $s_{fftt} \wedge s_{fftt}$	\\
$s_{ftft}$ & $\sortdef$
	& $v_{ftft}$
	& $\mid$ & $s_{ffff} \vee s_{ftft}$
	& $\mid$ & $s_{ffft} \vee s_{ftft}$
	& $\mid$ & $s_{ftft} \vee s_{ffff}$
	& $\mid$ & $s_{ftft} \vee s_{ffft}$
	& $\mid$ & $s_{ftft} \vee s_{ftft}$	\\
	&&& $\mid$ & $s_{fttt} \wedge s_{ftft}$
	& $\mid$ & $s_{tttt} \wedge s_{ftft}$
	& $\mid$ & $s_{ftft} \wedge s_{fttt}$
	& $\mid$ & $s_{ftft} \wedge s_{tttt}$
	& $\mid$ & $s_{ftft} \wedge s_{ftft}$	\\
$s_{fttt}$ & $\sortdef$
	&&& $s_{fttt} \vee s_{fttt}$
	& $\mid$ & $s_{fttt} \vee s_{ffff}$
	& $\mid$ & $s_{fttt} \vee s_{ffft}$
	& $\mid$ & $s_{fttt} \vee s_{fftt}$
	& $\mid$ & $s_{fttt} \vee s_{ftft}$
	& $\mid$ & $s_{ffff} \vee s_{fttt}$
	& $\mid$ & $s_{ffft} \vee s_{fttt}$	\\
	&&& $\mid$ & $s_{fftt} \vee s_{fttt}$
	& $\mid$ & $s_{fftt} \vee s_{ftft}$
	& $\mid$ & $s_{ftft} \vee s_{fttt}$
	& $\mid$ & $s_{ftft} \vee s_{fftt}$
	& $\mid$ & $s_{fttt} \wedge s_{fttt}$
	& $\mid$ & $s_{fttt} \wedge s_{tttt}$
	& $\mid$ & $s_{tttt} \wedge s_{fttt}$	\\
$s_{tttt}$ & $\sortdef$
	& $t$
	& $\mid$ & $s_{fttt} \vee s_{tttt}$
	& $\mid$ & $s_{tttt} \vee s_{fttt}$
	& $\mid$ & $s_{tttt} \vee s_{tttt}$
	& $\mid$ & $s_{tttt} \vee s_{ffff}$
	& $\mid$ & $s_{tttt} \vee s_{ffft}$
	& $\mid$ & $s_{tttt} \vee s_{fftt}$	\\
	&&& $\mid$ & $s_{tttt} \vee s_{ftft}$
	& $\mid$ & $s_{ffff} \vee s_{tttt}$
	& $\mid$ & $s_{ffft} \vee s_{tttt}$
	& $\mid$ & $s_{fftt} \vee s_{tttt}$
	& $\mid$ & $s_{ftft} \vee s_{tttt}$
	& $\mid$ & $s_{tttt} \wedge s_{tttt}$	\\
\end{tabular}
\end{center}
\caption{Sort Definitions in Exm.\ \protect\eqr{y5}}
\label{Sort Definitions in Exm. \eqR{y5}}
\end{figure}

\begin{figure}
\begin{center}
\begin{tabular}[t]{@{}rcl@{}}
$f \wedge x$ & $:$ & $f$	\\
%$x \wedge f$ & $:$ & $f$	\\
%$x \wedge (x \wedge y)$ & $:$ & $x \wedge y$	\\
$y \wedge x$ & $:$ & $x \wedge y$	\\
%$y \wedge (x \wedge y)$ & $:$ & $x \wedge y$	\\
%$(x \vee y) \wedge (x \wedge y)$ & $:$ & $x \wedge y$	\\
%$x \wedge y \wedge x$ & $:$ & $x \wedge y$	\\
$x \wedge y \wedge y$ & $:$ & $x \wedge y$	\\
$x \wedge y \wedge (x \vee y)$ & $:$ & $x \wedge y$	\\
%$x$ & $:$ & $x$	\\
$f \vee x$ & $:$ & $x$	\\
%$x \vee f$ & $:$ & $x$ \\
$x \vee x$ & $:$ & $x$	\\
$x \vee (x \wedge y)$ & $:$ & $x$ \\
%$x \wedge y \vee x$ & $:$ & $x$	\\
$t \wedge x$ & $:$ & $x$	\\
%$x \wedge t$ & $:$ & $x$	\\
$x \wedge x$ & $:$ & $x$	\\
$x \wedge (x \vee y)$ & $:$ & $x$	\\
%$(x \vee y) \wedge x$ & $:$ & $x$	\\
%$y \vee x \wedge y$ & $:$ & $y$	\\
%$x \wedge y \vee y$ & $:$ & $y$	\\
%$y \wedge (x \vee y)$ & $:$ & $y$	\\
%$(x \vee y) \wedge y$ & $:$ & $y$	\\
%$x \vee (x \vee y)$ & $:$ & $x \vee y$ \\
$y \vee x$ & $:$ & $x \vee y$	\\
%$y \vee (x \vee y)$ & $:$ & $x \vee y$ \\
%$x \vee y \vee x$ & $:$ & $x \vee y$	\\
$x \vee y \vee y$ & $:$ & $x \vee y$	\\
$x \vee y \vee (x \wedge y)$ & $:$ & $x \vee y$	\\
%$x \wedge y \vee (x \vee y)$ & $:$ & $x \vee y$ \\
$t \vee x$ & $:$ & $t$	\\
%$x \vee t$ & $:$ & $t$ \\
\end{tabular}
\end{center}
\caption{Equational Theory of $Bool$ with $\wedge$ and $\vee$}
\label{Equational Theory of Bool with wedge and vee}
\end{figure}

\begin{figure}
\begin{tabular}[t]{@{}rcl@{}}
$x\1+x$ & $:$ & $0$	\\
$x$ & $:$ & $x$ \\
$0\1+x$ & $:$ & $x$	\\
$x\1+0$ & $:$ & $x$	\\
$y\1+(x\1+y)$ & $:$ & $x$	\\
$x\1+y\1+y$ & $:$ & $x$	\\
$y\1+z\1+1\1+(y\1+z\1+(x\1+1))$ & $:$ & $x$ \\
$y\1+z\1+(x\1+1)\1+(y\1+z\1+1)$ & $:$ & $x$ \\
$x\1+(x\1+y)$ & $:$ & $y$	\\
$x\1+1\1+(x\1+(1\1+y))$ & $:$ & $y$	\\
$x\1+y\1+x$ & $:$ & $y$	\\
$x\1+z\1+(y\1+z\1+x)$ & $:$ & $y$	\\
$x\1+(1\1+y)\1+(x\1+1)$ & $:$ & $y$	\\
$x\1+(1\1+z)\1+(y\1+z\1+(x\1+1))$ & $:$ & $y$ \\
$y\1+z\1+x\1+(x\1+z)$ & $:$ & $y$	\\
$y\1+z\1+(x\1+1)\1+(x\1+(1\1+z))$ & $:$ & $y$ \\
$1\1+(x\1+(1\1+y))$ & $:$ & $x\1+y$	\\
$y\1+x$ & $:$ & $x\1+y$	\\
$z\1+(y\1+z\1+x)$ & $:$ & $x\1+y$	\\
$1\1+z\1+(y\1+z\1+(x\1+1))$ & $:$ & $x\1+y$ \\
$x\1+z\1+(y\1+z)$ & $:$ & $x\1+y$	\\
$x\1+(1\1+y)\1+1$ & $:$ & $x\1+y$	\\
$x\1+(1\1+z)\1+(y\1+z\1+1)$ & $:$ & $x\1+y$ \\
$y\1+z\1+(x\1+z)$ & $:$ & $x\1+y$	\\
$y\1+z\1+1\1+(x\1+(1\1+z))$ & $:$ & $x\1+y$ \\
$y\1+z\1+x\1+z$ & $:$ & $x\1+y$ \\
$y\1+z\1+(x\1+1)\1+(1\1+z)$ & $:$ & $x\1+y$ \\
$x\1+y\1+(y\1+z\1+x)$ & $:$ & $z$	\\
$x\1+(1\1+y)\1+(y\1+z\1+(x\1+1))$ & $:$ & $z$ \\
$y\1+z\1+x\1+(x\1+y)$ & $:$ & $z$	\\
$y\1+z\1+(x\1+1)\1+(x\1+(1\1+y))$ & $:$ & $z$ \\
$y\1+(y\1+z\1+x)$ & $:$ & $x\1+z$	\\
$1\1+y\1+(y\1+z\1+(x\1+1))$ & $:$ & $x\1+z$ \\
$x\1+y\1+(y\1+z)$ & $:$ & $x\1+z$	\\
$x\1+(1\1+y)\1+(y\1+z\1+1)$ & $:$ & $x\1+z$ \\
\end{tabular}
\hspace*{-0.13cm}
\begin{tabular}[t]{@{}rcl@{}}
$y\1+z\1+(x\1+y)$ & $:$ & $x\1+z$	\\
$y\1+z\1+1\1+(x\1+(1\1+y))$ & $:$ & $x\1+z$ \\
$y\1+z\1+x\1+y$ & $:$ & $x\1+z$ \\
$y\1+z\1+(x\1+1)\1+(1\1+y)$ & $:$ & $x\1+z$ \\
$x\1+y\1+(x\1+z)$ & $:$ & $y\1+z$	\\
$x\1+z\1+(x\1+y)$ & $:$ & $y\1+z$	\\
$x\1+(1\1+y)\1+(x\1+(1\1+z))$ & $:$ & $y\1+z$	\\
$x\1+(1\1+z)\1+(x\1+(1\1+y))$ & $:$ & $y\1+z$	\\
$1\1+(y\1+z\1+(x\1+1))$ & $:$ & $y\1+z\1+x$ \\
$y\1+(x\1+z)$ & $:$ & $y\1+z\1+x$	\\
$z\1+(x\1+y)$ & $:$ & $y\1+z\1+x$	\\
$1\1+y\1+(x\1+(1\1+z))$ & $:$ & $y\1+z\1+x$ \\
$1\1+z\1+(x\1+(1\1+y))$ & $:$ & $y\1+z\1+x$ \\
$x\1+y\1+z$ & $:$ & $y\1+z\1+x$ \\
$x\1+z\1+y$ & $:$ & $y\1+z\1+x$ \\
$x\1+(1\1+y)\1+(1\1+z)$ & $:$ & $y\1+z\1+x$ \\
$x\1+(1\1+z)\1+(1\1+y)$ & $:$ & $y\1+z\1+x$ \\
$y\1+z\1+(x\1+1)\1+1$ & $:$ & $y\1+z\1+x$	\\
$1\1+(y\1+z\1+x)$ & $:$ & $y\1+z\1+(x\1+1)$ \\
$x\1+(y\1+z\1+1)$ & $:$ & $y\1+z\1+(x\1+1)$ \\
$y\1+(x\1+(1\1+z))$ & $:$ & $y\1+z\1+(x\1+1)$	\\
$z\1+(x\1+(1\1+y))$ & $:$ & $y\1+z\1+(x\1+1)$	\\
$1\1+y\1+(x\1+z)$ & $:$ & $y\1+z\1+(x\1+1)$ \\
$1\1+z\1+(x\1+y)$ & $:$ & $y\1+z\1+(x\1+1)$ \\
$x\1+y\1+(1\1+z)$ & $:$ & $y\1+z\1+(x\1+1)$ \\
$x\1+z\1+(1\1+y)$ & $:$ & $y\1+z\1+(x\1+1)$ \\
$x\1+(1\1+y)\1+z$ & $:$ & $y\1+z\1+(x\1+1)$ \\
$x\1+(1\1+z)\1+y$ & $:$ & $y\1+z\1+(x\1+1)$ \\
$y\1+z\1+1\1+x$ & $:$ & $y\1+z\1+(x\1+1)$	\\
$y\1+z\1+x\1+1$ & $:$ & $y\1+z\1+(x\1+1)$	\\
$x\1+(y\1+z\1+(x\1+1))$ & $:$ & $y\1+z\1+1$ \\
$x\1+y\1+(x\1+(1\1+z))$ & $:$ & $y\1+z\1+1$ \\
$x\1+z\1+(x\1+(1\1+y))$ & $:$ & $y\1+z\1+1$ \\
$x\1+(1\1+y)\1+(x\1+z)$ & $:$ & $y\1+z\1+1$ \\
$x\1+(1\1+z)\1+(x\1+y)$ & $:$ & $y\1+z\1+1$ \\
\end{tabular}
\hspace*{-0.13cm}
\begin{tabular}[t]{@{}rcl@{}}
$y\1+z\1+(x\1+1)\1+x$ & $:$ & $y\1+z\1+1$	\\
$1\1+(x\1+z)$ & $:$ & $x\1+(1\1+z)$	\\
$y\1+(y\1+z\1+(x\1+1))$ & $:$ & $x\1+(1\1+z)$	\\
$z\1+(x\1+1)$ & $:$ & $x\1+(1\1+z)$	\\
$1\1+y\1+(y\1+z\1+x)$ & $:$ & $x\1+(1\1+z)$ \\
$x\1+1\1+z$ & $:$ & $x\1+(1\1+z)$	\\
$x\1+y\1+(y\1+z\1+1)$ & $:$ & $x\1+(1\1+z)$ \\
$x\1+z\1+1$ & $:$ & $x\1+(1\1+z)$	\\
$x\1+(1\1+y)\1+(y\1+z)$ & $:$ & $x\1+(1\1+z)$	\\
$y\1+z\1+(x\1+(1\1+y))$ & $:$ & $x\1+(1\1+z)$	\\
$y\1+z\1+1\1+(x\1+y)$ & $:$ & $x\1+(1\1+z)$ \\
$y\1+z\1+x\1+(1\1+y)$ & $:$ & $x\1+(1\1+z)$ \\
$y\1+z\1+(x\1+1)\1+y$ & $:$ & $x\1+(1\1+z)$ \\
$x\1+y\1+(y\1+z\1+(x\1+1))$ & $:$ & $1\1+z$ \\
$x\1+(1\1+y)\1+(y\1+z\1+x)$ & $:$ & $1\1+z$ \\
$y\1+z\1+x\1+(x\1+(1\1+y))$ & $:$ & $1\1+z$ \\
$y\1+z\1+(x\1+1)\1+(x\1+y)$ & $:$ & $1\1+z$ \\
$z\1+(y\1+z\1+(x\1+1))$ & $:$ & $x\1+(1\1+y)$	\\
$1\1+z\1+(y\1+z\1+x)$ & $:$ & $x\1+(1\1+y)$ \\
$x\1+z\1+(y\1+z\1+1)$ & $:$ & $x\1+(1\1+y)$ \\
$x\1+(1\1+z)\1+(y\1+z)$ & $:$ & $x\1+(1\1+y)$	\\
$y\1+z\1+(x\1+(1\1+z))$ & $:$ & $x\1+(1\1+y)$	\\
$y\1+z\1+1\1+(x\1+z)$ & $:$ & $x\1+(1\1+y)$ \\
$y\1+z\1+x\1+(1\1+z)$ & $:$ & $x\1+(1\1+y)$ \\
$y\1+z\1+(x\1+1)\1+z$ & $:$ & $x\1+(1\1+y)$ \\
$x\1+z\1+(y\1+z\1+(x\1+1))$ & $:$ & $1\1+y$ \\
$x\1+(1\1+z)\1+(y\1+z\1+x)$ & $:$ & $1\1+y$ \\
$y\1+z\1+x\1+(x\1+(1\1+z))$ & $:$ & $1\1+y$ \\
$y\1+z\1+(x\1+1)\1+(x\1+z)$ & $:$ & $1\1+y$ \\
$y\1+(x\1+(1\1+y))$ & $:$ & $x\1+1$	\\
$x\1+(1\1+y)\1+y$ & $:$ & $x\1+1$	\\
$x\1+y\1+(x\1+(1\1+y))$ & $:$ & $1$	\\
$x\1+(1\1+y)\1+(x\1+y)$ & $:$ & $1$	\\
$y\1+z\1+x\1+(y\1+z\1+(x\1+1))$ & $:$ & $1$ \\
$y\1+z\1+(x\1+1)\1+(y\1+z\1+x)$ & $:$ & $1$ \\
\end{tabular}
\caption{Equational Theory of $(\N \;mod\; 2)$ with $+$}
\label{Equational Theory of N mod 2 with +}
\end{figure}

\COROLLARY{ \eqd{y3}
If $\T /_E$ is finite,
for each finite $V_1 \subset V_2 \subset \V$ and
$V_3 \subset V_4 \subset \V$,
the set of all universally valid equations $t_1:t_2$
such that $V_1 \subset vars(t_1) \subset V_2$
and $V_3 \subset vars(t_2) \subset V_4$
is a regular tree language.
}
\PROOF{
Apply Thm.~\eqr{y2} to $V := V_2 \cup V_4$;
using the variable filter sorts $s_{V_1}^{V_2}$, $s_{V_3}^{V_4}$
from Sect.~\ref{Filter Sorts},
define
$s_{result} \sortdef
\bigmid_{\eqc{\vec{a}} \in A} \;
(rsg_V(V,\eqc{\vec{a}}) \cap s_{V_1}^{V_2})
:(rsg_V(V,\eqc{\vec{a}}) \cap s_{V_3}^{V_4})$.
}

\COROLLARY{ \eqd{y4}
If $\T /_E$ is finite and $n \in \N$ arbitrary,
the set $\jTH_n(\T/_E)$
of universally valid equations in $n$ variables can be
represented as the deductive closure of finitely many equations.
}
\PROOF{
Using Thm.~\eqr{y2},
$\jTH_n(\T/_E) = \L(s_{valid})$
where
$s_{valid} \sortdef \bigmid_{\eqc{\vec{a}} \in A}
\; rsg_V(V,\eqc{\vec{a}}):rsg_V(V,\eqc{\vec{a}})$,
and $V = \{x_1,\ldots,x_n\}$.
For each $\eqc{\vec{a}} \in A$,
choose some arbitrary normal form
$nf_{\vec{a}} \in \L(rsg_V(V,\eqc{\vec{a}}))$.

By Cor.~\eqr{7.5} and Lemma \eqr{6},
the sort definition of each $\eqc{a_i}$
is in head normal form and refers only to sort names corresponding to
some $\eqc{a_j}$.
Hence, by construction of $rsg_V$,
the sort definition of each $rsg_V(V,\eqc{\vec{a}})$
is in head normal form and refers only to sort names corresponding to
some $rsg_V(V,\eqc{\vec{a}'})$.

For $\eqc{\vec{a}} \in A$, let
$rsg_V(V,\eqc{\vec{a}}) \sortdef
\bigmid_{i=1}^m \;
f_i(rsg_V(V,\eqc{\vec{a}_{i1}}),\ldots,rsg_V(V,\eqc{\vec{a}_{in_i}}))$
for some $f_i$, $\eqc{\vec{a}_{ij}}$.
\\
Define
$TH_\eqc{\vec{a}} := \{
f_i(nf_{\vec{a}_{i1}},\ldots,nf_{\vec{a}_{in_i}}):nf_{\vec{a}}
\mid i=1,\ldots,m \}$.
Define $TH := \bigcup_{\eqc{\vec{a}} \in A} TH_\eqc{\vec{a}}$.

Let $t \in \L(rsg_V(V,\eqc{\vec{a}}))$ for some $\eqc{\vec{a}} \in A$;
we show by induction on $t$ that the equation $t:nf_{\vec{a}}$
is a deductive consequence of $TH$.
Consider the alternative in the sort definition of
$rsg_V(V,\eqc{\vec{a}})$ that leads to $t$:
\bi
\item If $t = f_i(t_1,\ldots,t_{n_i})$,
	where $t_j \in \L(rsg_V(V,\eqc{\vec{a}_{ij}}))$,
	then by induction hypothesis
	the equations $t_j:nf_{\vec{a}_{ij}}$
	are deductive consequences of $TH$.
	Since
	$(f_i(nf_{\vec{a}_{i1}},\ldots,nf_{\vec{a}_{in_i}})
	:nf_{\vec{a}})
	\in TH$,
	we have finished.
\item If $t = f_i$ is a constant or a variable,
	we immediately have $(f_i:nf_{\vec{a}}) \in TH$.
\ei

If we now take an arbitrary equation $(t_1:t_2) \in \L(s_{valid})$,
then $t_1,t_2 \in \L(rsg_V(V,\eqc{\vec{a}}))$
for some $\eqc{\vec{a}} \in A$;
the equation is a consequence of $t_1:nf_{\vec{a}}$
and $t_2:nf_{\vec{a}}$.
}

\begin{remark} \eqd{y4.5}
Note that no proper instances of the equations in $TH$ are needed to
derive any equation $t_1:t_2$ in $\L(s_{valid})$.
Hence, if we consider all variables in $V$ as constants
and choose $nf_{\vec{a}}$ to be of minimal size
within $\L(rsg_V(V,\eqc{\vec{a}}))$,
we obtain a noetherian ground--rewriting system
for the set of all universally valid equations in variables from $V$.
Moreover, this rewriting system assigns a unique normal form to each
term $t$ that may occur in a universally valid equation,
viz.\ $nf_{\vec{a}}$ if $t \in \L(rsg_V(V,\eqc{\vec{a}}))$.

Of course, when permitting proper instantiations, we lose these
properties, since the commutativity law, for example,
could be among the
universally valid equations.
On the other hand, we may delete equations that are instances of
others, thus reducing the number of equations significantly.
To find such subsumed equations, an appropriate indexing technique may
be used, see e.g.\ \cite{Graf.1992,Graf.Meyer.1993,Graf.1994}.
\end{remark}

\EXAMPLE{ \eqd{y5}
Consider $Bool = \{f,t\}$ with the operations $\wedge$ and $\vee$:

\hspace*{\fill}
\begin{tabular}[t]{@{}c|cc@{}}
$\wedge$ & $f$ & $t$	\\
\hline
$f$ & $f$ & $f$	\\
$\;t\;$ & $\;f\;$ & $\;t\;$	\\
\end{tabular}
\hspace*{\fill}
\begin{tabular}[t]{@{}c|cc@{}}
$\vee$ & $f$ & $t$	\\
\hline
$f$ & $f$ & $t$	\\
$\;t\;$ & $\;t\;$ & $\;t\;$	\\
\end{tabular}
\hspace*{\fill}

Computing $s_{valid}$ for $n=2$ variables, we obtain the six
non--empty sorts shown in Fig.~\ref{Sort Definitions in Exm.
\eqR{y5}}. Proceeding as described in the proof of
Cor.~\eqr{y4}, we obtain $76$ equations in $TH$; $43$ of them
are instances of others and can therefore be deleted. Of the
remaining $33$, a further $17$ have been manually deleted since
they were subsumed modulo commutativity. 
The remaining equations are given in Fig.~\ref{Equational Theory
of Bool with wedge and vee}. 
}

\EXAMPLE{ \eqd{y5a}
Figure~\ref{Equational Theory of N mod 2 with +}
shows an axiom system for
the equational theory of $(\N \;mod\; 2)$ with $+$ in 3
variables. The 105 equations have
been extracted from a system of 256 sort definitions 240 of which where
empty. Only equations that were syntactically subsumed have been
removed (automatically).
}

\begin{remark}
In order to estimate the complexity of computing an
axiomatization of a finite algebra $\T/E$,
observe the following facts.
Using an appropriate caching mechanism when computing 
	$rsg_V(V,\eqc{\vec{a}})$ for all $\eqc{\vec{a}} \in A_T$,
	we can ensure that $rsg_V$ is called exactly once for each such
	$\eqc{\vec{a}}$.
If all sort definitions are in head normal form,
	the cost of one $rsg_V$ call without its recursive subcalls is
	precisely the cost of its grouping algorithm.
In Sect.~\ref{Optimized Argument Selection},
	we estimated the latter to be
	$\O(N' \cdot m' \cdot (m'+n') \cdot g'^{N'})$,
	where $N'$ denotes the number of sorts to be anti--unified
	simultaneously,
	$m'$ is the maximal number of alternatives of a
	sort definition's right--hand side,
	$n'$ is the number of variables in $V$, and
	$g'$ is the maximum number of disjuncts of a sort--definition's
	right--hand side that start with the same function symbol.
We have $N' = N^n$ and $n' = n$,
	leading to an overall complexity of
	$\O((N \cdot g')^{(N^n)} \cdot N^n \cdot m' \cdot (m'+n))$ 
	for computing
	all $N^{(N^n)}$ $rsg_V$ calls.
Transformation of
	the sort definitions into an axiom system can be done in
	linear time.
\end{remark}

\subsection{Typed Equational Theories in $n$ Variables}
\label{Typed Equational Theories in $n$ Variables}

Theorem \eqr{y2} and Corollaries \eqr{y3} and \eqr{y4}
can be generalized to
typed algebras where different variables may
have different --~disjoint~-- domains.
We will do this in this section.
The definitions given below generalize Def.~\eqr{y0.5} to $n$
typed variables.

\DEFINITION{ \eqd{z0.4}
$\;$	\\
Let $\TY$ be a finite set of types.
Assume each $x \in \V$ has a fixed type $type(x) \in \TY$
such that for each $T \in \TY$ there are infinitely many $x \in \V$
with $type(x) = T$.
Assume each $f \in \F$ has a fixed signature
$f: T_1 \times \ldots \times T_n \ra T$
where $T_1,\ldots,T_n,T \in \TY$.

For $T \in \TY$, let $\T_T$ be the set of well--typed terms of type $T$,
which is defined as usual.
Let $\T := \bigcup_{T \in \TY} \T_T$;
for $t \in \T$ define $type(t) := T$ if $t \in \T_T$;
$type(t)$ is uniquely determined since any $f$ has only one signature.
A substitution $\{x_1 \la t_1,\ldots,x_n \la t_n\}$ is called
well--typed if $type(x_i) = type(t_i)$ for $i=1,\ldots,n$.
Consequently, we always have $type(\sigma t) = type(t)$
for well--typed $\sigma$, $t$.

We call a term of the form $t_1:t_2$ a typed (formal) equation
if $t_1,t_2 \in \T_T$ for some $T \in \TY$;
in this case we define $type(t_1:t_2) := T$.
An equation $t_1:t_2$ is called universally valid
if $\sigma t_1 =_E \sigma t_2$ for every well--typed
substitution $\sigma$.
For $x_1,\ldots,x_n \in \V$ and $T \in \TY$,
define $\jTH^T_{x_1,\ldots,x_n}(\T /_E)$ as the set of all typed formal
equations of type $T$ with
variables $x_1,\ldots,x_n$ that are universally valid over $\T /_E$.
Define 
$\jTH_{x_1,\ldots,x_n}(\T /_E) 
:= \bigcup_{T \in \TY} \;\; \jTH^T_{x_1,\ldots,x_n}(\T /_E)$
as the set of all possible typed formal equations with
variables $x_1,\ldots,x_n$
that are universally valid over $\T /_E$.
Note that not all $x_i$ must occur in such an equation,
but no other variables may occur.

We still assume that $\T/_E$ is finite;
hence $\T_T/_E$ is finite for each $T \in \TY$.
We assume in this section that we are given $n$ fixed variables
$x_1,\ldots,x_n$ of types $T_1,\ldots,T_n$, respectively,
such that $\T_{T_i}/_E$ is of cardinality $N_i$.
Let $N := N_1 \cdot \ldots \cdot N_n$.
}

\DEFINITION{ \eqd{z0.5}
Modifying Defs.~\eqr{x0.5} and \eqr{y0.5}, define

$$B :=
\left\{
\left(
\begin{array}{ccc}
b_{1,1} & \ldots & b_{1,N}	\\
\vdots && \vdots	\\
b_{n,1} & \ldots & b_{n,N}	\\
\end{array}
\right)
\raisebox{-4.5ex}{\rule{0.03cm}{10ex}} \;\;
\forall t_1 \in \T_{T_1},\ldots,t_n \in \T_{T_n} \;\;
\exists j \in \{1,\ldots,N\} \;\;
\forall i \in \{1,\ldots,n\} \;\;
b_{i,j} =_E t_i
\right\}$$

$B$ is non--empty since
$\T_{T_1} /_E \times \ldots \times \T_{T_n}/_E$ is finite, viz.\ of
cardinality $N$.
Intuitively, for each matrix $\vec{b} \in B$
there is a one--to--one correspondence between its
column vectors and the elements of
$\T_{T_1} /_E \times \ldots \times \T_{T_n}/_E$.

In this section,
let $x_1 = \phi(b_{1,1},\ldots,b_{1,N})$, \ldots,
$x_n = \phi(b_{n,1},\ldots,b_{n,N})$
for some $\vec{b} \in B$,
and let $V := \{ x_1,\ldots,x_n \}$.
Let
$A_T := \{ \eqc{\vec{a}} \in (\T_T /_E)^N \mid
\L(rsg_V(V,\eqc{\vec{a}})) \neq \{\} \}$
for $T \in \TY$.
}

\LEMMA{ \eqd{z1}
Let $t_1,t_2 \in \T_T$ for some $T \in \TY$
with $vars(t_1) \cup vars(t_2) \subset V$.
\\
The equation $t_1 : t_2$ is universally valid
iff there exists an $\vec{a} \in A_T$ such that
$t_1,t_2 \in \L(rsg_V(V,\eqc{\vec{a}}))$.
}
\PROOF{
The proof is similar to that of Lemma \eqr{y1}, with the following
observations:

``$\Ra$'':
The $\sigma_i$ from Lemma \eqr{vrsg1}
in Sect.~\ref{Variable--Restricted E--Anti--Unification}
are well--typed, since $type(x_j) = T_j = type(b_{j,i})$.
\\
Moreover, $\eqc{\vec{a}} \in A_T$,
since $type(a_i) = type(\sigma_i t_1) = type(t_1) = T$ for all $i$.

``$\La$'':
It is sufficient to
consider an arbitrary well--typed substitution
$\{x_1 \la t'_1,\ldots,x_n \la t'_n\}$
with $type(t'_i) = type(x_i) = T_i$.
}

\THEOREM{ \eqd{z2}
If $\T /_E$ is finite,
the set $\jTH^T_{x_1,\ldots,x_n}(\T/_E)$
of all universally valid equations between terms of type $T$
with variables $x_1,\ldots,x_n$
of types $T_1,\ldots,T_n \in \TY$ is a regular tree language.
}
\PROOF{
Let $\phi$ be
such that $\phi(b_{1,1},\ldots,b_{1,N}) = x_1$, \ldots,
$\phi(b_{n,1},\ldots,b_{n,N}) = x_n$ for some $\vec{b} \in B$;
let $V := \{ x_1,\ldots,x_n \}$.
By Cor.~\eqr{7.5}, $\eqc{a_i}$ is a regular tree language
for each $a \in \T_T$, $T \in \TY$.
Define
$s^T_{valid} \sortdef \bigmid_{\eqc{\vec{a}} \in A_T}
\; rsg_V(V,\eqc{\vec{a}}):rsg_V(V,\eqc{\vec{a}})$.
Let $t_1,t_2 \in \T_T$ with $vars(t_1) \cup vars(t_2) \subset V$.
Then,
\\
\begin{tabular}[t]{@{}ll@{\hspace*{0.5cm}}l@{}}
& $t_1 : t_2$ is universally valid \\
$\Lra$ & exists $\vec{a} \in A_T$
	such that $t_1,t_2 \in \L(rsg_V(V,\eqc{\vec{a}}))$
	& by Lemma \eqr{z1}	\\
$\Lra$ & exists $\eqc{\vec{a}} \in A_T$ such that
	$(t_1:t_2) \in
	\L(rsg_V(V,\eqc{\vec{a}}):rsg_V(V,\eqc{\vec{a}}))$ \\
$\Lra$ & $(t_1:t_2) \in \L(s^T_{valid})$	\\
\end{tabular}
}

\COROLLARY{ \eqd{z4}
If $\T /_E$ is finite and $n \in \N$ arbitrary,
the set $\jTH_{x_1,\ldots,x_n}(\T/_E)$
of universally valid equations
in variables $x_1,\ldots,x_n$ of types
$T_1,\ldots,T_n \in \TY$
can be represented as the deductive closure of finitely many equations.
}
\PROOF{
The proof is similar to that of Cor.~\eqr{y4}:

Using Thm.~\eqr{z2},
$\jTH^T_{x_1,\ldots,x_n}(\T/_E) = \L(s^T_{valid})$
where
$s^T_{valid} \sortdef \bigmid_{\eqc{\vec{a}} \in A_T}
\; rsg_V(V,\eqc{\vec{a}}):rsg_V(V,\eqc{\vec{a}})$.
Let $s_{valid} \sortdef \bigmid_{T \in \TY} \;\; s^T_{valid}$.
For each $\eqc{\vec{a}} \in A_T$
choose some arbitrary normal form
$nf_{\vec{a}} \in \L(rsg_V(V,\eqc{\vec{a}}))$.
For $\eqc{\vec{a}} \in A_T$, let
$rsg_V(V,\eqc{\vec{a}}) \sortdef
\bigmid_{i=1}^m \;
f_i(rsg_V(V,\eqc{\vec{a}_{i1}}),\ldots,rsg_V(V,\eqc{\vec{a}_{in_i}}))$
be in head normal form.
\\
Define
$TH_\eqc{\vec{a}} := \{
f_i(nf_{\vec{a}_{i1}},\ldots,nf_{\vec{a}_{in_i}}):nf_{\vec{a}}
\mid i=1,\ldots,m \}$.
Define $TH_T := \bigcup_{\eqc{\vec{a}} \in A_T} TH_\eqc{\vec{a}}$
and $TH := \bigcup_{T \in \TY} TH_T$.

By analogy with the proof of Thm.~\eqr{y4}, we can show that,
for arbitrary $t \in \L(rsg_V(V,\eqc{\vec{a}}))$,
$\eqc{\vec{a}} \in A_T$, and $T \in \TY$,
the equation $t:nf_{\vec{a}}$ is a deductive consequence of $TH$.
An arbitrary equation $(t_1:t_2) \in \L(s_{valid})$
follows from $t_1:nf_{\vec{a}}$ and $t_2:nf_{\vec{a}}$.
Note that an arbitrary equation $(t_1:t_2)$
in $\L(s^T_{valid})$ is generally not a
deductive consequence of $TH_T$ only, since rewritings on subterms,
e.g.\ of $t_1$, that are not of type $T$ may be necessary.
}

\COROLLARY{ \eqd{z5}
If $\T /_E$ is finite and $n \in \N$ arbitrary,
the set of universally valid quantifier--free
formulas in $n$ variables is a regular
tree language and can be represented as the
deductive closure of finitely
many axioms that may use the equality predicate.
The same holds if the set of admitted junctors is arbitrarily
restricted, as long as it contains logical equivalence ($\lra$).
}
\PROOF{
This follows immediately from Thm.~\eqr{z2} and Cor.~\eqr{z4}
by adding a type $Bool$ to $\TY$, coding each predicate as a
function into $Bool$, and coding
logical junctors as functions from $Bool$ to $Bool$.
Equations $(t_1:t_2)$ of type $Bool$ are read as logical
equivalences $(t_1 \lra t_2)$;
all other equations as equality axioms $(t_1=t_2)$.
}

\begin{figure}
\begin{tabular}[t]{@{}rcl@{}}
$1\1+x\1+(x\1+x\1+2)$ &$:$ &$ 0$	\\
$x\1+x\1+x$ &$:$ &$ 0$	\\
$0\1+x$ &$:$ &$ x$	\\
$1\1+x\1+2$ &$:$ &$ x$	\\
$x\1+x\1+(x\1+x)$ &$:$ &$ x$	\\
$x\1+x\1+1\1+(x\1+x\1+2)$ &$:$ &$ x$	\\
$x\1+x\1+1\1+2$ &$:$ &$ x\1+x$	\\
$1\1+x\1+x$ &$:$ &$ x\1+x\1+1$	\\
$x\1+x\1+2\1+2$ &$:$ &$ x\1+x\1+1$	\\
$2\1+x\1+(x\1+x\1+2)$ &$:$ &$ 1$	\\
$x\1+x\1+1\1+x$ &$:$ &$ 1$	\\
$x\1+1$ &$:$ &$ 1\1+x$	\\
$2\1+x\1+2$ &$:$ &$ 1\1+x$	\\
$x\1+x\1+1\1+(x\1+x)$ &$:$ &$ 1\1+x$	\\
$x\1+x\1+2\1+(x\1+x\1+2)$ &$:$ &$ 1\1+x$	\\
$x\1+2$ &$:$ &$ 2\1+x$	\\
$1\1+x\1+1$ &$:$ &$ 2\1+x$	\\
$x\1+x\1+(x\1+x\1+2)$ &$:$ &$ 2\1+x$	\\
$x\1+x\1+1\1+(x\1+x\1+1)$ &$:$ &$ 2\1+x$	\\
$2\1+x\1+x$ &$:$ &$ x\1+x\1+2$	\\
$x\1+x\1+1\1+1$ &$:$ &$ x\1+x\1+2$	\\
$x\1+x\1+1\1+(1\1+x)$ &$:$ &$ 2$	\\
$x\1+x\1+2\1+x$ &$:$ &$ 2$	\\
\hline
$(0<x) \wedge (x<1)$ &$:$ &$ f$	\\
$(1<x) \wedge (x<1)$ &$:$ &$ f$	\\
$(1<x) \wedge (x<2)$ &$:$ &$ f$	\\
$(1<x) \wedge (1=x)$ &$:$ &$ f$	\\
$(x<1) \wedge (1=x)$ &$:$ &$ f$	\\
$(x\1+x<2) \wedge (1=x)$ &$:$ &$ f$	\\
$(2<x)$ &$:$ &$ f$	\\
$(x<0)$ &$:$ &$ f$	\\
$(x<x)$ &$:$ &$ f$	\\
$(x=1\1+x)$ &$:$ &$ f$	\\
$(x=2\1+x)$ &$:$ &$ f$	\\
$(1\1+x=2\1+x)$ &$:$ &$ f$	\\
$(x\1+x=x\1+x\1+1)$ &$:$ &$ f$	\\
$(x\1+x=x\1+x\1+2)$ &$:$ &$ f$	\\
$(x\1+x\1+1=x\1+x\1+2)$ &$:$ &$ f$	\\
$(0<x) \wedge (1<x)$ &$:$ &$ 1<x$	\\
$(0<x) \wedge (x\1+x<2)$ &$:$ &$ 1<x$	\\
$(1<x) \wedge (x\1+x<2)$ &$:$ &$ 1<x$	\\
$(1<x\1+x\1+1)$ &$:$ &$ 1<x$	\\
$(1\1+x<1)$ &$:$ &$ 1<x$	\\
$(1\1+x<x)$ &$:$ &$ 1<x$	\\
$(1\1+x<x\1+x)$ &$:$ &$ 1<x$	\\
$(1\1+x<x\1+x\1+1)$ &$:$ &$ 1<x$	\\
$(2\1+x<x\1+x\1+1)$ &$:$ &$ 1<x$	\\
\end{tabular}
\hfill
\begin{tabular}[t]{@{}rcl@{}}
$(x\1+x<x)$ &$:$ &$ 1<x$	\\
$(x\1+x\1+2<1)$ &$:$ &$ 1<x$	\\
$(x\1+x\1+2<x)$ &$:$ &$ 1<x$	\\
$(x\1+x\1+2<2\1+x)$ &$:$ &$ 1<x$	\\
$(x\1+x\1+2<x\1+x\1+1)$ &$:$ &$ 1<x$	\\
$(x=2)$ &$:$ &$ 1<x$	\\
$(1\1+x=0)$ &$:$ &$ 1<x$	\\
$(1\1+x=x\1+x\1+2)$ &$:$ &$ 1<x$	\\
$(2\1+x=1)$ &$:$ &$ 1<x$	\\
$(2\1+x=x\1+x)$ &$:$ &$ 1<x$	\\
$(x\1+x=1)$ &$:$ &$ 1<x$	\\
$(x\1+x=2\1+x)$ &$:$ &$ 1<x$	\\
$(x\1+x\1+1=2)$ &$:$ &$ 1<x$	\\
$(x\1+x\1+1=x)$ &$:$ &$ 1<x$	\\
$(x\1+x\1+2=0)$ &$:$ &$ 1<x$	\\
$(0<x) \wedge (x<2)$ &$:$ &$ 1=x$	\\
$(0<x) \wedge (1=x)$ &$:$ &$ 1=x$	\\
$(x<2) \wedge (1=x)$ &$:$ &$ 1=x$	\\
$(1<1\1+x)$ &$:$ &$ 1=x$	\\
$(1<x\1+x)$ &$:$ &$ 1=x$	\\
$(x<x\1+x)$ &$:$ &$ 1=x$	\\
$(2\1+x<1)$ &$:$ &$ 1=x$	\\
$(2\1+x<1\1+x)$ &$:$ &$ 1=x$	\\
$(2\1+x<x\1+x)$ &$:$ &$ 1=x$	\\
$(2\1+x<x\1+x\1+2)$ &$:$ &$ 1=x$	\\
$(x\1+x\1+1<1)$ &$:$ &$ 1=x$	\\
$(x\1+x\1+1<x)$ &$:$ &$ 1=x$	\\
$(x\1+x\1+1<1\1+x)$ &$:$ &$ 1=x$	\\
$(x\1+x\1+1<x\1+x)$ &$:$ &$ 1=x$	\\
$(x\1+x\1+2<1\1+x)$ &$:$ &$ 1=x$	\\
$(x=1)$ &$:$ &$ 1=x$	\\
$(1\1+x=2)$ &$:$ &$ 1=x$	\\
$(1\1+x=x\1+x)$ &$:$ &$ 1=x$	\\
$(2\1+x=0)$ &$:$ &$ 1=x$	\\
$(2\1+x=x\1+x\1+1)$ &$:$ &$ 1=x$	\\
$(x\1+x=2)$ &$:$ &$ 1=x$	\\
$(x\1+x\1+1=0)$ &$:$ &$ 1=x$	\\
$(x\1+x\1+1=2\1+x)$ &$:$ &$ 1=x$	\\
$(x\1+x\1+2=1)$ &$:$ &$ 1=x$	\\
$(x\1+x\1+2=x)$ &$:$ &$ 1=x$	\\
$(0<x\1+x)$ &$:$ &$ 0<x$	\\
$(2\1+x<2)$ &$:$ &$ 0<x$	\\
$(2\1+x<x)$ &$:$ &$ 0<x$	\\
$(x\1+x\1+2<2)$ &$:$ &$ 0<x$	\\
$(x\1+x\1+2<x\1+x)$ &$:$ &$ 0<x$	\\
$(0<x) \vee (1<x)$ &$:$ &$ 0<x$	\\
$(0<x) \vee (1=x)$ &$:$ &$ 0<x$	\\
\end{tabular}
\hfill
\begin{tabular}[t]{@{}rcl@{}}
$(1<x) \vee (1=x)$ &$:$ &$ 0<x$	\\
$(x<1) \wedge (x<2)$ &$:$ &$ x<1$	\\
$(x<1) \wedge (x\1+x<2)$ &$:$ &$ x<1$	\\
$(x<2) \wedge (x\1+x<2)$ &$:$ &$ x<1$	\\
$(1<2\1+x)$ &$:$ &$ x<1$	\\
$(1<x\1+x\1+2)$ &$:$ &$ x<1$	\\
$(x<2\1+x)$ &$:$ &$ x<1$	\\
$(x<x\1+x\1+1)$ &$:$ &$ x<1$	\\
$(x<x\1+x\1+2)$ &$:$ &$ x<1$	\\
$(1\1+x<x\1+x\1+2)$ &$:$ &$ x<1$	\\
$(x\1+x<1)$ &$:$ &$ x<1$	\\
$(x\1+x<1\1+x)$ &$:$ &$ x<1$	\\
$(x\1+x<2\1+x)$ &$:$ &$ x<1$	\\
$(x\1+x<x\1+x\1+2)$ &$:$ &$ x<1$	\\
$(x\1+x\1+1<2\1+x)$ &$:$ &$ x<1$	\\
$(x=0)$ &$:$ &$ x<1$	\\
$(1\1+x=1)$ &$:$ &$ x<1$	\\
$(1\1+x=x\1+x\1+1)$ &$:$ &$ x<1$	\\
$(2\1+x=2)$ &$:$ &$ x<1$	\\
$(2\1+x=x\1+x\1+2)$ &$:$ &$ x<1$	\\
$(x\1+x=0)$ &$:$ &$ x<1$	\\
$(x\1+x=x)$ &$:$ &$ x<1$	\\
$(x\1+x\1+1=1)$ &$:$ &$ x<1$	\\
$(x\1+x\1+2=2)$ &$:$ &$ x<1$	\\
$(0<2\1+x)$ &$:$ &$ x\1+x<2$	\\
$(0<x\1+x\1+1)$ &$:$ &$ x\1+x<2$	\\
$(1\1+x<2)$ &$:$ &$ x\1+x<2$	\\
$(1\1+x<2\1+x)$ &$:$ &$ x\1+x<2$	\\
$(x\1+x<x\1+x\1+1)$ &$:$ &$ x\1+x<2$	\\
$(1<x) \vee (x<1)$ &$:$ &$ x\1+x<2$	\\
$(1<x) \vee (x\1+x<2)$ &$:$ &$ x\1+x<2$	\\
$(x<1) \vee (x\1+x<2)$ &$:$ &$ x\1+x<2$	\\
$(0<1\1+x)$ &$:$ &$ x<2$	\\
$(0<x\1+x\1+2)$ &$:$ &$ x<2$	\\
$(x<1\1+x)$ &$:$ &$ x<2$	\\
$(x\1+x\1+1<2)$ &$:$ &$ x<2$	\\
$(x\1+x\1+1<x\1+x\1+2)$ &$:$ &$ x<2$	\\
$(x<1) \vee (x<2)$ &$:$ &$ x<2$	\\
$(x<1) \vee (1=x)$ &$:$ &$ x<2$	\\
$(x<2) \vee (1=x)$ &$:$ &$ x<2$	\\
$(x=x)$ &$:$ &$ t$	\\
$(0<x) \vee (x<1)$ &$:$ &$ t$	\\
$(0<x) \vee (x<2)$ &$:$ &$ t$	\\
$(0<x) \vee (x\1+x<2)$ &$:$ &$ t$	\\
$(1<x) \vee (x<2)$ &$:$ &$ t$	\\
$(x<2) \vee (x\1+x<2)$ &$:$ &$ t$	\\
$(1=x) \vee (x\1+x<2)$ &$:$ &$ t$	\\
\end{tabular}
\caption{Theory of $\N \;mod\; 3$ with $+$, $<$, $=$,
	and $Bool$ with $\wedge$ and $\vee$}
\label{Theory of N mod 3 with +, <, =, and Bool with wedge and vee}
\end{figure}

\EXAMPLE{ \eqd{z6}
Consider $(\N \;mod\; 3)$ with function $(+)$ 
and predicates $(<)$ and $(=)$ :

\hspace*{\fill}
\begin{tabular}[t]{@{}c|ccc@{}}
$+$ & $0$ & $1$ & $2$	\\
\hline
$0$ & $0$ & $1$ & $2$	\\
$1$ & $1$ & $2$ & $0$	\\
$\;2\;$ & $\;2\;$ & $\;0\;$ & $\;1\;$	\\
\end{tabular}
\hspace*{\fill}
\begin{tabular}[t]{@{}c|ccc@{}}
$<$ & $0$ & $1$ & $2$	\\
\hline
$0$ & $f$ & $t$ & $t$	\\
$1$ & $f$ & $f$ & $t$	\\
$\;2\;$ & $\;f\;$ & $\;f\;$ & $\;f\;$	\\
\end{tabular}
\hspace*{\fill}
\begin{tabular}[t]{@{}c|ccc@{}}
$=$ & $0$ & $1$ & $2$	\\
\hline
$0$ & $t$ & $f$ & $f$	\\
$1$ & $f$ & $t$ & $f$	\\
$\;2\;$ & $\;f\;$ & $\;f\;$ & $\;t\;$	\\
\end{tabular}
\hspace*{\fill}
\begin{tabular}[t]{@{}c|cc@{}}
$\wedge$ & $f$ & $t$	\\
\hline
$f$ & $f$ & $f$	\\
$\;t\;$ & $\;f\;$ & $\;t\;$	\\
\end{tabular}
\hspace*{\fill}
\begin{tabular}[t]{@{}c|cc@{}}
$\vee$ & $f$ & $t$	\\
\hline
$f$ & $f$ & $t$	\\
$\;t\;$ & $\;t\;$ & $\;t\;$	\\
\end{tabular}
\hspace*{\fill}

The deductive closure of the formulas given in Fig.~\ref{Theory
of N mod 3 with +, <, =, and Bool with wedge and vee} yields the
set of all valid formulas in one variable $x$ of type $\N
\;mod\; 3$ and with $\wedge$ and $\vee$ (and $\lra$)
as the only logical
junctors. Pure ground formulas and formulas that are instances
of others have been deleted, as well as variants modulo
commutativity or idempotency of $\wedge$ and $\vee$. 
Equations of type $(\N \;mod\;3)$ are listed first, followed
by equations of type $Bool$.
Note that the former are not redundant;
for example, reducing $x+1=1+x$ to $t$ requires the equations:

$$
\begin{tabular}[t]{@{}rcl@{\hspace*{0.5cm}}l@{}}
$x+1$ & $:$ & $1+x$ & and	\\
$1+x=1+x$ & $:$ & $t$, & which has been subsumed by	\\
$x=x$ & $:$ & $t$.	\\
\end{tabular}
$$

As another example, the formula $x=0 \vee x=1 \vee x=2$ reduces to $t$
via  the equations:

$$
\begin{tabular}[t]{@{}rcl@{\hspace*{0.5cm}}l@{}}
$(x=0)$ & $:$ & $(x<1)$,	\\
$(x=2)$ & $:$ & $(1<x)$,	\\
$(1<x) \vee (x<1)$ & $:$ & $x+x<2$, & and	\\
$(1=x) \vee (x+x<2)$ & $:$ & $t$.	\\
\end{tabular}
$$

}

\begin{remark}
Since $(=)$ and $(\lra)$ are required among the predicates and
junctors, respectively,
Cor.~\eqr{z5} does not apply to the set of universally valid Horn
formulas.
\end{remark}

\begin{remark}
By analogy with
the final remark in Sect.~\ref{Equational Theories in $n$
Variables}, we can estimate the complexity of computing all $N_1^N +
\ldots + N_n^N$ $rsg_V$ calls as $\O((N_1^N + \ldots + N_n^N) \cdot
g'^N \cdot N \cdot m' \cdot (m'+n))$,
where $m'$ denotes the maximal number of alternatives of a
sort definition's right--hand side, and
$g'$ is the maximum number of disjuncts of a sort definition's
right--hand side that start with the same function symbol.
Figure~\ref{Axiomatization Runtimes and Results} 
shows the runtimes and result statistics for some examples
(optimizations ``bdgsv'' used).
Column ``Dom'' shows the involved domains, ``F'' shows the functions,
``P'' the predicates, ``J'' the junctors, and ``V'' the number of
variables.
Columns ``S'', ``A'', ``M'', and ``I''
show the time for setting--up, anti--unifying, enumerating the minimal
terms, and removing the redundant instances, respectively.
``$>n$'' means running out of memory after $n$ seconds.
Column ``$\Sigma$'' shows the total time; column ``Sz'' shows the
number of non--redundant equations (without manual deletions).
\end{remark}

\begin{remark}
The above results refer to finite algebras only,
i.e.\ models with finite $\T /_E$.
It remains to be uninvestigated whether,
given an arbitrary equational theory,
we can find a finite ``test set''
$\T_{test} /_E \subset \T /_E$
of ground instances
such that an equation is universally valid iff it holds for all
instances from the test set.
\end{remark}

\begin{figure}
\begin{center}
\begin{tabular}{@{}|*{5}{l@{\hspace*{0.5cm}}}|*{5}{r@{\hspace*{0.5cm}}}|r|@{}}
\hline
Dom & F & P & J & V & S & A & M & I & $\Sigma$ & Sz	\\
\hline
$\N \;mod\; 2$ & $+$ &&& 2 & 4 & 0 & 1 & 5 & 6 & 27	\\
$\N \;mod\; 2$ & $+$ &&& 3 & 0 & 3881 & 0 & 11 & 3892 & 105	\\
$\N \;mod\; 2$ , $Bool$ & $+$ & $<$ , $=$ & $\wedge$ , $\vee$ & 2 , 0 
	& 0 & 18 & 0 & 41 & 59 & 452	\\
$\N \;mod\; 3$ , $Bool$ & $+$ & $<$ , $=$ & $\wedge$ , $\vee$ & 1 , 0 
	& 0 & 30 & 1 & 17 & 48 & 307	\\
$\N \;mod\; 3$ , $Bool$ & $+$ & $<$ , $=$ & $\wedge$ , $\vee$ & 2 , 0 
	& 20 & $>$143 &&&&	\\
\hline
\end{tabular}
\end{center}
\caption{Axiomatization Runtimes and Results}
\label{Axiomatization Runtimes and Results}
\end{figure}

\clearpage

\nocite{Burghardt.1993}
\nocite{Heinz.1994}

\bibliographystyle{alpha}
\bibliography{lit}

\addtolength{\textheight}{1.0cm}

\renewcommand{\baselinestretch}{0.753}

\clearpage

{\LARGE\bf Appendix}

\appendix

\section{PROLOG Source Code}
\label{PROLOG Source Code}

\footnotesize

\input{auw}

\end{document}

%% file: auw
%